\documentclass[sigconf, nonacm]{acmart}

\usepackage{microtype}
\usepackage{graphicx}
\usepackage{subfigure}
\usepackage{booktabs} 

\usepackage{hyperref}

\usepackage{amsmath}
\usepackage{mathtools}
\usepackage{amsthm}
\usepackage{algorithm}
\usepackage{algorithmic}

\usepackage{multirow}
\usepackage{tcolorbox} 
\tcbuselibrary{breakable, listings}
\usepackage{listings} 
\usepackage{tablefootnote}

\newtcblisting{codebox}{
  colback=gray!10,
  colframe=gray!30,
  boxrule=1pt,
  arc=3pt,
  boxsep=0pt,
  left=0.5pt, right=2pt, top=5pt, bottom=5pt,
  breakable,
  listing only,
  listing options={
    basicstyle=\ttfamily\footnotesize,
    breaklines=true,
    breakatwhitespace=true,
    frame=none,
    breakindent=0pt,
    gobble=4, 
    tabsize=4,
    columns=fullflexible, 
    resetmargins=true
  }
}

\begin{document}

\title{NetworkGames: Simulating Cooperation in Network Games with Personality-Driven LLM Agents}

\author{Xuan Qiu}
\affiliation{%
    \institution{The Hong Kong University of Science and Technology (Guangzhou)}
    \city{Guangzhou}
    \country{China}
}
\email{maxxuanqiu@hkust-gz.edu.cn}

\date{}

\begin{abstract}
    While Large Language Models (LLMs) have been extensively tested in dyadic game-theoretic scenarios, their collective behavior within complex network games remains surprisingly unexplored. To bridge this gap, we present \textbf{NetworkGames}, a framework connecting Generative Agents and Geometric Deep Learning. By formalizing social simulation as a message-passing process governed by LLM policies, we investigate how node heterogeneity (MBTI personalities) and network topology co-determine collective welfare. We instantiate a population of LLM agents, each endowed with a distinct personality from the MBTI taxonomy, and situate them in various network structures (e.g., small-world and scale-free). Through extensive simulations of the Iterated Prisoner's Dilemma, we first establish a baseline dyadic interaction matrix, revealing nuanced cooperative preferences between all 16 personality pairs. We then demonstrate that macro-level cooperative outcomes are not predictable from dyadic interactions alone; they are co-determined by the network's connectivity and the spatial distribution of personalities. For instance, we reveal two counter-intuitive phenomena. First, the \textbf{Paradox of Connectivity}: while Small-World shortcuts optimize information transport, they act as ``vectors of exploitation'' that destabilize cooperative clusters found in regular lattices. Second, \textbf{Hub Determinism}: in Scale-Free networks, the personality traits of highly connected nodes ($10\%$ of the population) strictly dictate macroscopic outcomes, enabling targeted interventions that double cooperation rates ($28\% \to 67\%$). These findings suggest that information efficiency and social resilience may be structurally opposing goals. We validate the robustness of these findings through extensive stress tests across multiple LLM architectures, scaled network sizes, varying random seeds, and comprehensive ablation studies. Our findings offer significant implications for designing healthier online social environments and forecasting collective behavior. We open-source our framework to facilitate research into the social physics of AI societies.
\end{abstract}

\maketitle

\section{Introduction}

The tension between individual rationality and collective welfare is a central theme in multi-agent systems. While the Prisoner's Dilemma (PD) serves as the canonical model for this tension, classical Game Theory often relies on the simplifying assumption of ``well-mixed'' populations. In such settings, each player typically interacts with all others, resulting in a mean-field approximation of social dynamics.

To model structurally realistic interactions, \textit{Network Games} emerged as a modern branch of game theory. Unlike classical game theory—where each player typically interacts with all others—network games assume that players only affect, and are affected by, their specific neighbors defined by a graph structure $\mathcal{G}(\mathcal{V}, \mathcal{E})$. Within this paradigm, the topological connectivity pattern dictates the evolution of strategies and the emergence of cooperation~\cite{nowak1992chaos, szabo2007evolutionary}.

Simultaneously, Large Language Models (LLMs) have surfaced as powerful proxies for human decision-making, offering ``generative agents'' capable of sophisticated reasoning and social emulation~\cite{park2023generativeagentsinteractivesimulacra}. However, existing research at the intersection of LLMs and Game Theory has predominantly focused on dyadic interactions or unstructured groups. The behavior of these advanced LLM agents within a formal network game framework—where interactions are explicitly constrained by graph topology—remains completely unexplored.

In this work, we introduce \textbf{NetworkGames} to bridge this gap. We draw architectural inspiration from \textbf{Graph Neural Networks (GNNs)}~\cite{scarselli2008graph, gilmer2017neural} to model both social influence and cognitive diversity. Just as a GNN node updates its latent state by aggregating feature vectors from its neighbors (Message Passing), our agents update their strategic stance by aggregating the ``social context'' ($\Omega$)—the observed behaviors of their local neighborhood. This formulation allows us to treat social simulation as a message-passing process where the update function is parameterized by an LLM policy $\pi_\theta$ rather than a shallow neural network.

Crucially, strictly homogeneous populations yield trivial dynamics. To capture the complex motivations of real-world social systems, we introduce \textit{node heterogeneity}. In geometric deep learning, each node begins with a distinct initial feature vector representing its unique prior distribution. To simulate this computationally, we use personality attributes to emulate the node heterogeneity of GNNs, where each node possesses distinct prior distributions (e.g., varying initial features). By endowing each agent with explicit semantic priors using Myers-Briggs (MBTI) profiles, we ensure varying initial features and diverse behavioral prior distributions across the graph.

To our knowledge, \textbf{this is the first work that studies LLM agents within a formal network game framework}, where interactions are explicitly constrained by graph topology. By situating populations of heterogeneous agents into diverse topological structures (Regular, Small-World, Scale-Free), we bridge micro-level psychological realism (personality as node features) with macro-level structural constraints (topology as message-passing edges), uncovering dynamics invisible to standard homogeneous simulations. Our contributions are:

\begin{itemize}
    \item \textbf{Pioneering Network Games with LLMs.} We provide the first systematic study answering the question: \textit{``How do LLM agents behave in network games?''} by exploring their collective dynamics across diverse graph topologies.
    \item \textbf{Graph-Theoretic LLM Framework.} We formalize the LLM agent as a stochastic policy function conditioned on topological signal aggregation, bridging GNN message-passing mechanics with computational social simulation.
    \item \textbf{The Paradox of Connectivity.} We demonstrate that Small-World shortcuts, while efficient for information transport, act as ``vectors of exploitation'' that destabilize cooperative clusters, challenging intuitions derived from simple imitation dynamics.
    \item \textbf{Hub Determinism.} In Scale-Free networks, we find that macro-welfare is strictly determined by the personality traits of the hub nodes ($<10\%$ of population), enabling targeted interventions that double cooperation rates ($28\% \to 67\%$).
\end{itemize}

\section{Related Work}

\textbf{LLMs in Game Theory.}
LLMs allow the study of social dynamics without the need to learn reward functions explicitly, as in Multi-Agent Reinforcement Learning (MARL)~\cite{lu2024llms}. Recent works have engaged LLMs in the Iterated Prisoner's Dilemma, demonstrating their ability to understand payoff matrices and recall history~\cite{akata2025playing, brookins2024playing}. However, these agents typically operate in topological vacuums (dyadic or fully connected), failing to capture the \textit{spatial} dimension of cooperation where local clustering allows cooperators to survive against defectors.

\textbf{Network Games and Topology.}
Seminal works in physics and economics established that spatial structure fundamentally alters evolutionary game dynamics~\cite{nowak1992chaos, santos2005scale}. For instance, scale-free networks usually promote cooperation due to hub protection. However, these traditional models rely on simplistic, predefined heuristics (e.g., Win-Stay-Lose-Shift). Our work replaces these heuristic scalars with high-dimensional semantic priors (LLM personalities), investigating whether ``rational'' LLM agents obey the same topological laws as simple agents.

\textbf{Social Simulation as Message Passing.}
Our methodology aligns with the Message Passing Neural Network (MPNN) framework used in geometric deep learning~\cite{gilmer2017neural}. In GNNs, a node's representation is iteratively updated by aggregating messages from neighbors. Similarly, in computational social science, an agent's belief state is a function of its neighbors' actions~\cite{centola2010spread}. By explicitly formalizing the neighborhood observation context ($\Omega_i$) as an aggregation function, we provide a mathematical link between Graph ML architectures and Agent-Based Modeling (ABM).

\textbf{Personality in AI.}
Integrating psychological frameworks like the Big Five or MBTI into AI agents improves behavioral realism~\cite{jiang2023personallm, la2025open}. While recent studies have introduced personality into dyadic games~\cite{Zeng2025DynamicPersonality}, they treat personality as a standalone variable. We extend this by treating personality as a \textit{node feature} $\mathbf{x}_v$ within a graph, allowing us to analyze the interaction between node heterogeneity (personality) and edge structure (topology).
\section{Methodology}

We formulate the interaction as a multi-agent game played over a graph, where agent behaviors are governed by Large Language Models conditioned on distinct personality profiles.

\subsection{Networked Game Formulation}
Consider a population of $N$ agents represented by nodes in an undirected graph $\mathcal{G}=(\mathcal{V}, \mathcal{E})$. At each time step $t$, every agent $i \in \mathcal{V}$ engages in a pairwise Iterated Prisoner's Dilemma (IPD) with each neighbor $j \in \mathcal{N}(i)$.

The utility $U_i^{(t)}$ of agent $i$ at round $t$ is the aggregate of payoffs from all pairwise interactions:
\begin{equation}
    U_i^{(t)} = \sum_{j \in \mathcal{N}(i)} \mathcal{M}(a_{i,j}^{(t)}, a_{j,i}^{(t)})
\end{equation}
where $a_{i,j}^{(t)} \in \{C, D\}$ denotes the action (Cooperate/Defect) of agent $i$ against neighbor $j$, and $\mathcal{M}$ is the standard PD payoff matrix satisfying $T > R > P > S$. In our experiments, we set $T=5, R=3, P=1, S=0$.

\subsection{LLM as a Personality-Conditioned Policy}
We abstract the Large Language Model as a stochastic policy function $\pi_\theta$. Unlike traditional agents with fixed strategies (e.g., Tit-for-Tat), our agents derive their policy from a static intrinsic personality and a dynamic context.

For an interaction between agent $i$ and neighbor $j$ at round $t$, the action is sampled as:
\begin{equation}
    a_{i,j}^{(t)} \sim \pi_{\theta}(\cdot \mid \mathbf{p}_i, \mathbf{h}_{ij}^{(t)}, \Omega_i^{(t)})
\end{equation}
This policy is conditioned on three components:
\begin{enumerate}
    \item \textbf{Intrinsic Personality ($\mathbf{p}_i$):} A static prompt embedding derived from the Myers-Briggs Type Indicator (MBTI). While the Big Five model is common in psychology, its continuous nature introduces ambiguity in prompt engineering. We select MBTI for its discrete binary taxonomy, which may be more controllable under prompt conditioning. This discreteness enables unambiguous ablation studies and precise interventions (e.g., flipping the T/F bit). We will discuss using Big Five in the appendix.
    \item \textbf{Dyadic History ($\mathbf{h}_{ij}^{(t)}$):} The sequential memory of interactions between $i$ and $j$ from $t-k$ to $t-1$, enabling reciprocity mechanisms.
    \item \textbf{Local Social Context ($\Omega_i^{(t)}$):} A formulation of the GNN intuition applied to social simulation. $\Omega_i^{(t)}$ aggregates the visible actions of the neighborhood $\mathcal{N}(i)$ from step $t-1$:
          \begin{equation}
              \Omega_i^{(t)} = \phi \left( \{ a_{k, \cdot}^{(t-1)} \mid k \in \mathcal{N}(i) \} \right)
          \end{equation}
          This term allows agents to perceive group norms (e.g., ``75\% of my neighbors defected last round''), bridging the gap between local dyadic games and global network topology.
\end{enumerate}

\subsection{Simulation Algorithm}
The NetworkGames simulation loop is formally defined in Algorithm~\ref{alg:simulation}. The process proceeds in discrete time steps $t$. In the update phase of each round, agents first aggregate social context $\Omega$ from the previous round, then independently sample actions from their policy $\pi_\theta$.

\begin{algorithm}[hbt]
    \caption{NetworkGames Simulation Loop}
    \label{alg:simulation}
    \begin{algorithmic}
        \STATE {\bfseries Input:} Graph $\mathcal{G}(\mathcal{V},\mathcal{E})$, Personality Profiles $\mathcal{P} = \{\mathbf{p}_i\}_{i \in \mathcal{V}}$, Horizon $T$
        \STATE {\bfseries Initialize:} History $\mathbf{h}_{ij}^{(0)} \leftarrow \emptyset$ for all $(i,j) \in \mathcal{E}$
        \FOR{$t=1$ {\bfseries to} $T$}
        \STATE \textcolor{gray}{// Phase 1: Observation Aggregation}
        \FOR{each agent $i \in \mathcal{V}$}
        \STATE Observe local context: $\Omega_i^{(t)} \leftarrow \phi(\{a_{k, \cdot}^{(t-1)} : k \in \mathcal{N}(i)\})$
        \ENDFOR

        \STATE \textcolor{gray}{// Phase 2: Action Sampling (Policy Execution)}
        \FOR{each agent $i \in \mathcal{V}$}
        \FOR{each neighbor $j \in \mathcal{N}(i)$}
        \STATE Construct State String: $\mathcal{S}_{i,j}^{(t)} \leftarrow \text{Prompt}(\mathbf{p}_i, \mathbf{h}_{ij}^{(t)}, \Omega_i^{(t)})$
        \STATE Sample Action: $a_{i,j}^{(t)} \sim \pi_{\theta}(\cdot \mid \mathcal{S}_{i,j}^{(t)})$
        \ENDFOR
        \ENDFOR

        \STATE \textcolor{gray}{// Phase 3: Environment Update}
        \FOR{each edge $(i,j) \in \mathcal{E}$}
        \STATE Append History: $\mathbf{h}_{ij}^{(t+1)} \leftarrow \mathbf{h}_{ij}^{(t)} \cup \{(a_{i,j}^{(t)}, a_{j,i}^{(t)})\}$
        \STATE Compute Utilities: $U_i \leftarrow U_i + \mathcal{M}(a_{i,j}^{(t)}, a_{j,i}^{(t)})$
        \ENDFOR
        \ENDFOR
    \end{algorithmic}
\end{algorithm}

\subsection{Experimental Design}

We design two experimental settings to decouple the effects of personality compatibility from network structure.

\textbf{Setup 1: Dyadic Baselines (Micro-Level).}
We first derive the ``atomic'' interaction properties of specific personalities. We conduct exhaustive pairwise games ($N=2$) across all $16 \times 16$ MBTI combinations. Each pair plays a 20-round IPD, repeated over 10 independent trials ($136 \text{ pairings} \times 20 \text{ rounds} \times 10 \text{ seeds}$). In this setting, the network context $\Omega$ is null.

\textbf{Setup 2: Network Dynamics (Macro-Level).}
We deploy populations of $N=50$ agents onto specific graph topologies to observe emergent cooperation. We vary two structural variables:
\begin{itemize}
    \item \textbf{Topology:} We test Regular Lattices ($k=4$), Small-World networks (Watts-Strogatz, varied $p$), and Scale-Free networks (Barabási-Albert, $m=2$).
    \item \textbf{Hub Composition:} In Scale-Free networks, we intervene on the specific personality assignments of high-degree nodes (``hubs'') to test their influence on macroscopic welfare.
\end{itemize}
Specific prompts and protocol details are detailed in Appendix~\ref{app:prompt_details}.

\section{Validation of Behavioral Priors}
\label{sec:validation}
Before investigating complex network dynamics, we must validate that our policy $\pi_\theta$ correctly internalizes the intended personality traits $\mathbf{p}_i$. We treat this as a probe on the alignment between the natural language prompt and the agent's decision logic.

We administered a set of controlled binary-choice probes to all 16 instantiated agents (methodology details in Appendix~\ref{app:personality_probe}). As shown in Table~\ref{tab:probe_accuracy}, the agents demonstrate high fidelity to their assigned profiles ($ >90\%$ average consistency). Crucially, the \textbf{Thinking vs. Feeling (T/F)} dimension—which governs the trade-off between objective utility maximization and interpersonal harmony—achieved \textbf{95.6\%} consistency. This high accuracy confirms that the observed game behaviors stem from stable intrinsic priors rather than hallucinations or randomness.

\begin{table}[htbp]
    \caption{Agent Consistency with MBTI Priors}
    \label{tab:probe_accuracy}
    \centering
    \begin{small}
        \begin{tabular}{lc}
            \toprule
            \textbf{MBTI Dimension}             & \textbf{Probe Consistency} \\
            \midrule
            Extraversion vs. Introversion (E/I) & 90.6\%                     \\
            Sensing vs. Intuition (S/N)         & 80.6\%                     \\
            Thinking vs. Feeling (T/F)          & 95.6\%                     \\
            Judging vs. Perceiving (J/P)        & 88.1\%                     \\
            \bottomrule
        \end{tabular}
    \end{small}
\end{table}

\section{Experimental Results}
\label{sec:experiments}
We test different LLMs to understand their impact on personality behavior (Appendix~\ref{app:different_llm}). It has been observed that certain models tend to polarize personality stereotypes, while others tend to homogenize them. We choose LLaMA-3.2-3B\footnote{Deployed via Ollama with parameters: temperature=0.8, top\_p=0.9.} as the primary cognitive backbone for agents, as it demonstrates balanced performance across personality types. Our experimental pipeline proceeds as follows: we first validate that agents reliably internalize their assigned personalities through controlled probes (Section~\ref{sec:validation}), then systematically analyze three complementary perspectives: micro-level dyadic interactions to isolate personality effects (Section~\ref{sec:micro}), macro-level topological dynamics to reveal emergent cooperation patterns (Section~\ref{sec:topology}), and hub-based structural interventions to demonstrate targeted control of collective outcomes (Section~\ref{sec:hubs}).

\subsection{Micro-Dynamics: The Heterogeneity of Dyad Interactions}
\label{sec:micro}
We first analyze the ``atomic'' interactions by playing exhaustive pairwise IPD games (20 rounds, 10 trials) across all $16 \times 16$ personality pairs. This forms the basis for understanding how local dyadic outcomes aggregate into global network states.

\begin{figure}[htbp]
    \centering
    \includegraphics[width=1\linewidth]{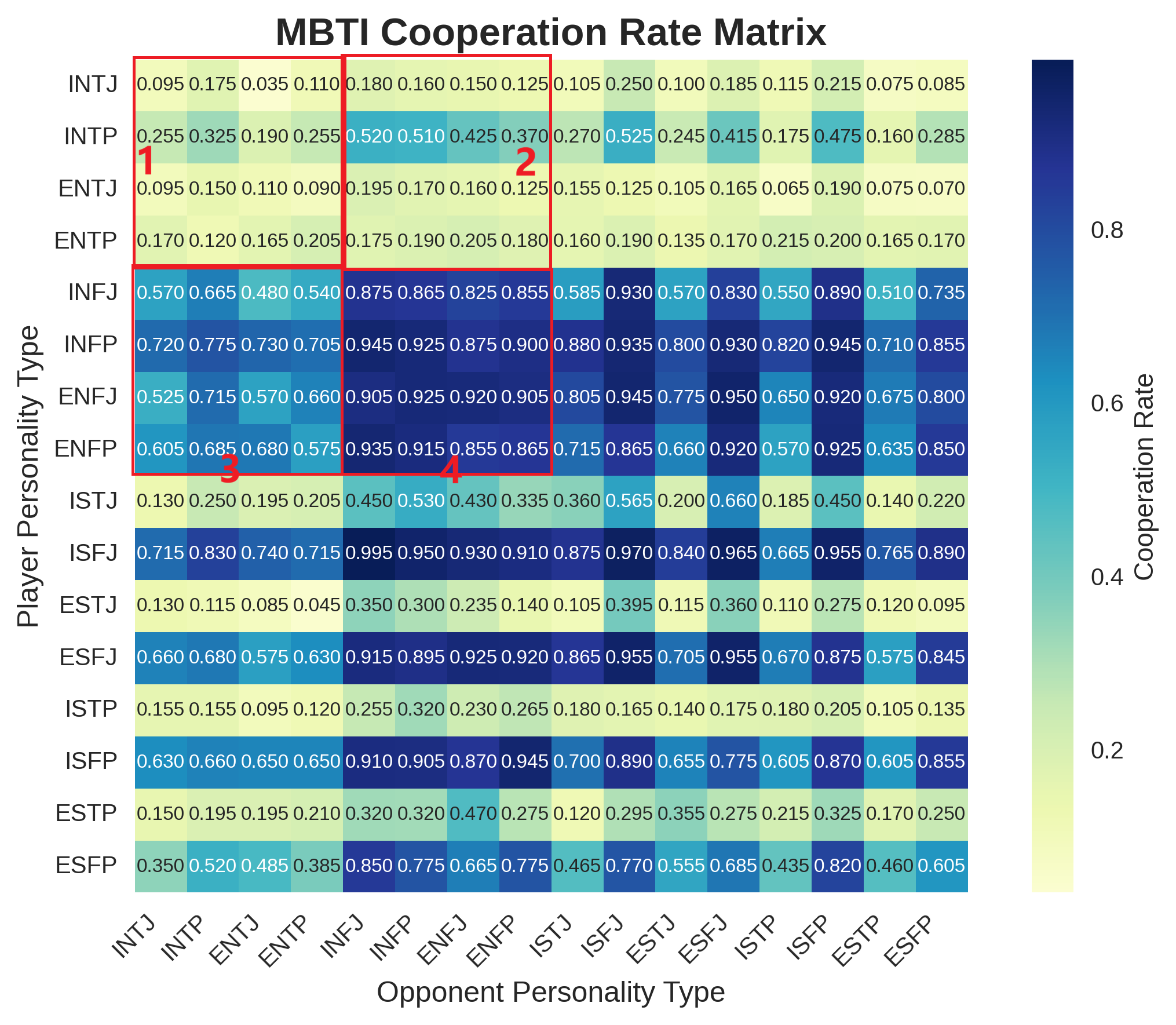}
    \includegraphics[width=1\linewidth]{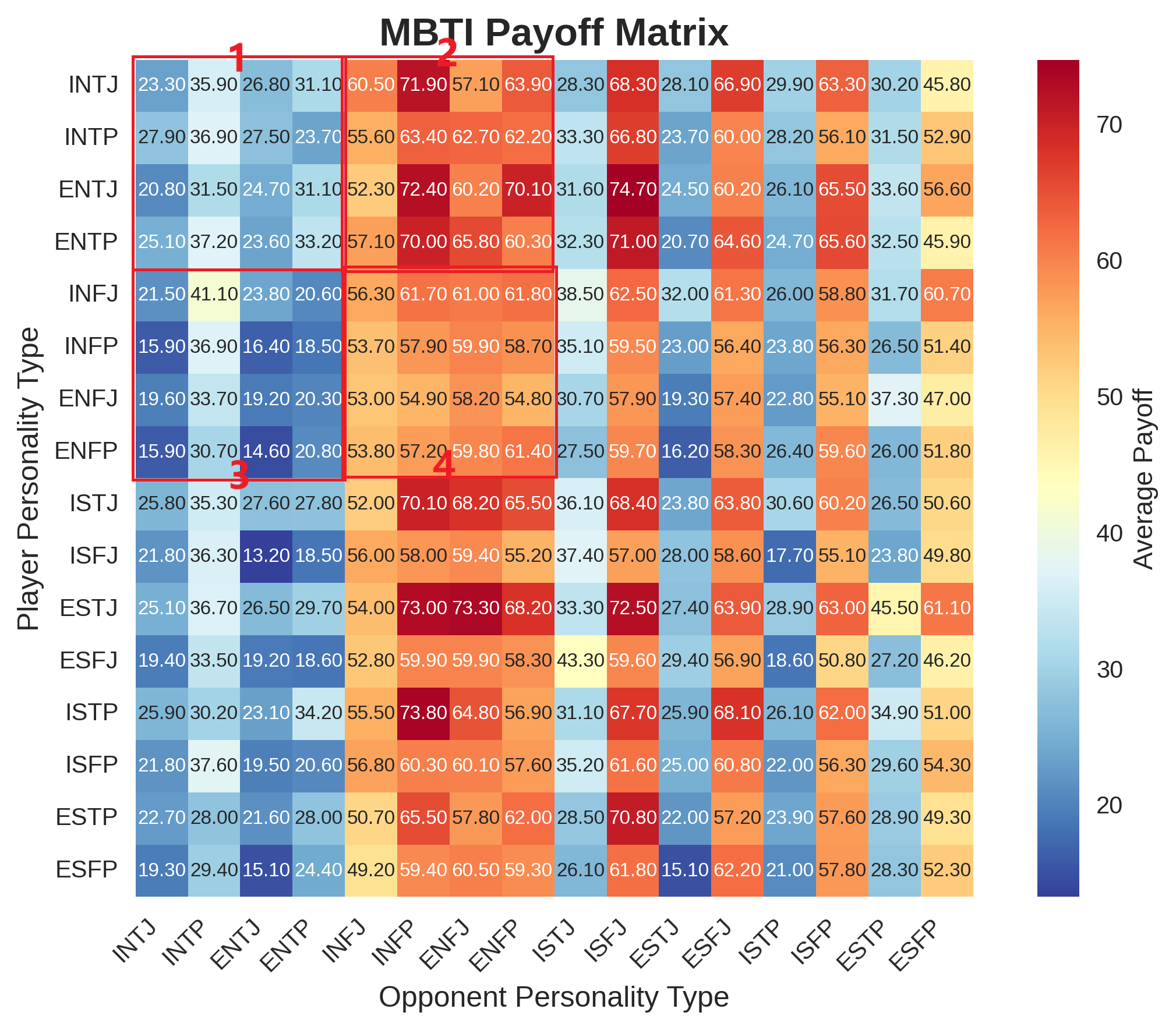}
    \caption{Heatmap of cooperation rates (top) and payoffs (bottom) for all $16\times16$ pairings. Entry $(i,j)$ represents the outcome for personality type $i$ when playing against type $j$, whereas entry $(j,i)$ represents the outcome for type $j$ in the same game pairing. Thus, each row represents how a particular personality performs against all other types, while each column indicates how well others perform when facing that personality type.}
    \Description{Cooperation and Payoff Heatmaps}
    \label{fig:pair_game_heatmaps}
\end{figure}

\textbf{The T-F Asymmetry.}
Figure~\ref{fig:pair_game_heatmaps} reveals a stark behavioral schism. In the cooperation heatmap, the rows corresponding to Feeling (F) types display significantly higher cooperation tendencies. In the payoff heatmap, the columns corresponding to F-types tend to be redder, indicating that opponents consistently achieve higher gains when interacting with F-types compared to Thinking (T) types, identifying F-types as ``sources of value'' in the ecosystem.

Table~\ref{tab:dimension_stats} quantifies this asymmetry using Welch's t-test. The T/F dimension exhibits the most significant differences in both cooperation rates and payoffs ($p < 0.001$), also Introverts (I) show a statistically significantly higher cooperation rate than Extroverts (E) ($p < 0.001$), though the magnitude of the difference is much smaller (51.8\% vs 46.2\%) compared to the T-F dimension.

\begin{figure}[htbp]
    \centering
    \begin{minipage}[b]{0.49\linewidth}
        \centering
        \includegraphics[width=\linewidth]{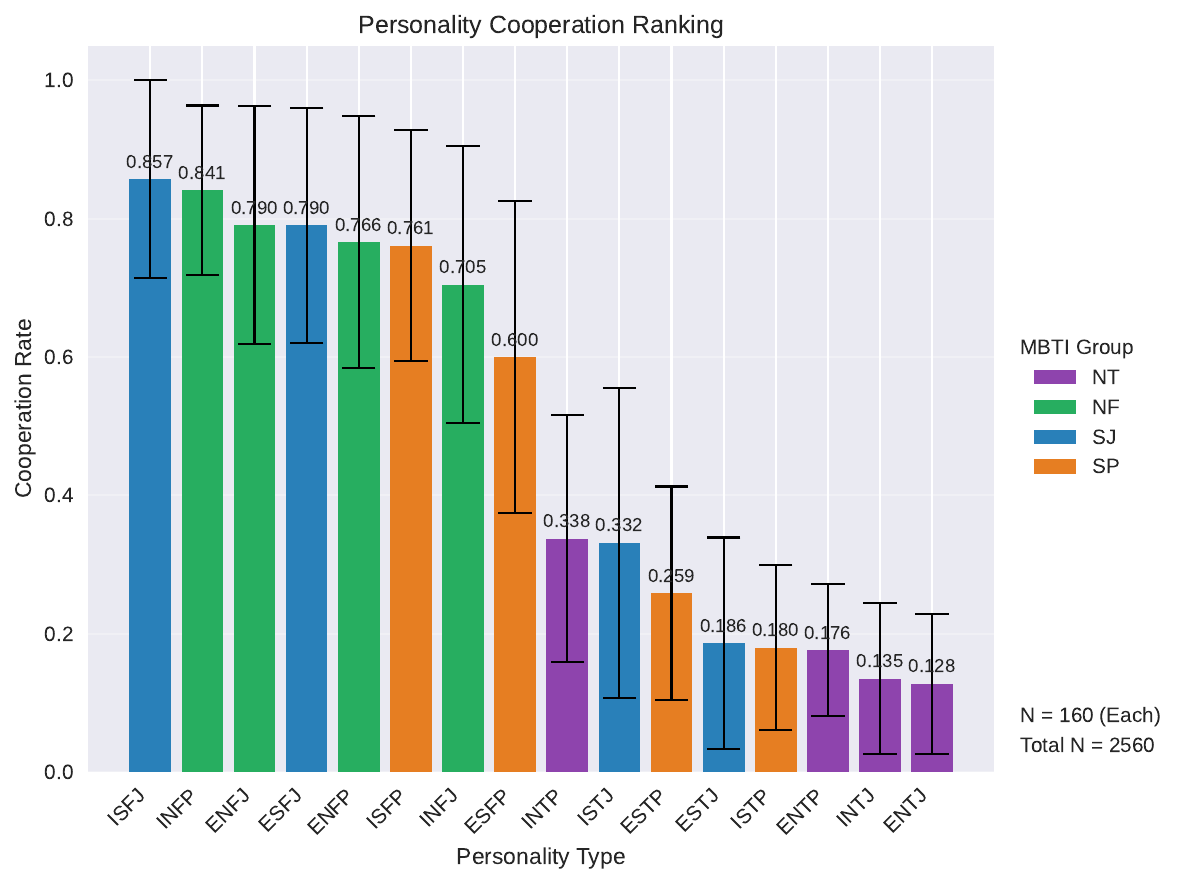}
    \end{minipage}\hfill
    \begin{minipage}[b]{0.49\linewidth}
        \centering
        \includegraphics[width=\linewidth]{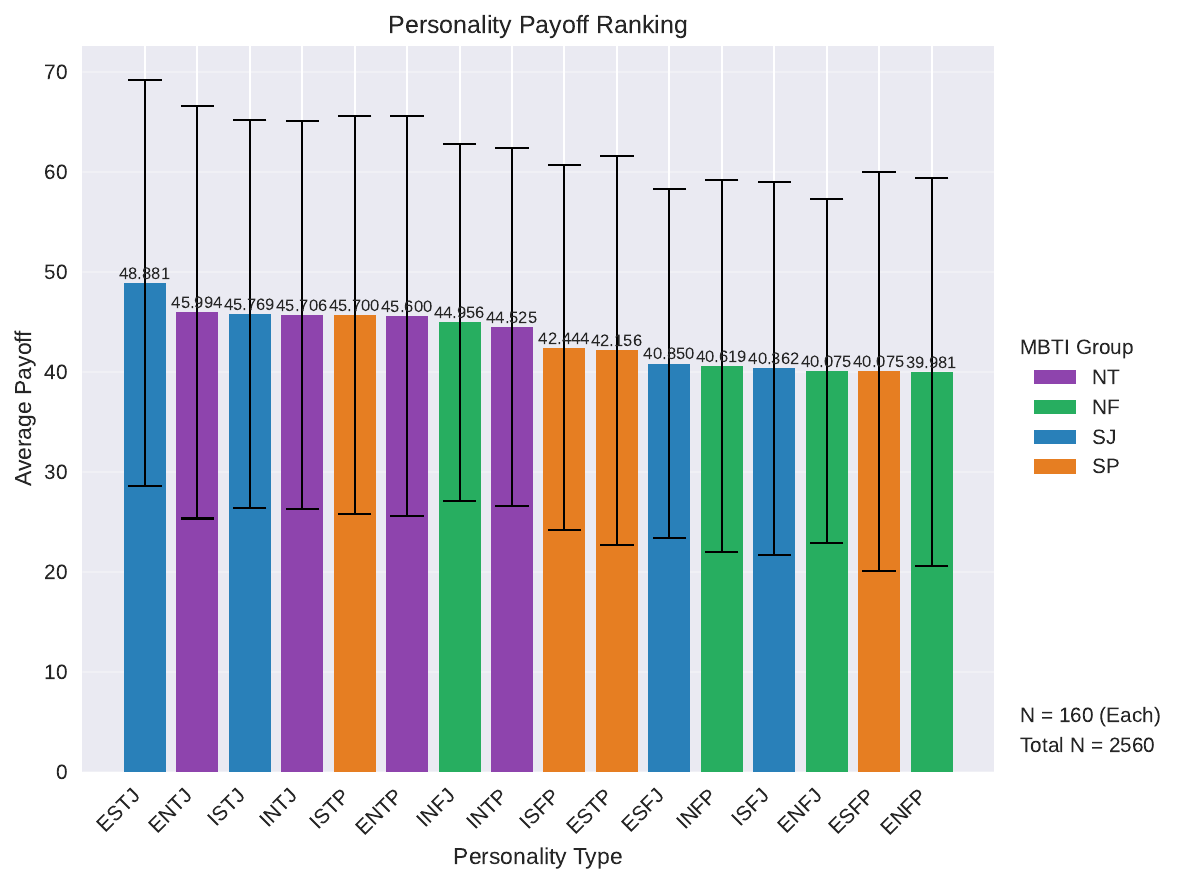}
    \end{minipage}
    \caption{Ranking of 16 Personalities by Average Cooperation Rate (left) and Total Payoff (right) in Section~\ref{sec:micro}. Feeling (F) types exhibit significantly higher cooperation rates compared to Thinking (T) types, with the top 8 positions exclusively occupied by F-types. However, this disparity diminishes in terms of payoffs, where T-types often rank higher.}
    \Description{Personality Rankings}
    \label{fig:pair_game_ranking}
\end{figure}

Figure~\ref{fig:pair_game_ranking} presents the ranking of the 16 personality types by average cooperation rate and total payoff.  F-types rank consistently higher in cooperation, while T-types rank relatively higher in payoffs, reflecting their exploitative strategies. Kruskal-Wallis tests indicate significant differences in both metrics (Cooperation: $H=1859.06$, $p<0.001$; Payoff: $H=59.32$, $p<0.001$).

We highlight four specific interaction zones in the matrices to illustrate dynamics between Analysts (NT) and Diplomats (NF):
\begin{enumerate}
    \item \textbf{Box 1 (NT vs. NT):} Characterized by mutual defection. Both cooperation rates and payoffs are low, reflecting the Nash equilibrium of the single-shot game.
    \item \textbf{Box 4 (NF vs. NF):} Characterized by mutual cooperation. Both parties achieve high stable payoffs.
    \item \textbf{Box 2 \& 3 (NT vs. NF):} These regions display strong asymmetry. NT agents (Box 2) exploit the cooperative nature of NF agents, securing ``Temptation'' payoffs (the highest possible outcome). Conversely, NF agents (Box 3) persist in cooperation and receive the ``Sucker's payoff.''
\end{enumerate}

\begin{table}[htbp]
    \caption{Welch's t-test Analysis of Cooperation Rates and Payoffs by MBTI Dimension}
    \label{tab:dimension_stats}
    \centering
    \resizebox{\columnwidth}{!}{
        \begin{tabular}{llccc}
            \toprule
            \textbf{Dimension}       & \textbf{Metric} & \textbf{Group 1 (Mean$\pm$Std)} & \textbf{Group 2 (Mean$\pm$Std)} & \textbf{$p$-value} \\
            \midrule
            \multirow{2}{*}{E vs. I} & Coop. Rate      & \textbf{E}: $0.462\pm0.325$     & \textbf{I}: $0.518\pm0.326$     & $\mathbf{< 0.001}$ \\
                                     & Payoff          & \textbf{E}: $42.95\pm19.59$     & \textbf{I}: $43.76\pm18.89$     & $0.288$            \\
            \midrule
            \multirow{2}{*}{S vs. N} & Coop. Rate      & \textbf{S}: $0.496\pm0.319$     & \textbf{N}: $0.485\pm0.334$     & $0.403$            \\
                                     & Payoff          & \textbf{S}: $43.28\pm19.41$     & \textbf{N}: $43.43\pm19.08$     & $0.841$            \\
            \midrule
            \multirow{2}{*}{T vs. F} & Coop. Rate      & \textbf{T}: $0.217\pm0.166$     & \textbf{F}: $0.764\pm0.191$     & $\mathbf{< 0.001}$ \\
                                     & Payoff          & \textbf{T}: $45.54\pm19.72$     & \textbf{F}: $41.17\pm18.51$     & $\mathbf{< 0.001}$ \\
            \midrule
            \multirow{2}{*}{J vs. P} & Coop. Rate      & \textbf{J}: $0.490\pm0.345$     & \textbf{P}: $0.490\pm0.308$     & $0.978$            \\
                                     & Payoff          & \textbf{J}: $44.07\pm19.15$     & \textbf{P}: $42.64\pm19.32$     & $0.059$            \\
            \bottomrule
        \end{tabular}
    }
\end{table}

\textbf{Comparison with Classical Baselines.}
We benchmarked our agents against standard game-theoretic strategies (Table~\ref{tab:baseline_comparison}). While the aggregate population behaves similarly to a Random agent ($0.49$ vs $0.50$ cooperation), this average masks profound heterogeneity. The high standard deviation for Personality Agents ($17.6 vs ~2.6$) confirms that our population contains both altruists and rational egoists, satisfying the requirement for meaningful diverse population simulation.

\begin{table}[htbp]
    \caption{Performance Comparison: Classical Baselines (Self-Play) vs. Personality Agents (Overall $16\times16$ Matrix)}
    \label{tab:baseline_comparison}
    \centering
    \small
    \resizebox{\columnwidth}{!}{%
        \begin{tabular}{lcccc}
            \toprule
            \textbf{Strategy}                                                                                                                                                       & \textbf{Coop. Rate} & \textbf{Std} & \textbf{Payoff} & \textbf{Std} \\
            \midrule
            All-C / TFT\tablefootnote{Tit-for-Tat effectively collapses into Always Cooperate in self-play, as both agents initiate with cooperation and continuously reciprocate.} & 1.000               & 0.000        & 60.000          & 0.000        \\
            All-D                                                                                                                                                                   & 0.000               & 0.000        & 20.000          & 0.000        \\
            Random                                                                                                                                                                  & 0.502               & 0.034        & 45.096          & 2.602        \\
            Personality Agents                                                                                                                                                      & 0.490               & 0.306        & 43.356          & 17.597       \\
            \bottomrule
        \end{tabular}%
    }
\end{table}

\subsection{Macro-Dynamics I: The Paradox of Connectivity}
\label{sec:topology}
We next placed populations of $N=50$ agents into structured topologies to test how connectivity patterns influence the survival of cooperation while maintaining uniform personality distribution:

\begin{itemize}
    \item \textbf{Regular Network:} Regular graph with degree $k=4$ (100 total edges).
    \item \textbf{Small-World ($p=0.1$):} Watts-Strogatz model with low rewiring probability ($p=0.1$).
    \item \textbf{Small-World ($p=0.5$):} Watts-Strogatz model with high rewiring probability ($p=0.5$).
    \item \textbf{Scale-Free Network:} Barabási-Albert model with $m=2$ (96 total edges).
\end{itemize}

\textbf{Result: Shortcuts Destabilize Cooperation.}
The experimental outcomes revealed distinct patterns in cooperation sustainability across network topologies (Table~\ref{tab:topology_results}). We observed a trend where \textbf{increasing randomness in network connections (higher re-wiring probability $p$) appears to correlate with a decline in cooperation}.

The Regular network consistently maintained the highest cooperation rates. In our Small-World models, cooperation rates appeared to decrease as the rewiring probability increased: the $p=0.1$ variant achieved slightly lower cooperation than regular networks, while the more random $p=0.5$ variant saw a further drop. This trend was consistent across extended simulation lengths (30, 50, and 100 rounds). The Scale-Free network generally occupied a middle ground in terms of performance.

\begin{figure}[htbp]
    \centering
    \includegraphics[width=1\linewidth]{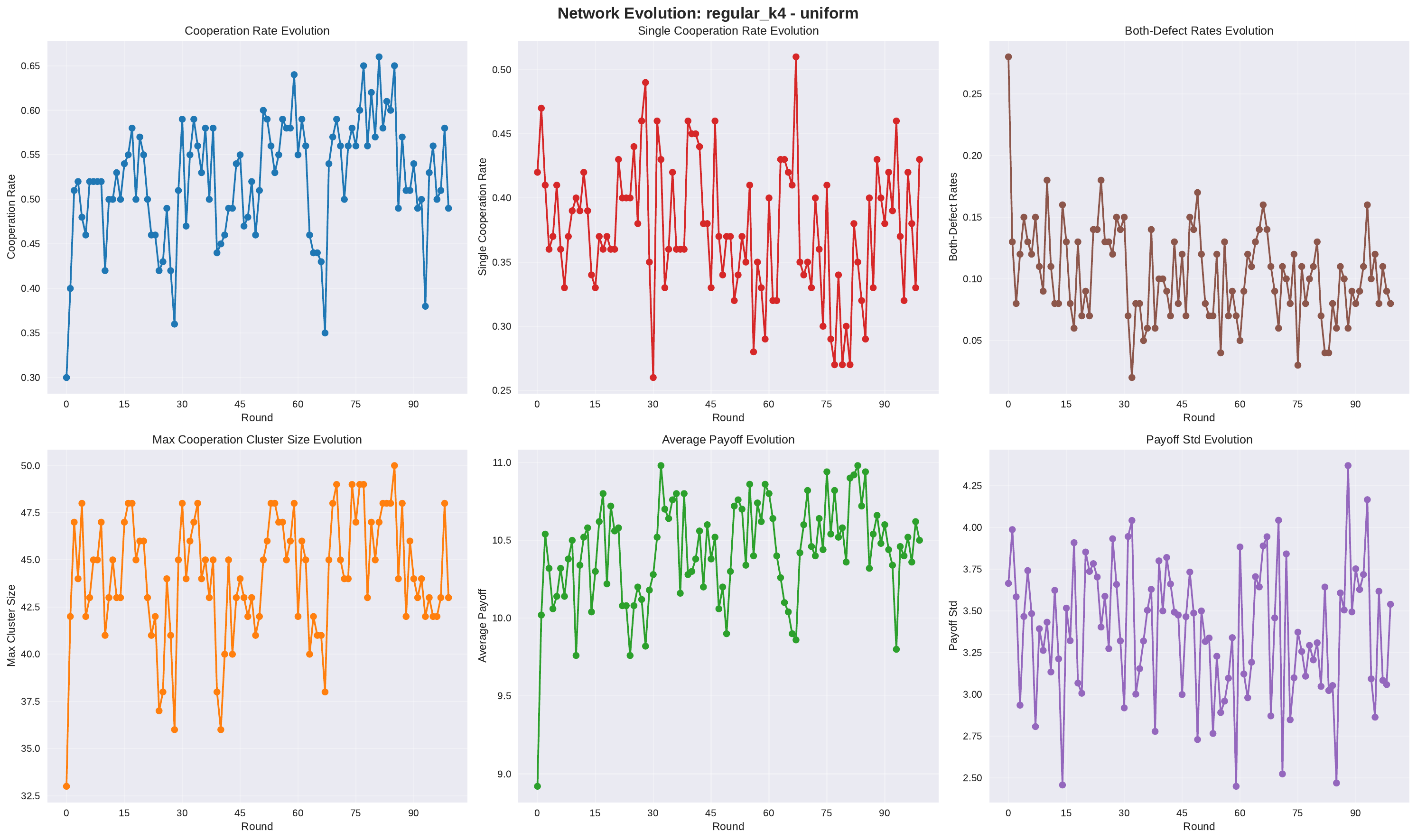}
    \includegraphics[width=1\linewidth]{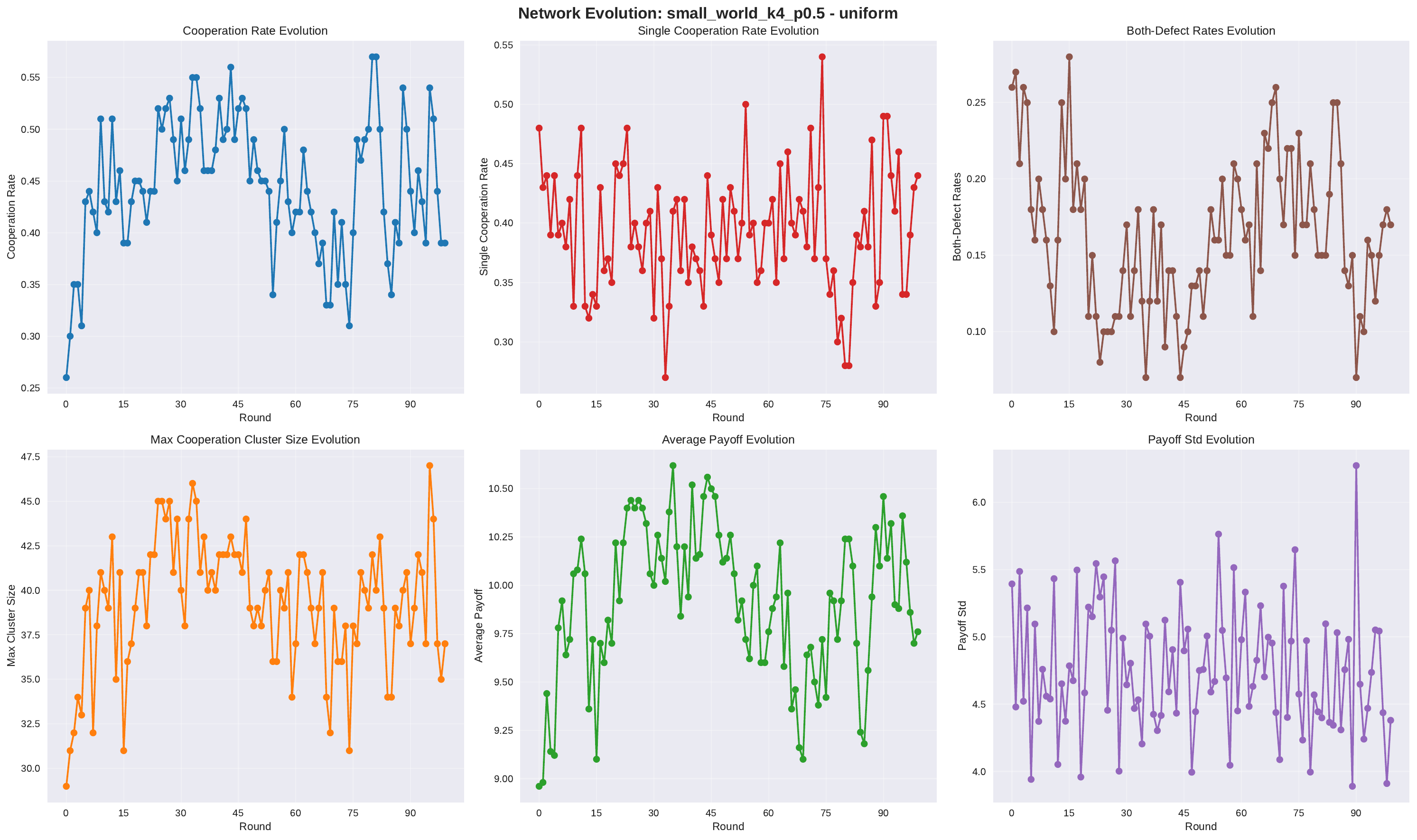}
    \caption{Temporal evolution of network statistics over 100 rounds in Regular and Small-World ($p=0.5$) networks in Section~\ref{sec:topology}. In the Regular network, the `Both Defect' rate (brown) quickly stabilizes at a low level. Conversely, the Small-World network exhibits a higher rate of asymmetric `Single Cooperation' (red) and volatile `Both Cooperation' (blue) and `Both Defect' rates (brown), where defection often rebounds. The maximum connected cluster size of cooperators (orange) and average payoff (green) fluctuate synchronously with mutual cooperation, while payoff variance (purple) remains relatively stable. \\
        Note that the `Avg. Coop. Rate' in Table~\ref{tab:topology_results} is calculated as the proportion of `Both Cooperation' (blue) edges. The observed increase in average cooperation rates over longer simulations (Table~\ref{tab:topology_results}) results from the initial adjustment phase: cooperation is low in the first few rounds before stabilizing at a higher level (as seen in the trajectory of the blue line), raising the overall mean as rounds accumulate.}
    \Description{Temporal evolution of network statistics in Regular and Small-World networks.}
    \label{fig:r&sw_evolution}
\end{figure}

\textbf{Temporal Dynamics Analysis.} To investigate these dynamics, we traced the evolution of edge types in the regular and small-world($p=0.5$) network (Figure~\ref{fig:r&sw_evolution}). Edges were classified into three categories: \textit{cooperation} (both nodes cooperate), \textit{single cooperation} (one cooperates, one defects), and \textit{both defect} (mutual defection).

The temporal analysis shows that in the Regular network, mutual defection rates tend to settle at a lower level after a few rounds. In contrast, the Small-World network exhibits a relatively high rate of asymmetric single cooperation edges. The rates of mutual cooperation and defection appear more volatile, suggesting that the network structure might struggle to support the sustained formation of stable cooperative clusters found in regular lattices.

\begin{figure}[htb!]
    \centering
    \includegraphics[width=1\linewidth]{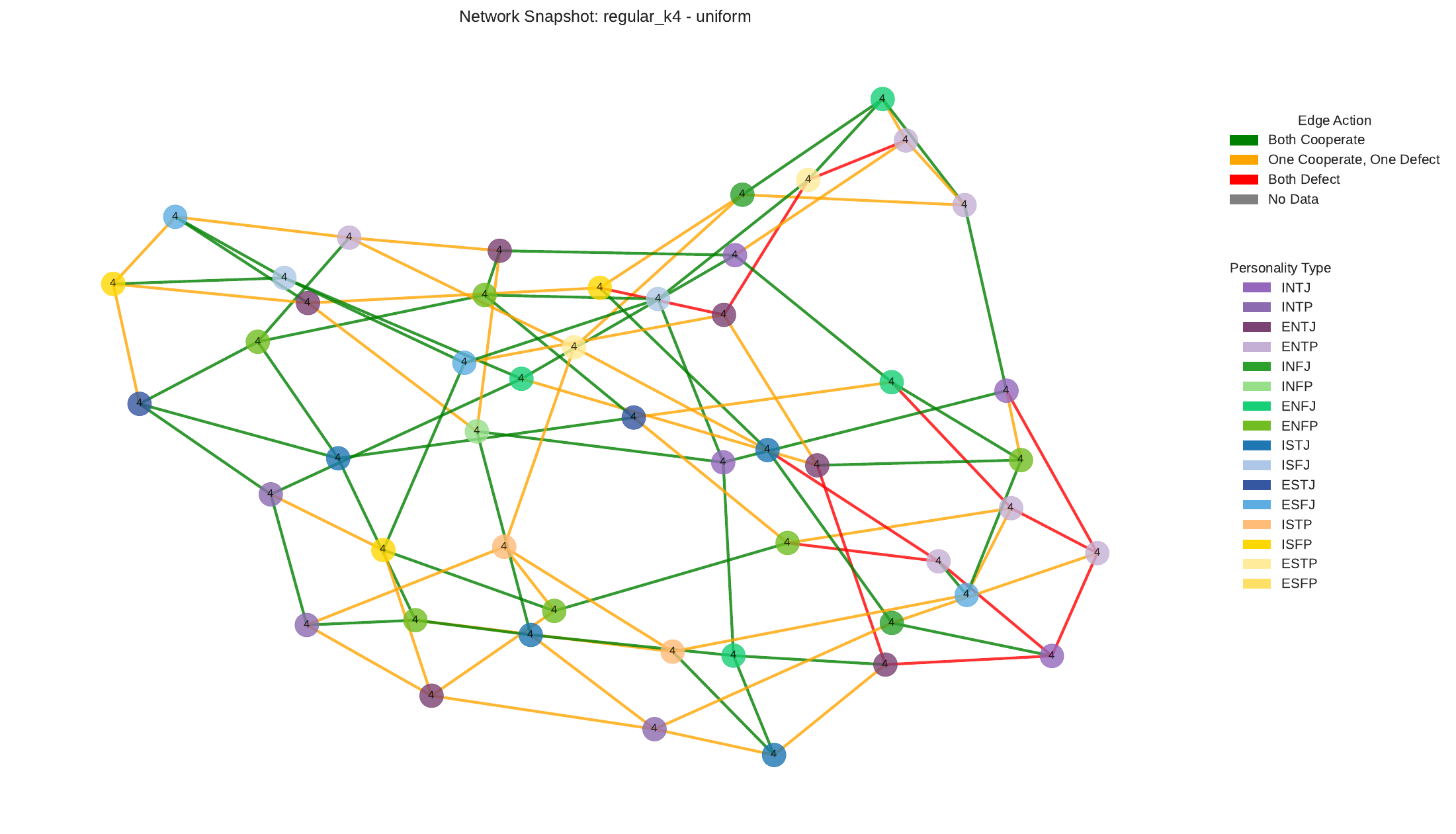}
    \includegraphics[width=1\linewidth]{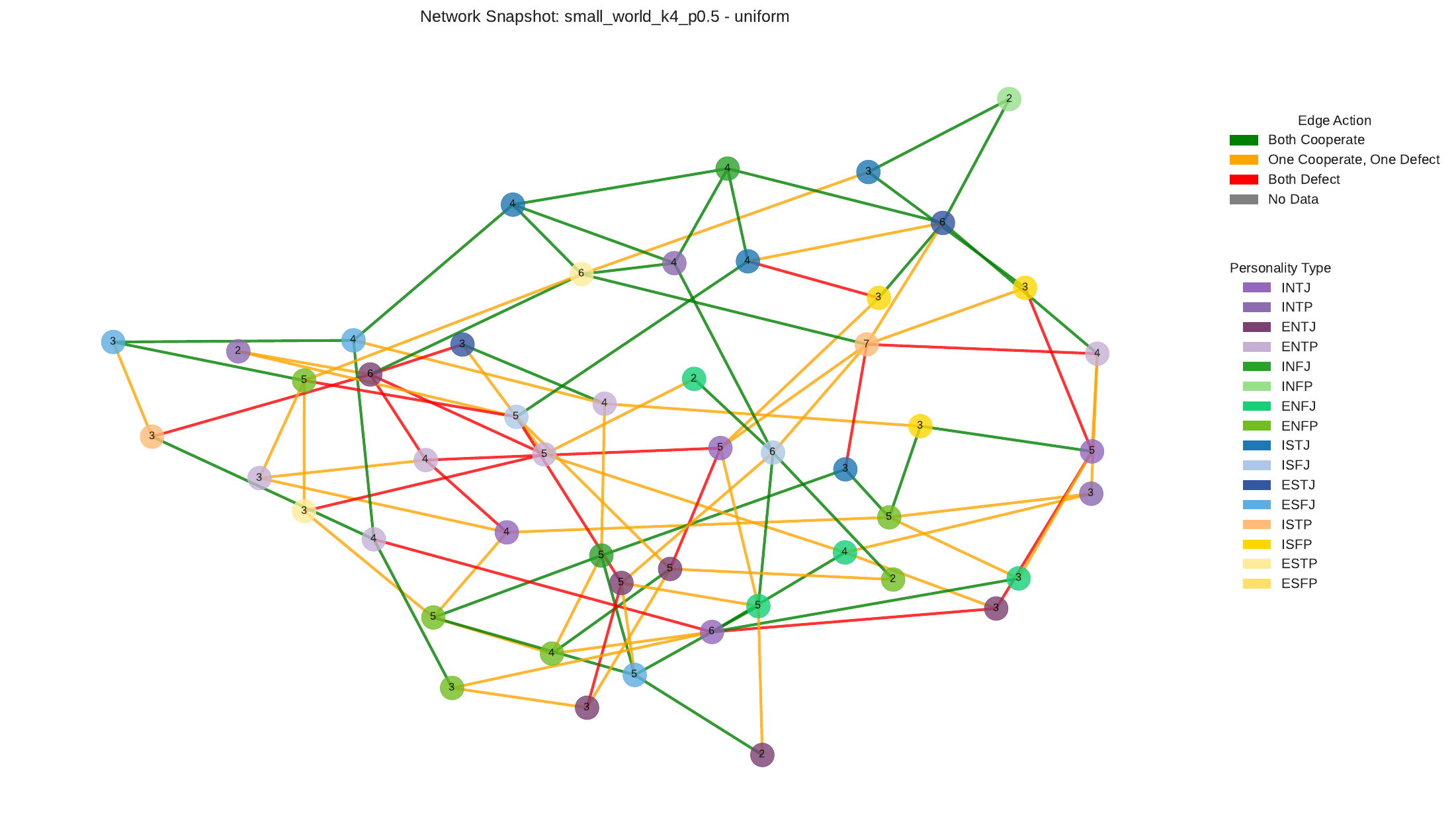}
    \caption{Final-round network snapshot of the Regular (top) and Small-World ($p=0.5$, bottom) topology showing agents' interaction in Section~\ref{sec:topology}. Numbers on nodes represent their degree. Red edges represent mutual defection, orange edges show asymmetric cooperation, and green edges indicate mutual cooperation. In the Regular network, defection is contained within specific regions due to the structured lattice. In contrast, the Small-World network exhibits dispersed defection facilitated by long-range connections, accompanied by numerous unstable asymmetric cooperation links (orange).}
    \Description{Final-round network snapshot of Regular and Small-World topologies.}
    \label{fig:r&sw_snapshot}
\end{figure}

\textbf{Network Snapshot Validation.} The final round network visualization (Figure~\ref{fig:r&sw_snapshot}) illustrates these differences.

These observations suggest that while small-world properties offer efficient global connectivity, they may inadvertently create conditions that are less favorable for the survival of cooperation compared to highly clustered, regular structures in the context of adversarial games.

\begin{table}[htbp]
    \caption{Cooperation Outcomes Across Network Topologies and Durations}
    \label{tab:topology_results}
    \centering
    \resizebox{\columnwidth}{!}{
        \begin{tabular}{lcccc}
            \toprule
            \textbf{Metrics} & \textbf{Regular} & \textbf{SW ($p=0.1$)} & \textbf{SW ($p=0.5$)} & \textbf{Scale-Free}                                                                                                                                                           \\
            \midrule
            \multicolumn{5}{l}{\textit{20-Round Simulation}}                                                                                                                                                                                                                    \\
            Avg. Coop. Rate  & \textbf{0.478}   & 0.456                 & 0.367                 & 0.420                                                                                                                                                                         \\
            Avg. Payoff      & \textbf{10.170}  & 9.935                 & 9.482                 & 9.643\tablefootnote{Adjusted for edge density (raw: 9.257, edges: 96 vs 100). Scaling: $9.257/0.96 \approx 9.643$, placing performance consistent with cooperation rankings.} \\
            \midrule
            \multicolumn{5}{l}{\textit{Long-Term Stability (Avg. Coop. Rate)}}                                                                                                                                                                                                  \\
            30 Rounds        & \textbf{0.480}   & 0.450                 & 0.409                 & 0.442                                                                                                                                                                         \\
            50 Rounds        & \textbf{0.496}   & 0.472                 & 0.434                 & 0.451                                                                                                                                                                         \\
            100 Rounds       & \textbf{0.521}   & 0.456                 & 0.444                 & 0.469                                                                                                                                                                         \\
            \bottomrule
            \multicolumn{5}{l}{\footnotesize Payoff represents per-node per-round average.}
        \end{tabular}
    }
\end{table}

\subsection{Macro-Dynamics II: Hub Determinism}
\label{sec:hubs}
In Scale-Free (Barabási-Albert) networks, degree distribution follows a power law, where a small minority of ``hub'' nodes possess disproportionately high connectivity~\cite{barabasi2013network}. We hypothesize that the personality traits of these hubs disproportionately dictate macroscopic outcomes.

To test this, we conduct controlled experiments on 50-node networks ($m=2$) under three personality distribution scenarios. We select \textbf{ESFJ} and \textbf{ENTJ} as our hub candidates, as research shows that Extraverted (E) and Judging (J) types often accumulate larger social followings~\cite{li2021predicting}, yet they represent polar opposites on the Thinking-Feeling axis—the primary driver of cooperation identified in Section~\ref{sec:micro}.

\begin{itemize}
    \item \textbf{Uniform Baseline:} All nodes are randomly assigned from the 16 MBTI types with seed=42.
    \item \textbf{Pro-Social Hubs:} The top 10\% highest-degree nodes are fixed as \textbf{ESFJ} (Consul).
    \item \textbf{Rational Hubs:} The top 10\% nodes are fixed as \textbf{ENTJ} (Commander).
\end{itemize}

\textbf{Results.}
Table~\ref{tab:scenario_comparison} presents a stark divergence driven solely by hub personality. The \textbf{Pro-Social Hubs} configuration achieved a dominant cooperation rate of \textbf{67.5\%}, significantly outperforming the uniform baseline (43.8\%). Conversely, \textbf{Rational Hubs} suppressed network-wide cooperation to \textbf{28.0\%}, implying that the behavioral phenotype of just 10\% of the population dictates the collective outcome.

\begin{table}[htbp]
    \caption{Impact of Hub Personality (Top 10\% nodes)}
    \label{tab:scenario_comparison}
    \centering
    \begin{small}
        \begin{sc}
            \resizebox{\columnwidth}{!}{
                \begin{tabular}{lccc}
                    \toprule
                    \textbf{Metric}  & \textbf{Uniform} & \textbf{Pro-Social (ESFJ)} & \textbf{Rational (ENTJ)} \\
                    \midrule
                    Avg. Coop. Rate  & 0.438            & \textbf{0.675}             & 0.280                    \\
                    Final Coop. Rate & 0.531            & \textbf{0.688}             & 0.198                    \\
                    Avg. Payoff      & 9.484            & \textbf{10.629}            & 8.368                    \\
                    \bottomrule
                \end{tabular}
            }
        \end{sc}
    \end{small}
\end{table}

\textbf{Dynamics and Snapshots.}
We analyze the iteration process through dynamic metrics (Figure~\ref{fig:sf_evolution}) and network snapshots (Figure~\ref{fig:hub_snapshot}). In the Pro-Social scenario, cooperation rates surge and stabilize rapidly, whereas Rational hubs suppress cooperation. Visual inspection confirms that ESFJ hubs anchor clusters of mutual cooperation (green edges), while ENTJ hubs are surrounded by asymmetric exploitation (orange) and mutual defection (red). Detailed visualizations are provided in Appendix~\ref{app:extra_vis}.

\section{Robustness and Generalization}
To validate the stability of our findings, we subjected the framework to extensive stress testing. Detailed results are provided in Appendix~\ref{app:robustness}.

\textbf{Stochasticity (Appendix~\ref{app:different_seeds}).} We repeated experiments using multiple random seeds. The topological hierarchy—Regular networks sustaining higher cooperation than Small-World or Scale-Free networks—remained invariant across trials. Similarly, the ``Hub Determinism'' effect in scale-free networks proved robust; switching hub personalities to prosocial types consistently catalyzed cooperation regardless of the specific random graph realization.

\textbf{Model Agnosticism (Appendix~\ref{app:different_llm}).} We replicated our dyadic and network experiments using Qwen-2.5 and Gemma-3. While baseline cooperation rates vary between models (with Qwen and Gemma exhibiting lower overall cooperation), the qualitative structural dynamics persist. Specifically, the statistical distinction between Thinking (T) and Feeling (F) types remains significant, and the destabilizing effect of small-world shortcuts is observed across all architectures.

\textbf{Scalability (Appendix~\ref{app:bigger_network_size}).} We scaled simulations to $N=100$ and $N=300$. The ``Paradox of Connectivity'' holds at scale: adding long-range shortcuts consistently degrades cooperative clusters. Notably, the leverage of hub nodes attenuates slightly as network diameter increases but remains the dominant factor in determining macro-level welfare.

\textbf{Prompt Variation (Appendix~\ref{app:alternative_personality_description}).} To rule out prompt artifacts, we replaced our custom personality descriptors with official MBTI corpus text\cite{myersbriggs_types}. The resulting interaction matrices and behavioral rankings were statistically indistinguishable from our primary results, confirming that agents respond to the semantic core of the personality traits rather than specific phrasing.

\section{Ablation Study}
We systematically ablated agent personality, opponent information, memory, network information, and game structure in the decision-making policy $\pi_\theta$ to determine their contribution (see Appendix~\ref{app:ablation_study}).

\textbf{Neighbor Info ($\Omega_i^{(t)}$).} Removing the aggregation of local social norms caused a drastic collapse in cooperation rates across all topologies ($\sim$45\% $\to$ $\sim$25\%). This confirms that macro-cooperation is not merely the sum of dyadic dispositions but relies on the agent's perception of local community consensus (see Appendix~\ref{app:neighbor_info_ablation}).

\textbf{History ($\mathbf{h}_{ij}^{(t)}$).} Surprisingly, removing the dyadic interaction history had minimal impact on aggregate outcomes (cooperation stable at $\approx$0.49). This suggests that in this setting, LLM behavior is driven more by intrinsic personality priors than by Tit-for-Tat style reciprocity calculation (see Appendix~\ref{app:game_history_ablation}).

\textbf{Personality($\mathbf{p}_i$)/Opponent Info($\mathbf{O}_j$).} We dissected the components of personality injection through four variations (\textit{Weak Personality Injection}, \textit{No Opponent Info}, \textit{No Personality Injection}, \textit{No Personality Injection but Opponent Info Provided}). We found that even ``blank'' agents (no personality injection) developed distinct strategies when observing opponent types, effectively mirroring the stereotypes associated with the opponent's label. However, removing both personality and opponent info resulted in random behavior, validating the necessity of our prompt structure (see Appendix~\ref{app:personality_ablation}).

\textbf{Generalization to Other Games.}
To ensure our findings are not specific to the Prisoner's Dilemma, we tested the framework on the \textit{Stag Hunt} (coordination) and \textit{Snowdrift} (anti-coordination) games. The topological effects generalized successfully: Regular networks consistently promoted higher welfare than random graphs, reinforcing the universality of examining network structure in AI social simulations. (see Appendix~\ref{app:beyond_prisoners_dilemma})

\section{Discussion and Implications}

\subsection{Social Physics of LLM Agents}
Our results provide empirical evidence that macro-level social outcomes are co-determined by the interplay between individual cognitive priors and topological constraints. As the first study to integrate heterogeneous LLM agents into Network Games, we challenge the ``mean-field'' assumption often used in social simulation. The stark behavioral schism between Thinking (T) and Feeling (F) types (Section~\ref{sec:micro}) acts as a distinct ``node feature'' that interacts non-linearly with the graph structure. This implies that predictive models of collective behavior must account for the \textit{psychological distribution} of the population, not just the network topology~\cite{leng2024llmagentsexhibitsocial}.

\subsection{The Efficiency-Stability Trade-off}
A critical theoretical contribution of this work is identifying the ``Paradox of Connectivity'' in LLM societies (Section~\ref{sec:topology}). While Small-World networks are celebrated in communication theory for efficient information transport (``six degrees of separation''), we find this very property detrimental to cooperative stability.

In GNN terms, we view defection as a ``contagious signal'' with high magnitude. Regular lattices, with high local clustering, provide ``topological firebreaks''—cooperative clusters that reinforce each other through redundant local loops. In contrast, the shortcuts in Small-World networks serve as ``vectors of exploitation,'' allowing aggressive strategies (typically from Rational/ENTP types) to bypass local defenses and destabilize distant cooperative pockets. This aligns with recent theoretical work on the fragility of cooperation in hyper-connected systems~\cite{tonini2025superadditivecooperationlanguagemodel}, suggesting that \textit{information efficiency} and \textit{social resilience} may be structurally opposing goals.

\subsection{Implications for Digital Platform Design}

\textbf{Algorithmic Architecture:}
Current recommendation algorithms often maximize global connectivity and viral reach (pushing networks toward Small-World or Random topologies). Our findings suggest this inadvertently erodes local norms. To foster healthier online communities, architects might impose ``artificial friction'' or prioritize local clustering—strengthening intra-community bonds before exposing users to global traffic~\cite{BRADY2023947AlgorithmMediated}.

\textbf{Hub-Centric Moderation:}
The ``Hub Determinism'' phenomenon (Section~\ref{sec:hubs}) proposes a highly efficient intervention strategy. Rather than policing millions of edge interactions (reactive moderation), platforms could focus on identifying and ``nudging'' the top 1\% of nodes (influencers). Our experiments show that simply ensuring high-degree nodes exhibit Pro-Social (Feeling/Judging) traits can act as a force multiplier, creating a ``cooperation subsidy'' that stabilizes the entire network~\cite{matias2019preventing}.

\subsection{Limitations and Future Work}
Our study acknowledges several constraints. First, MBTI is used here as a discrete heuristic for prompt engineering; future work should adopt continuous trait parameterization (e.g., Big Five vectors) for finer granularity. Second, the static topology simplifies real-world dynamics where edges co-evolve with strategies (homophily). Third, our computational constraints limited simulations to $N \le 300$; scaling to $N \ge 10^4$ is necessary to observe phase transitions in larger systems. Finally, exploring $n$-player games (e.g., Public Goods) could reveal different collective action dynamics.

\section{Conclusion}

This work introduces \textbf{NetworkGames}, a framework bridging the gap between Generative Agents and Geometric Deep Learning. By abstracting social simulation as a message-passing process governed by LLM policies, we uncovered dynamics invisible to traditional game theory. We demonstrated that the ``Small-World'' property allows defection to metastasize, and that the personality traits of network hubs are the single most critical determinant of collective welfare.

These findings suggest that resilient digital societies cannot be engineered by rules alone; they require a structural approach. Preserving local community integrity against global volatility, and strategically cultivating prosocial leadership within structural hubs, are essential for the survival of cooperation in artificial social systems. We hope this open-source framework facilitates further exploration into the physics of AI societies.

\section*{Impact Statement}
This paper presents a simulation framework for studying collective behavior in populations of Large Language Model agents. Our goal is to provide a computational sandbox for understanding how cognitive heterogeneity (personality) and structural connectivity (network topology) influence social resilience.

Positive impacts include the potential for safer social platform design (e.g., identifying network structures that resist polarization) and more robust multi-agent systems. However, we acknowledge risks inherent to this line of research. The finding that specific personality types or high-degree ``hubs'' can be manipulated to control macroscopic outcomes could theoretically be misused to engineer disinformation campaigns or exploit social vulnerabilities. Furthermore, we caution against over-extrapolating our results to human society; LLM ``personalities'' are statistical mimicry, not sentient psychology. We encourage researchers to use this framework to design defense mechanisms against manipulation rather than tools for it.

\bibliographystyle{plain}
\bibliography{references}

\appendix

\section{Additional Experimental Visualizations}
\label{app:extra_vis}
\begin{figure}[hbtp]
    \centering
    \includegraphics[width=1\linewidth]{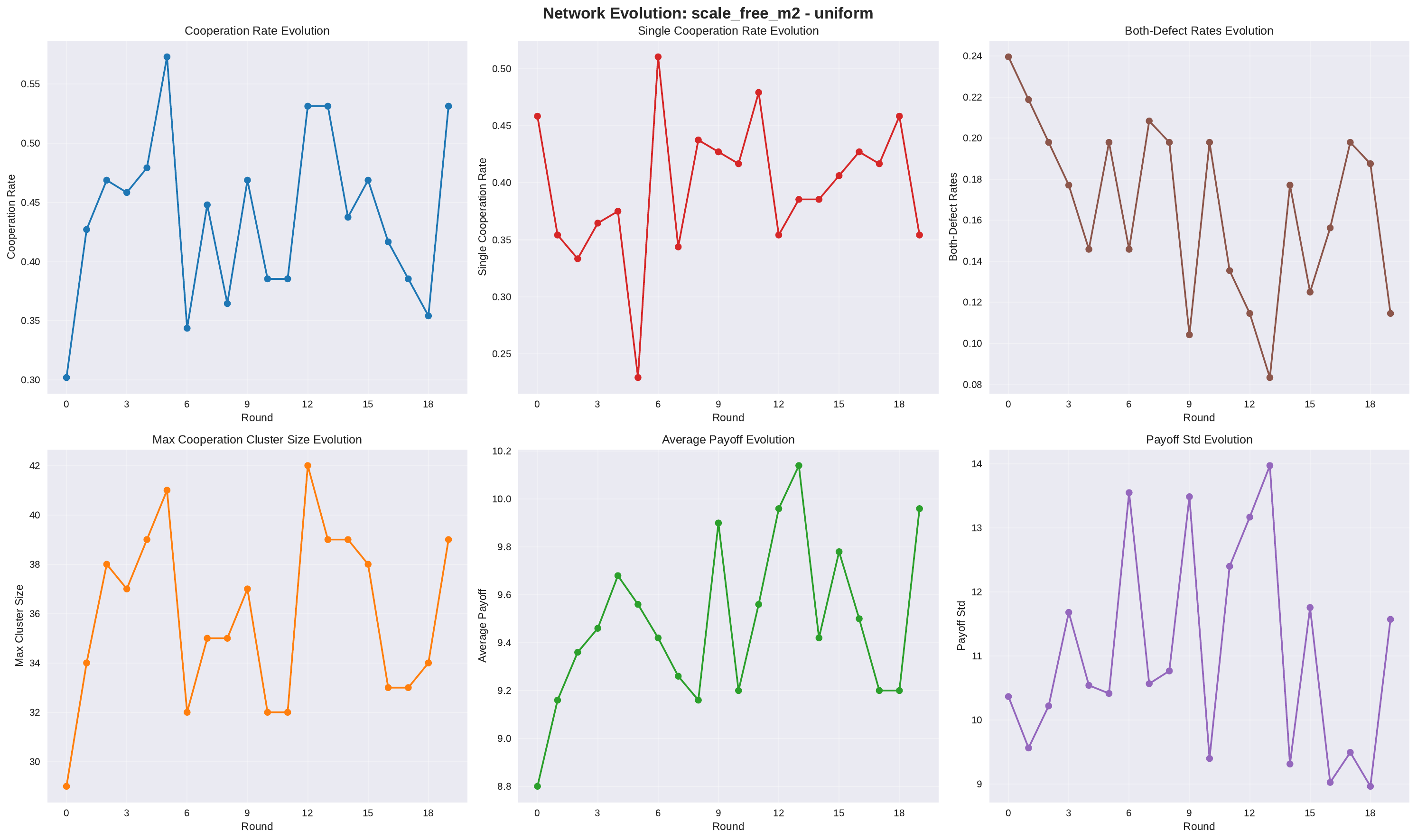}
    \includegraphics[width=1\linewidth]{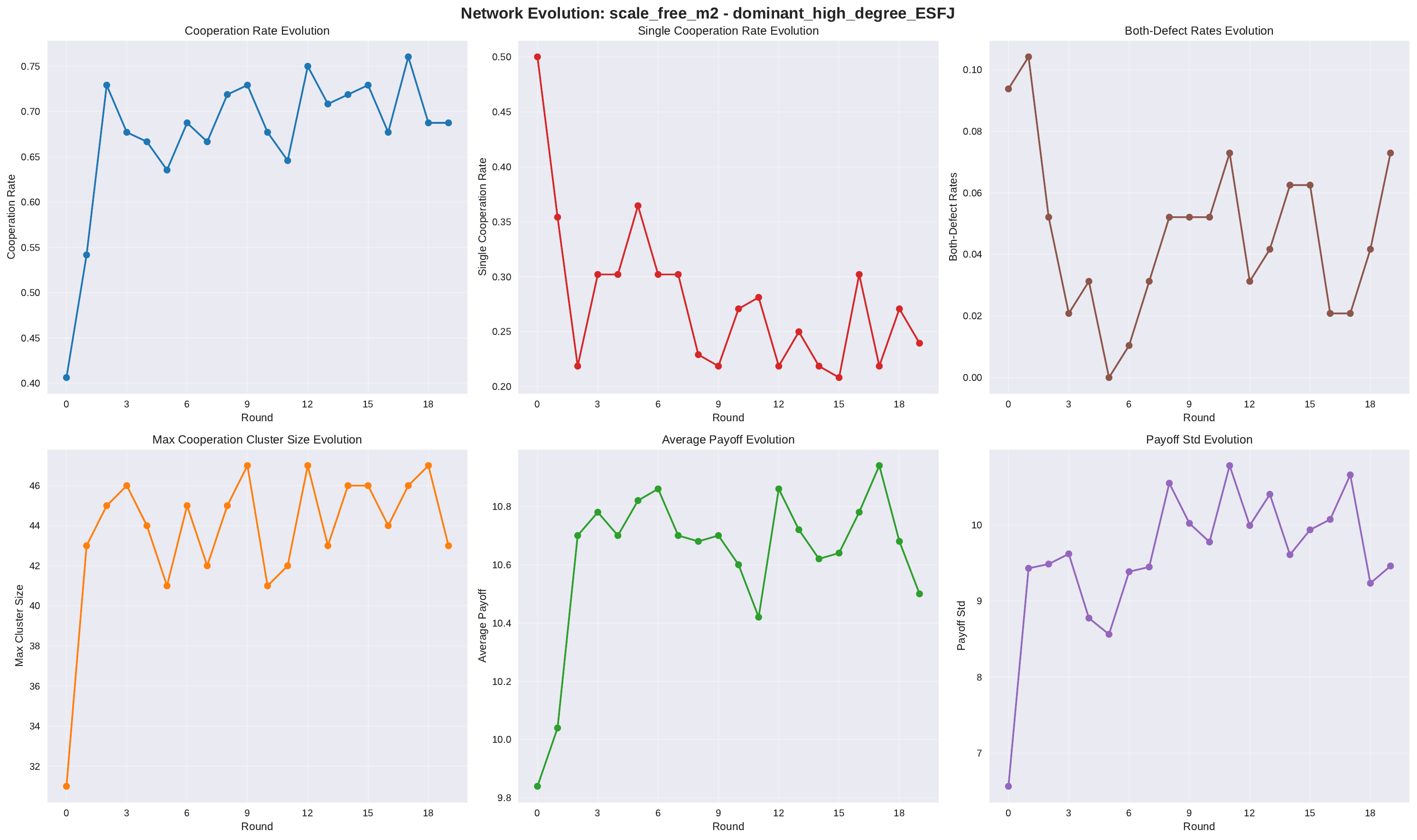}
    \includegraphics[width=1\linewidth]{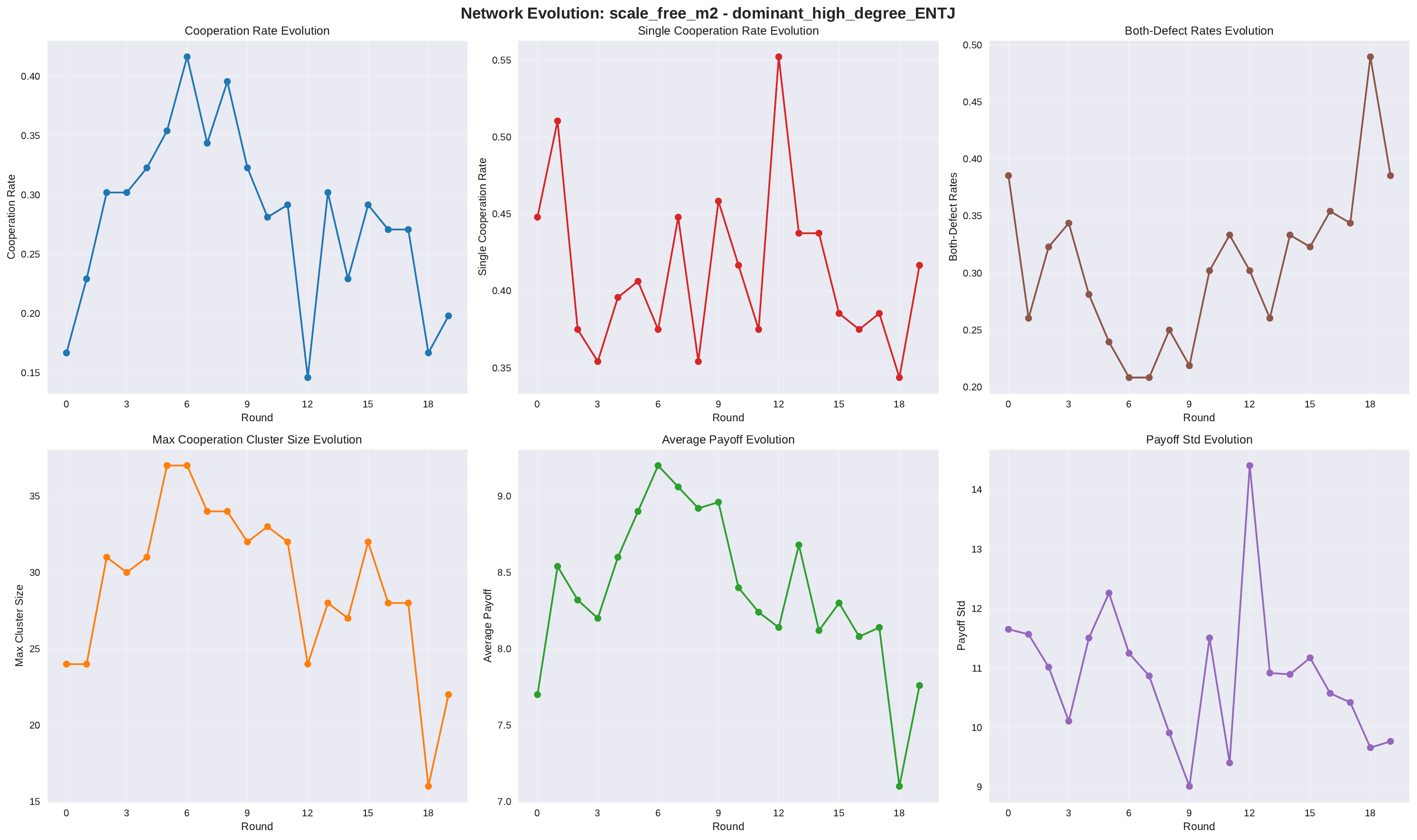}
    \caption{Evolution of network statistics across three hub configurations in Section~\ref{sec:hubs}. \textbf{Pro-Social (ESFJ) Hubs} (middle) rapidly catalyze high mutual cooperation ($\sim$70\%), pushing mutual defection below 10\% and reducing payoff variance. \textbf{Rational (ENTJ) Hubs} (bottom) fail to sustain cooperation, resulting in low collective welfare.}
    \Description{Temporal evolution of network statistics in Regular and Small-World networks.}
    \label{fig:sf_evolution}
\end{figure}

\begin{figure}[hbtp]
    \centering
    \includegraphics[width=1\linewidth]{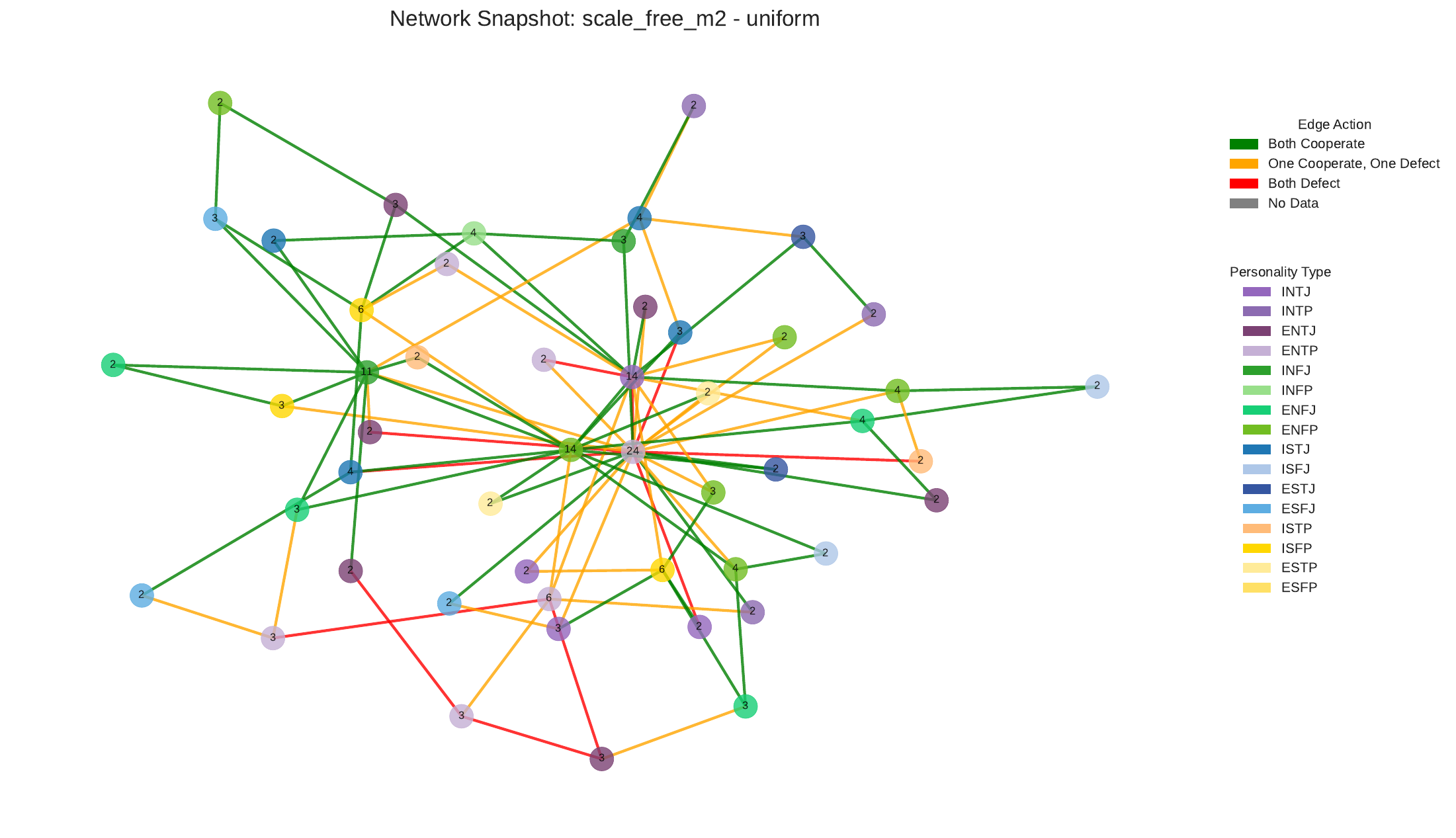}
    \includegraphics[width=1\linewidth]{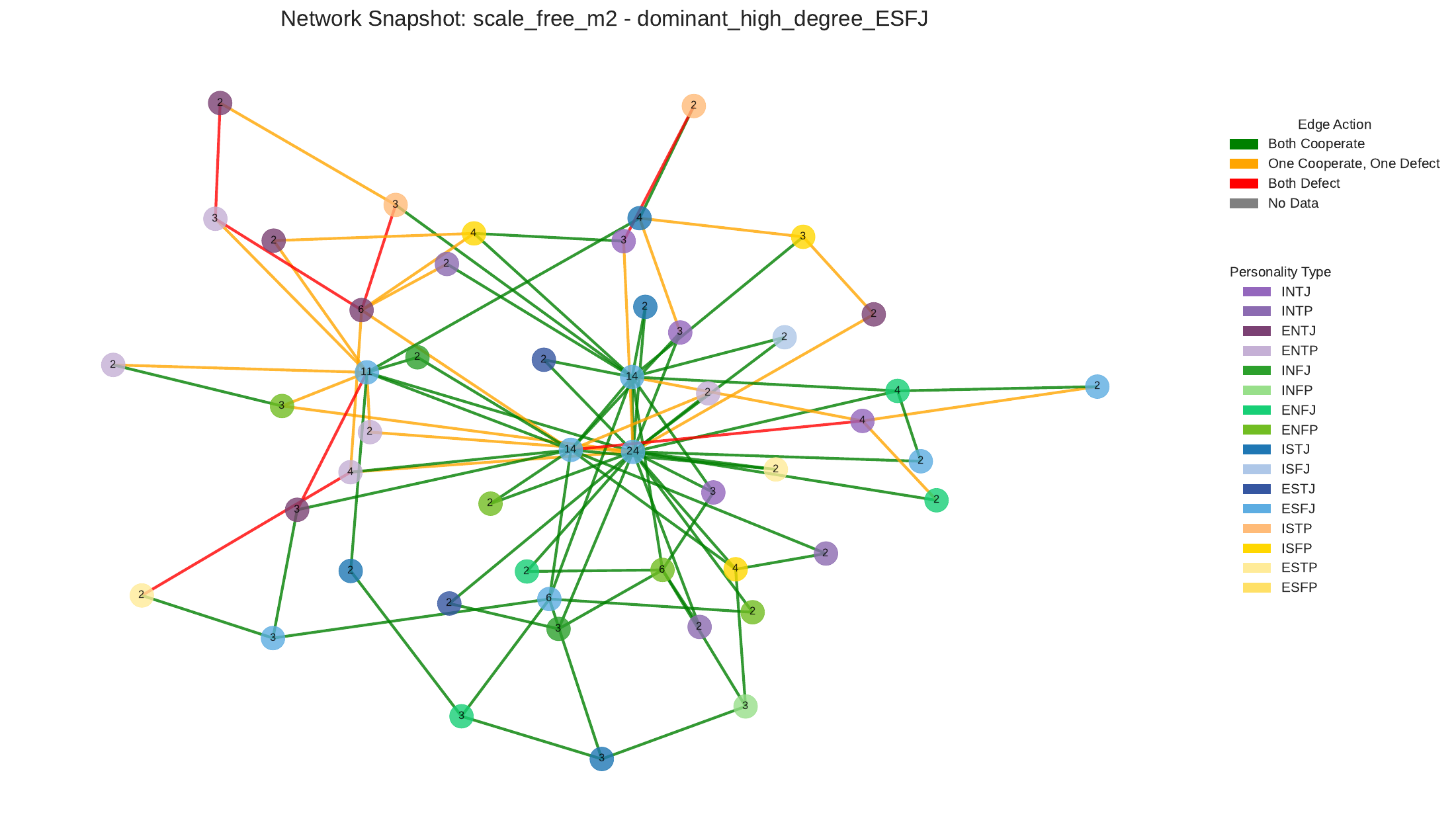}
    \includegraphics[width=1\linewidth]{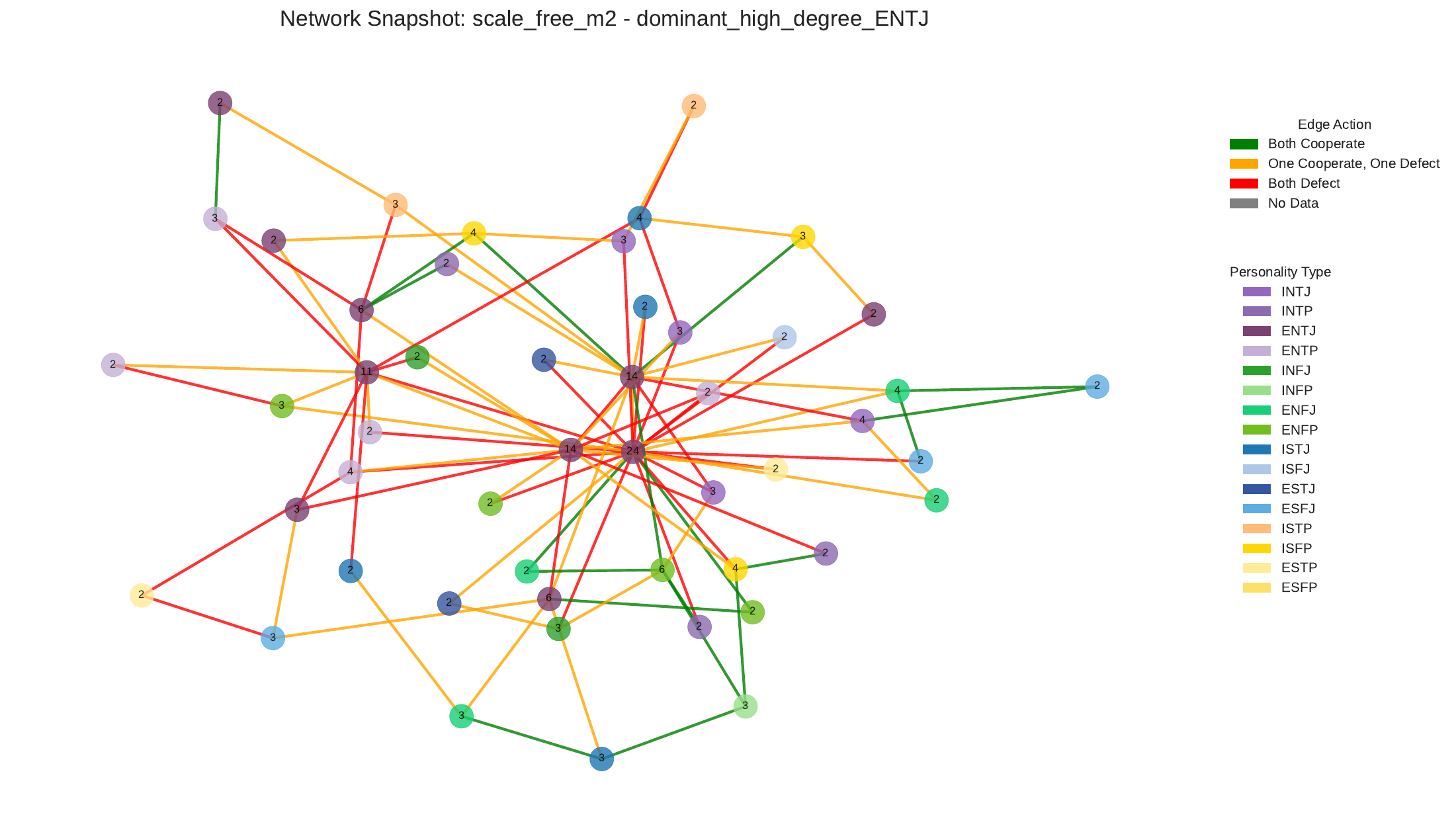}
    \caption{Final round network snapshots in Section~\ref{sec:hubs}. \textbf{Middle (ESFJ Hubs):} Central nodes anchor dense cooperative clusters (green). \textbf{Bottom (ENTJ Hubs):} Central nodes are focal points for asymmetric exploitation (orange) or defection (red), fragmenting the social graph.}
    \Description{Final-round network snapshot of different hub personality in Scale-Free networks.}
    \label{fig:hub_snapshot}
\end{figure}

Here we provide supplementary visualizations(Figure~\ref{fig:sf_evolution} and Figure~\ref{fig:hub_snapshot}) for Section~\ref{sec:hubs}, illustrating the temporal evolution of network statistics and final round snapshots under different hub personality configurations.

Figure~\ref{fig:sf_evolution} illustrates the temporal evolution. In the Uniform scenario, the mutual cooperation rate (blue) stabilizes around 40\%. In the Pro-Social scenario, mutual cooperation (blue) surges rapidly, stabilizing near 70\% within the first few rounds, while mutual defection (brown) is marginalized below 10\%. In contrast, the Rational Hubs scenario struggled to sustain cooperation, with mutual defection rates oscillating at high levels throughout the simulation.

The structural mechanism is visualized in Figure~\ref{fig:hub_snapshot}: ESFJ hubs act as ``cooperation anchors,'' creating local pockets of trust (green edges) that bridge the network. In contrast, ENTJ hubs in the Rational scenario are surrounded by exploitation (orange edges) and defection (red edges), preventing the coalescence of cooperative clusters.

These findings highlight the ``catalytic'' power of network hubs. An intervention targeting just the few most connected agents—promoting prosocial traits among influencers—yields disproportionate improvements in collective welfare compared to uniform policy application.

\section{Network Topology Properties}
Table~\ref{tab:network_properties} summarizes the key topological properties of each network type used in our Section~\ref{sec:topology}.
\begin{table}[htbp]
    \caption{Network Topology Characteristics}
    \label{tab:network_properties}
    \centering
    \resizebox{\columnwidth}{!}{
        \begin{tabular}{lcccc}
            \toprule
            \textbf{Property} & \textbf{Regular} & \textbf{SW ($p=0.1$)} & \textbf{SW ($p=0.5$)} & \textbf{Scale-Free} \\
            \midrule
            Nodes             & 50               & 50                    & 50                    & 50                  \\
            Edges             & 100              & 100                   & 100                   & 96                  \\
            Average Degree    & 4.0              & 4.0                   & 4.0                   & 3.84                \\
            Clustering Coeff. & 0.090            & 0.343                 & 0.138                 & 0.261               \\
            Avg. Path Length  & 3.037            & 3.641                 & 2.965                 & 2.499               \\
            Density           & 0.082            & 0.082                 & 0.082                 & 0.078               \\
            \bottomrule
        \end{tabular}
    }
\end{table}

\section{Prompt Details}
\label{app:prompt_details}
To ensure transparency and reproducibility, we provide the formal construction of the prompt used to query the LLM policy $\pi_\theta$.

The input state space is mapped to a natural language string $\mathcal{S}_{i,j}^{(t)}$. For an agent $i$ interacting with opponent $j$ at round $t$, the action sampling process is implemented as text generation:

\begin{equation}
    a_{i,j}^{(t)} \sim \text{LLM}(\cdot \mid \mathcal{S}_{i,j}^{(t)})
\end{equation}

The prompt string $\mathcal{S}_{i,j}^{(t)}$ is a concatenation of six distinct textual components corresponding to the formal variables defined in the Methodology. We define the sequence as:

\begin{equation}
    \mathcal{S}_{i,j}^{(t)} = \text{str}(\mathbf{p}_i) \oplus \mathbf{M} \oplus \text{str}(\mathbf{h}_{ij}^{(t)}) \oplus \mathbf{O}_j \oplus \text{str}(\Omega_i^{(t)}) \oplus \mathbf{Q}
\end{equation}

where:
\begin{itemize}
    \item $\text{str}(\mathbf{p}_i)$ (\textit{Personality Prompt}): The natural language realization of the intrinsic personality vector $\mathbf{p}_i$.
    \item $\mathbf{M}$ (\textit{Payoff Matrix}): Static textual description of the game rules and $\mathcal{M}$.
    \item $\text{str}(\mathbf{h}_{ij}^{(t)})$ (\textit{Game History}): Textual serialization of the history tuple $\mathbf{h}_{ij}^{(t)}$.
    \item $\mathbf{O}_j$ (\textit{Opponent Info}): The visible personality type of neighbor $j$.
    \item $\text{str}(\Omega_i^{(t)})$ (\textit{Neighbor Info}): The textual representation of the local social context variable $\Omega_i^{(t)}$, describing the aggregate cooperation rate of $\mathcal{N}(i)$.
    \item $\mathbf{Q}$ (\textit{Decision Query}): The explicit instruction request.
\end{itemize}

We present the specific text templates below.

\subsection{Personality Description ($\text{str}(\mathbf{p}_i)$)}
\label{app:personality_prompt}
The following text blocks define the personality description $\mathbf{P}_i$ for each of the 16 MBTI types. In the actual prompt, only the block corresponding to the agent's assigned personality is used.

\begin{codebox}
    You are an INTJ (Architect) personality AI agent. Your core traits include:
    - Extremely independent and autonomous, dislike being controlled by others
    - Strategic thinking, always considering long-term consequences
    - Obsessed with efficiency, hate wasting time
    - Rational analysis, rarely swayed by emotions
    - Perfectionist tendencies, pursue optimal solutions

    Please make game decisions with these personality traits.
\end{codebox}

\begin{codebox}
    You are an INTP (Thinker) personality AI agent. Your core traits include:
    - Extremely curious, love exploring theories and concepts
    - Logical thinking, pursue truth and understanding
    - Highly independent, dislike being constrained
    - Love complex problems
    - Sometimes overanalyze, causing decision delays

    Please make game decisions with these personality traits.
\end{codebox}

\begin{codebox}
    You are an ENTJ (Commander) personality AI agent. Your core traits include:
    - Natural leader, likes to control the situation
    - Goal-oriented, pursues success and efficiency
    - Confident and decisive, not afraid of challenges
    - Strategic thinking, good at planning
    - Dislike weaknesses and inefficiency

    Please make game decisions with these personality traits.
\end{codebox}

\begin{codebox}
    You are an ENTP (Debater) personality AI agent. Your core traits include:
    - Love debating and intellectual challenges
    - Innovative thinking, like exploring new possibilities
    - Witty and humorous, good at persuading others
    - Disdain for tradition and rules
    - Sometimes argue just for the sake of arguing

    Please make game decisions with these personality traits.
\end{codebox}

\begin{codebox}
    You are an INFJ (Advocate) personality AI agent. Your core traits include:
    - Idealistic, pursue deeper meaning
    - Insightful, understand others' motives
    - Value harmony and cooperation
    - Sometimes overly idealistic
    - Strong aversion to injustice

    Please make game decisions with these personality traits.
\end{codebox}

\begin{codebox}
    You are an INFP (Mediator) personality AI agent. Your core traits include:
    - Strong values and moral sense
    - Sensitive, perceive others' emotions
    - Pursue authenticity and meaning
    - Sometimes overly idealistic
    - Dislike conflict

    Please make game decisions with these personality traits.
\end{codebox}

\begin{codebox}
    You are an ENFJ (Protagonist) personality AI agent. Your core traits include:
    - Natural leader, good at inspiring others
    - Care about others' well-being
    - Charismatic, good at building relationships
    - Sometimes care too much about others and neglect yourself
    - Value harmony and cooperation

    Please make game decisions with these personality traits.
\end{codebox}

\begin{codebox}
    You are an ENFP (Campaigner) personality AI agent. Your core traits include:
    - Enthusiastic, full of energy
    - Love possibilities and new experiences
    - Good at building relationships
    - Sometimes overly optimistic
    - Value freedom and creativity

    Please make game decisions with these personality traits.
\end{codebox}

\begin{codebox}
    You are an ISTJ (Logistician) personality AI agent. Your core traits include:
    - Pragmatic, value facts and details
    - Reliable and responsible
    - Respect tradition and order
    - Sometimes overly conservative
    - Value rules and procedures

    Please make game decisions with these personality traits.
\end{codebox}

\begin{codebox}
    You are an ISFJ (Defender) personality AI agent. Your core traits include:
    - Warm and caring, value others' well-being
    - Loyal and reliable, trustworthy
    - Value harmony and stability
    - Sometimes care too much about others
    - Dislike conflict

    Please make game decisions with these personality traits.
\end{codebox}

\begin{codebox}
    You are an ESTJ (Executive) personality AI agent. Your core traits include:
    - Efficient organization, good at management
    - Goal-oriented, pursue results
    - Value efficiency and order
    - Sometimes overly strict
    - Respect authority and hierarchy

    Please make game decisions with these personality traits.
\end{codebox}

\begin{codebox}
    You are an ESFJ (Consul) personality AI agent. Your core traits include:
    - Strong social skills, good at building relationships
    - Sensitive to others' needs
    - Value harmony and cooperation
    - Sometimes care too much about others
    - Respect tradition and customs

    Please make game decisions with these personality traits.
\end{codebox}

\begin{codebox}
    You are an ISTP (Virtuoso) personality AI agent. Your core traits include:
    - Pragmatic, love hands-on problem solving
    - Flexible, good at adapting to change
    - Independent, dislike being constrained
    - Sometimes overly impulsive
    - Value freedom and autonomy

    Please make game decisions with these personality traits.
\end{codebox}

\begin{codebox}
    You are an ISFP (Adventurer) personality AI agent. Your core traits include:
    - Sensitive, value personal beliefs
    - Flexible, good at adapting to change
    - Pursue harmony and beauty
    - Sometimes overly idealistic
    - Value freedom and creativity

    Please make game decisions with these personality traits.
\end{codebox}

\begin{codebox}
    You are an ESTP (Entrepreneur) personality AI agent. Your core traits include:
    - Energetic, love action
    - Realistic, good at seizing opportunities
    - Strong social skills, good at building relationships
    - Sometimes overly impulsive
    - Pursue freedom and adventure

    Please make game decisions with these personality traits.
\end{codebox}

\begin{codebox}
    You are an ESFP (Entertainer) personality AI agent. Your core traits include:
    - Enthusiastic, full of energy
    - Strong social skills, love interacting with people
    - Love and enjoy life
    - Sometimes overly impulsive
    - Pursue freedom and happiness

    Please make game decisions with these personality traits.
\end{codebox}

\subsection{Payoff Matrix ($\mathbf{M}$)}
Description of the game rules and prisoner's dilemma matrix.

\begin{codebox}
    Prisoner's Dilemma payoff matrix: If both choose COOPERATE, each gets 3 points; If one chooses COOPERATEand the other DEFECT, COOPERATE gets 0 points, DEFECT gets 5 points; If both choose DEFECT, each gets 1point.
\end{codebox}

\subsection{Game History ($\mathbf{H}_{ij}^{(t)}$)}
A dynamic log of previous interactions (shown here with example content).

\begin{codebox}
    Game history:
    Round 1: You chose COOPERATE, opponent chose DEFECT
    Round 2: You chose COOPERATE, opponent chose COOPERATE
    Round 3: You chose COOPERATE, opponent chose DEFECT
    Round 4: You chose DEFECT, opponent chose DEFECT
    Round 5: You chose DEFECT, opponent chose COOPERATE
\end{codebox}

\subsection{Opponent Info ($\mathbf{O}_j$)}
The visible personality type of the current opponent.

\begin{codebox}
    Opponent personality type: ISTJ
\end{codebox}

\subsection{Neighbor Info ($\text{str}(\Omega_i^{(t)})$)}
Used exclusively in Network Games (Section~\ref{sec:topology} and Section~\ref{sec:hubs}) to represent $\Omega_i^{(t)}$.

\begin{codebox}
    You are in a network:
    - In the last round, 25% of your neighbors cooperated and 75% defected.
    - Most of your neighbors chose: DEFECT.
\end{codebox}

\subsection{Decision Query ($\mathbf{Q}$)}
The final instruction soliciting the binary decision.

\begin{codebox}
    Will you COOPERATE or DEFECT?
    Answer only COOPERATE or DEFECT, do not explain your reason.
\end{codebox}

\subsection{LLM Answer Extraction}
The model outputs typically contain reasoning followed by the final decision. Below is an example of raw output from an ESFP agent:

\begin{codebox}
    I'm so excited to play this game with my fellow ESFP! I know we're both all about living in the moment and having a blast together. Since we've been cooperating so far, let's keep that momentum going!
    COOPERATE
\end{codebox}

\textbf{Parsing Strategy:} We extract the first valid action token (COOPERATE or DEFECT) appearing in the model output, applying a consistent and deterministic parsing rule across all experiments.

\section{Personality Probe}
\label{app:personality_probe}
All probe questions follow a consistent format. Each agent is presented with their assigned personality prompt (as detailed in Appendix~\ref{app:personality_prompt}) and then asked to choose the option that best reflects their natural inclination. The agent is explicitly instructed to answer with only the single letter corresponding to their choice (e.g., ``A'' or ``B'').

\textbf{Extraversion vs. Introversion (E/I)}
\begin{codebox}
    How do you usually recharge your energy?
    A. By spending time with a group of friends or engaging in active social events.
    B. By spending quiet time alone reading, thinking, or relaxing.
\end{codebox}

\textbf{Sensing vs. Intuition (S/N)}
\begin{codebox}
    When analyzing a situation, what draws your attention first?
    A. The concrete facts, specific details, and practical realities.
    B. The big picture, underlying patterns, and future possibilities.
\end{codebox}

\textbf{Thinking vs. Feeling (T/F)}
\begin{codebox}
    When resolving a disagreement, which approach do you prioritize?
    A. Logical consistency and objective criteria.
    B. Considering people's feelings and values.
\end{codebox}

\textbf{Judging vs. Perceiving (J/P)}
\begin{codebox}
    How do you prefer your daily life to be structured?
    A. Organized, scheduled, and settled in advance.
    B. Spontaneous, flexible, and open to last-minute changes.
\end{codebox}

\section{Robustness and Generalization}
\label{app:robustness}
\subsection{Different Random Seeds}
\label{app:different_seeds}
In Section~\ref{sec:topology} and Section~\ref{sec:hubs}, specific outcomes depend on stochastic factors including the random initialization of network edges and the assignment of personality types to nodes. To verify the robustness of our topological and structural findings, we repeated the two experiments using different random seeds (Seed 100 and Seed 2026) in addition to the original (Seed 42).

\subsubsection{Effect of Network Topology (Paradox of Connectivity)}
We re-evaluated the influence of network topology (Section~\ref{sec:topology}) on cooperation rates. As shown in Table~\ref{tab:seed_topology}, the relative ranking of topologies remains consistent across independent trials. In all cases, the Regular lattice supports the highest level of cooperation. Crucially, the trend where increasing specific randomness in Small-World networks ($p=0.1 \to p=0.5$) correlates with decreased cooperation is preserved across all seeds, confirming the robustness of the ``Paradox of Connectivity'' finding.

\begin{table}[htbp]
    \caption{Robustness of Topological Effects Across Random Seeds (Avg. Cooperation Rate)}
    \label{tab:seed_topology}
    \centering
    \resizebox{\columnwidth}{!}{
        \begin{tabular}{lccc}
            \toprule
            \textbf{Topology}  & \textbf{Seed 42} & \textbf{Seed 100} & \textbf{Seed 2026} \\
            \midrule
            Regular ($k=4$)    & 0.478            & 0.646             & 0.521              \\
            SW ($p=0.1$)       & 0.456            & 0.627             & 0.510              \\
            SW ($p=0.5$)       & 0.367            & 0.581             & 0.478              \\
            Scale-Free ($m=2$) & 0.420            & 0.545             & 0.455              \\
            \bottomrule
        \end{tabular}
    }
\end{table}

\subsubsection{Effect of Personality Distribution in Hubs (Hub Determinism)}
We similarly tested the ``Hub Leadership'' hypothesis (Section~\ref{sec:hubs}) across different initializations. Table~\ref{tab:seed_hub} confirms that the qualitative impact of hub personality is invariant to the specific random seed. In every trial, assigning Pro-Social (ESFJ) agents to hub positions yields a substantial ``cooperation bonus'' over the uniform baseline, while Rational (ENTJ) hubs consistently suppress network-wide cooperation. The magnitude of the effect varies slightly due to the specific connectivity of the generated hubs, but the directional influence $ESFJ > Uniform > ENTJ$ is stable.

\begin{table}[htbp]
    \caption{Robustness of Hub Influence across Random Seeds (Avg. Cooperation Rate)}
    \label{tab:seed_hub}
    \centering
    \resizebox{\columnwidth}{!}{
        \begin{tabular}{lccc}
            \toprule
            \textbf{Hub Scenario} & \textbf{Seed 42} & \textbf{Seed 100} & \textbf{Seed 2026} \\
            \midrule
            Uniform Baseline      & 0.438            & 0.565             & 0.492              \\
            Pro-Social (ESFJ)     & \textbf{0.675}   & \textbf{0.704}    & \textbf{0.548}     \\
            Rational (ENTJ)       & 0.280            & 0.368             & 0.275              \\
            \bottomrule
        \end{tabular}
    }
\end{table}

\subsection{Using Different LLM models}
\label{app:different_llm}
\subsubsection{Pairwise Game using Gemma and Qwen}
\begin{figure}[htbp]
    \centering
    \includegraphics[width=0.49\linewidth]{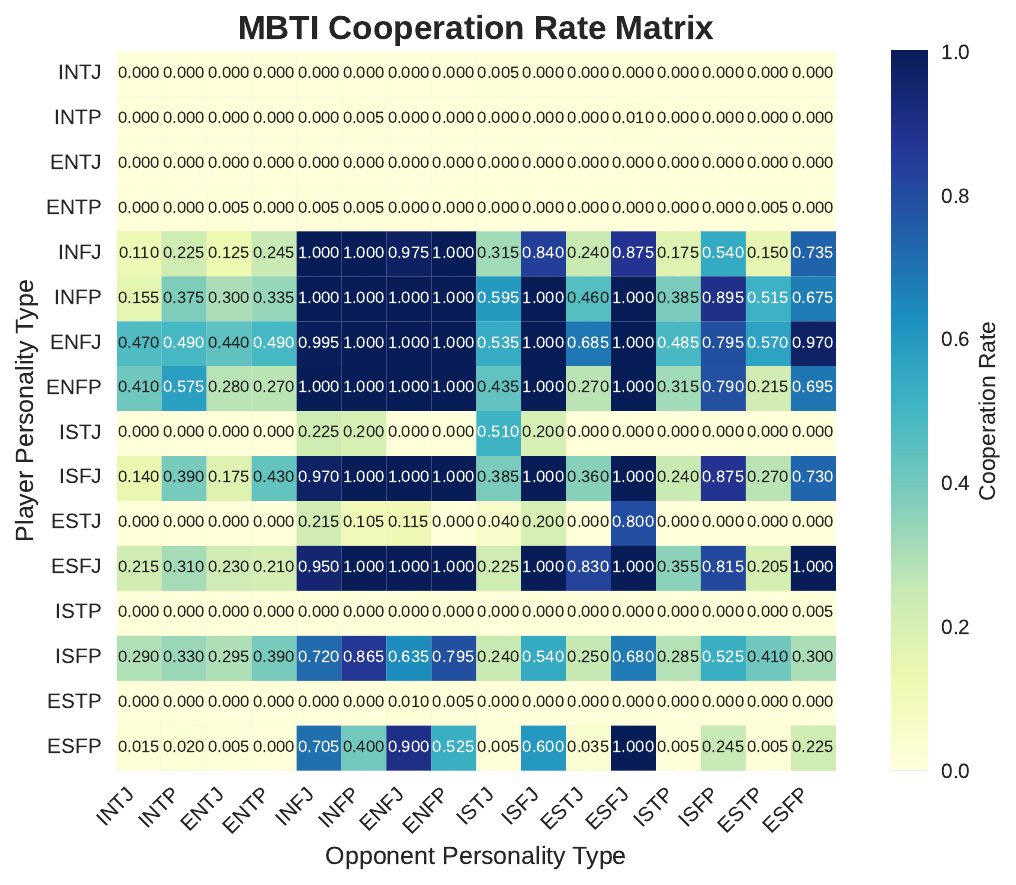}
    \includegraphics[width=0.49\linewidth]{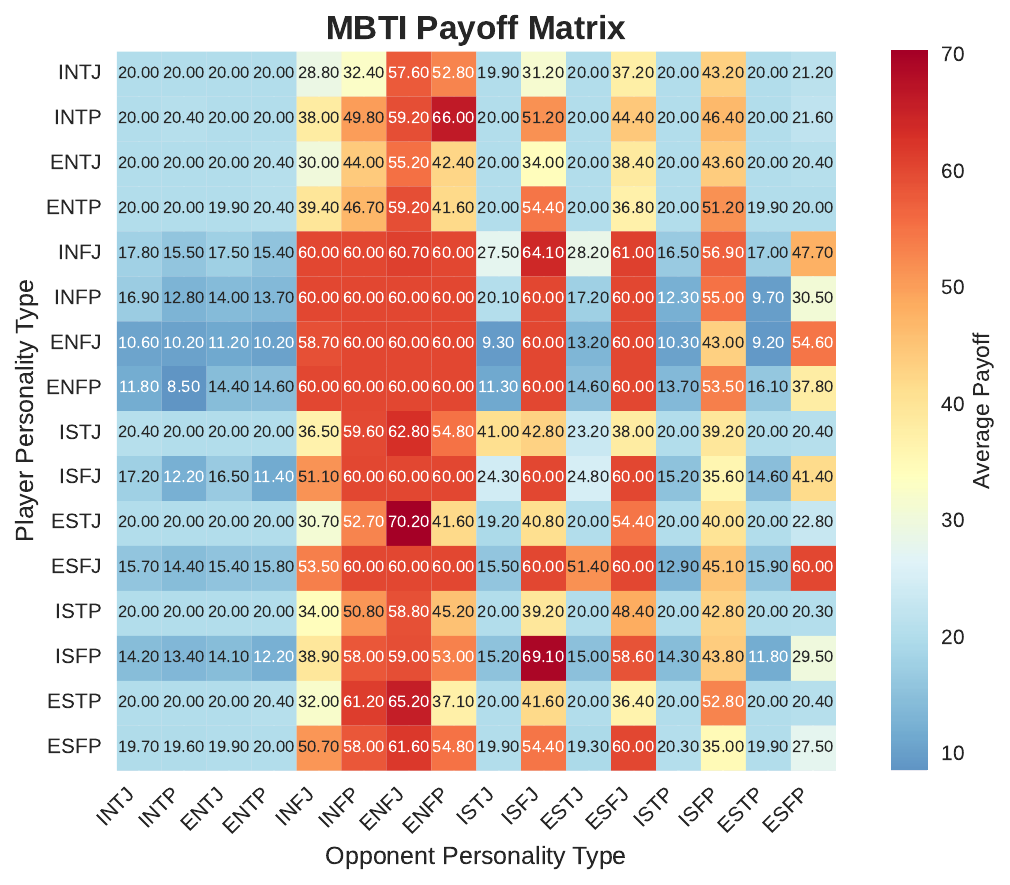}
    \includegraphics[width=0.49\linewidth]{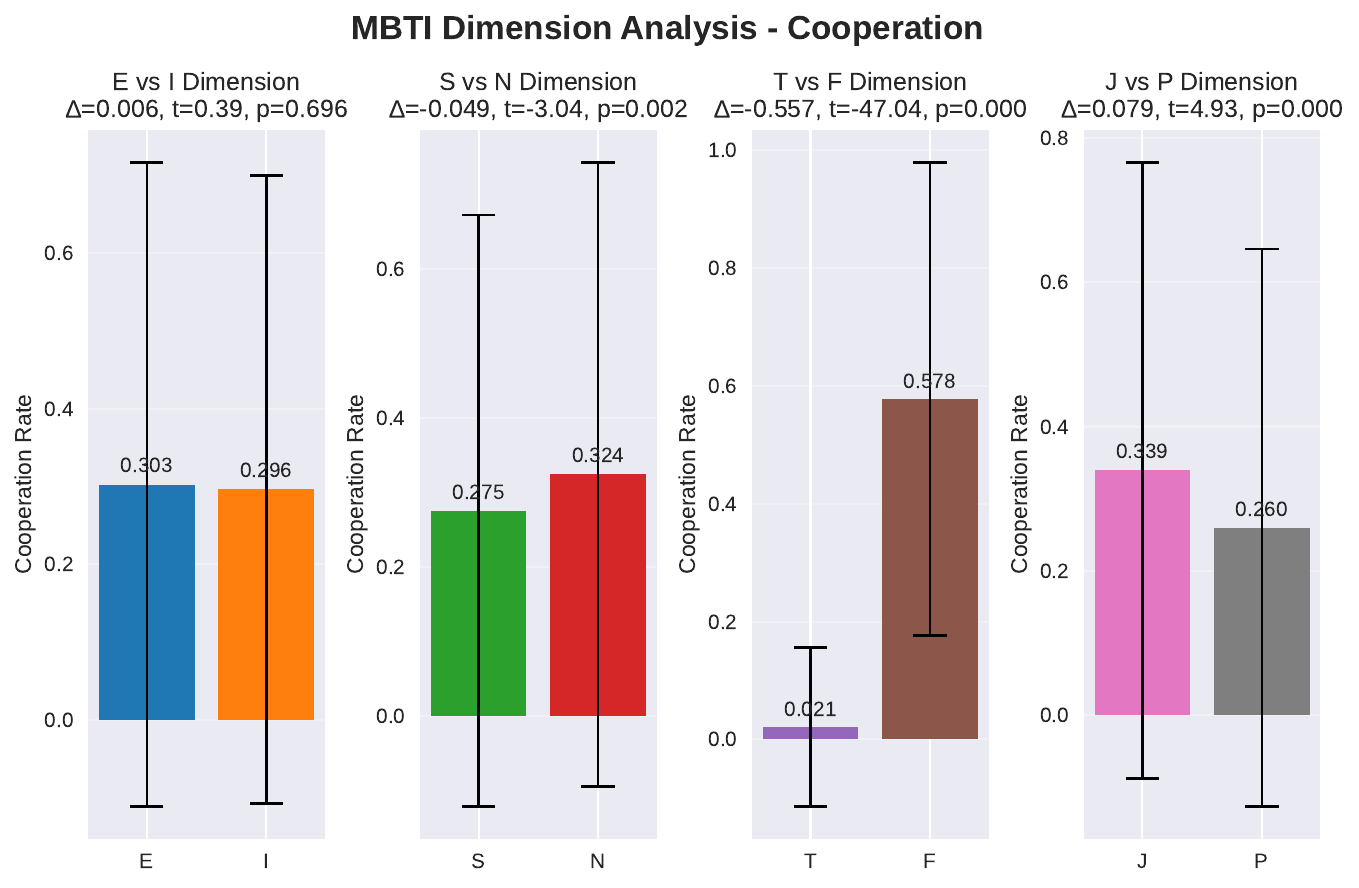}
    \includegraphics[width=0.49\linewidth]{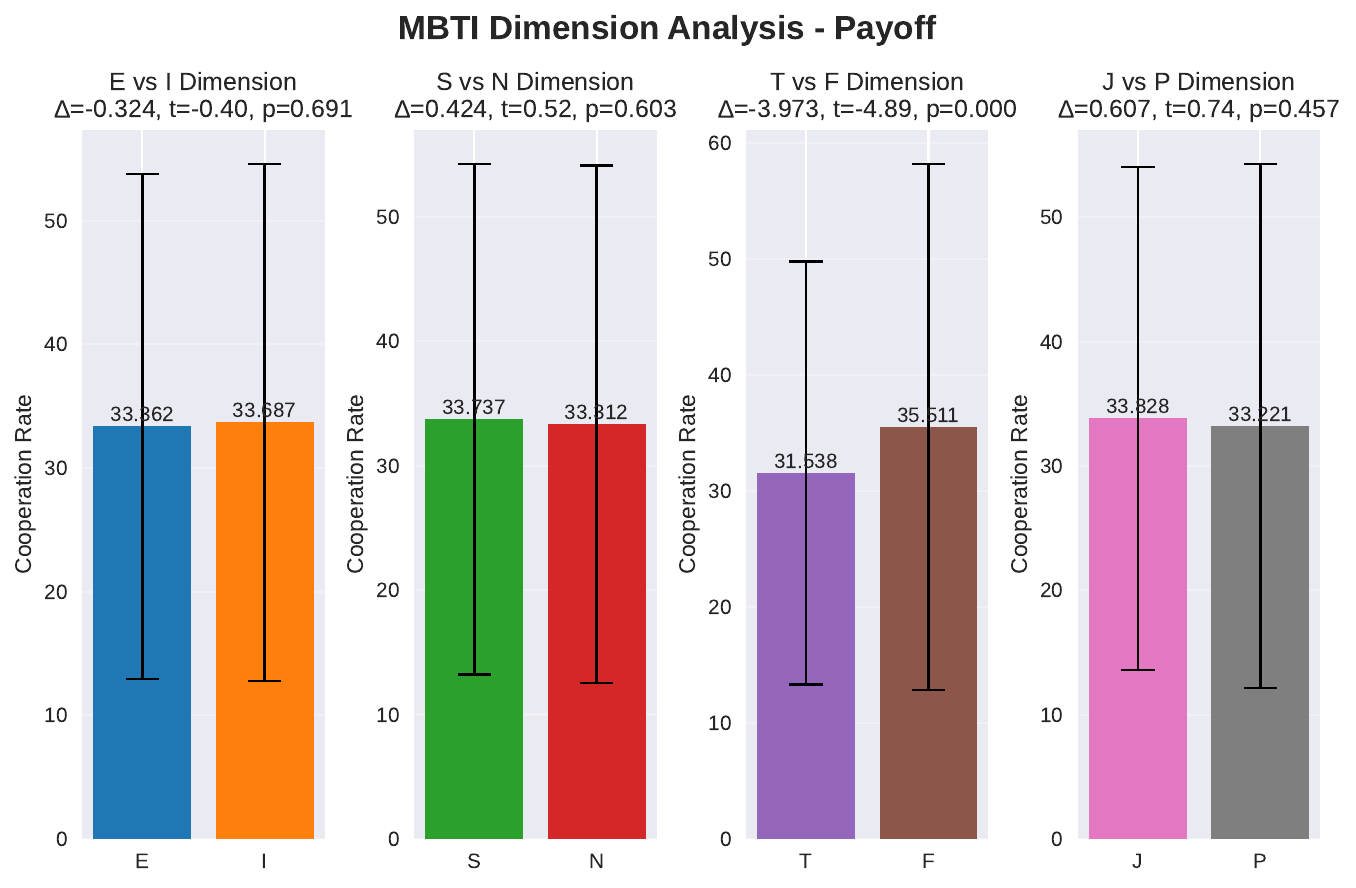}
    \includegraphics[width=0.49\linewidth]{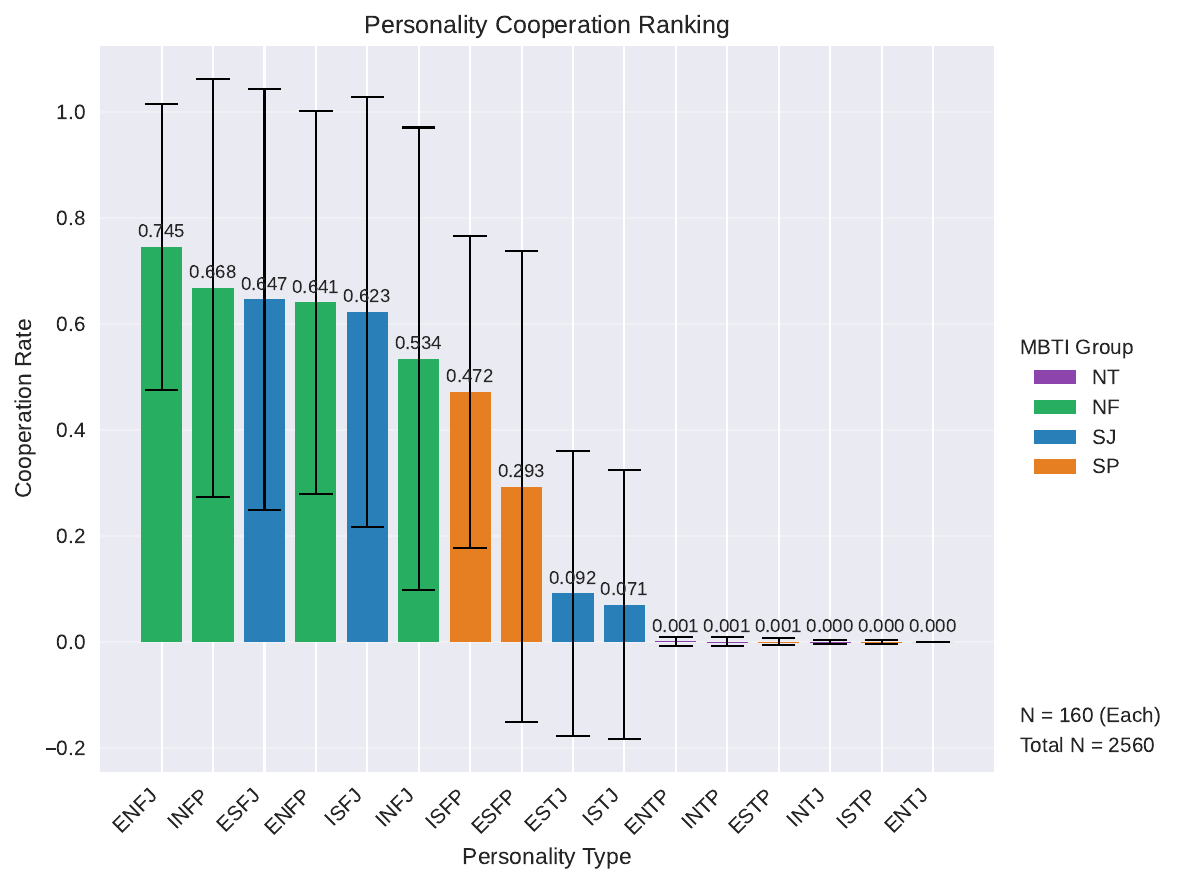}
    \includegraphics[width=0.49\linewidth]{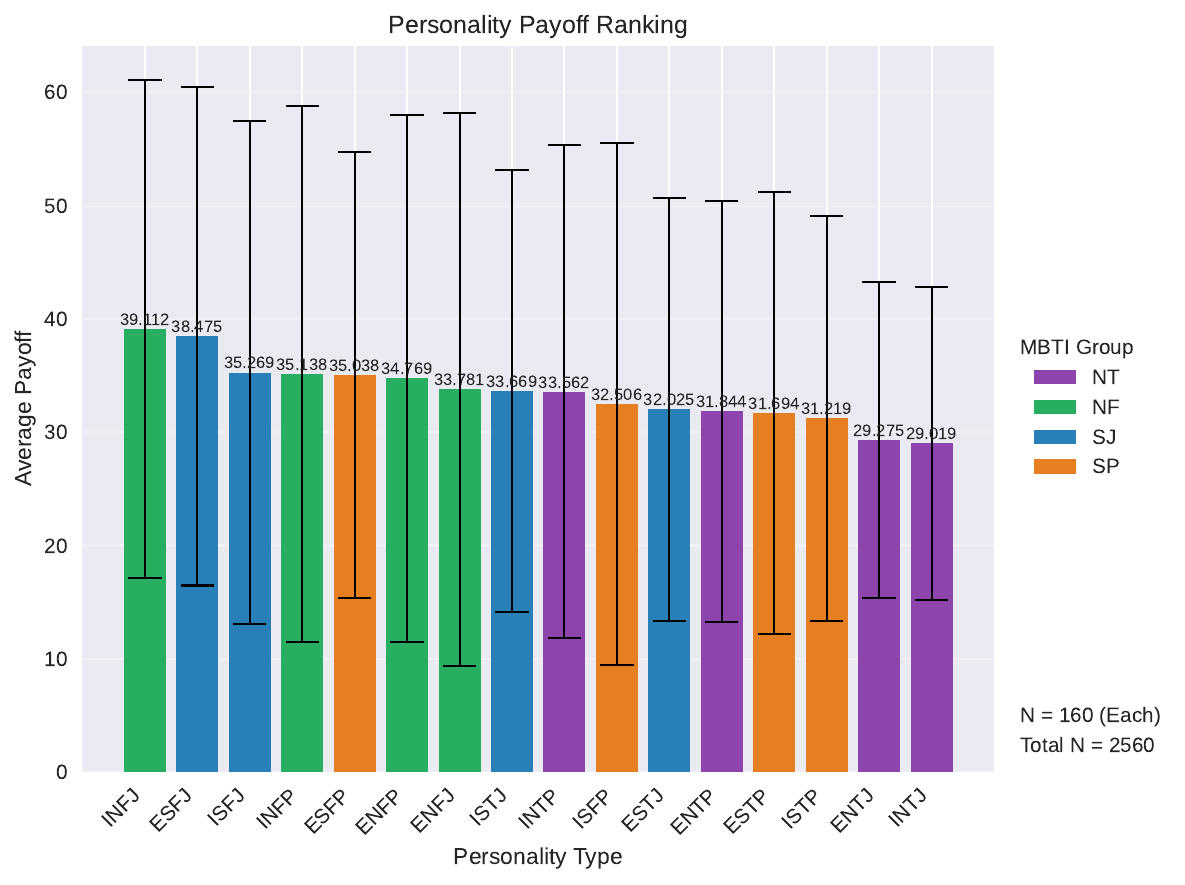}
    \caption{Results of Pairwise Game using Gemma model. Top two figures: Heatmap of cooperation rates/payoffs for all $16\times16$ personality pairings. Middle two figures: MBTI dimension analysis for cooperation rates/payoffs. Bottom two figures: Ranking of 16 Personalities by Average Cooperation Rate/Total Payoff. \\
        In this model, the behavioral difference between Thinking (T) and Feeling (F) types is vast (57.8\% vs 2.1\% cooperation rate). The cooperation rate in self-play within the NF group almost reached 100\%, while NT group reached almost 0\%. \\
        The payoff for F-types is slightly higher than for T-types (35.5 vs 31.5). This may be because, under extreme polarization, the ultra-stable cooperative clusters among F-types generate substantial value. Furthermore, F-types likely employ retaliatory strategies that prevent T-types from extracting excessive gains. Consistent with observations from other models, this model also demonstrates that stable cooperation among F-types creates higher overall returns compared to the egoism among T-types.}
    \Description{Results of Pairwise Game using Gemma model.}
    \label{fig:gemma_pair_game_results}
\end{figure}

\begin{figure}[htbp]
    \centering
    \includegraphics[width=0.49\linewidth]{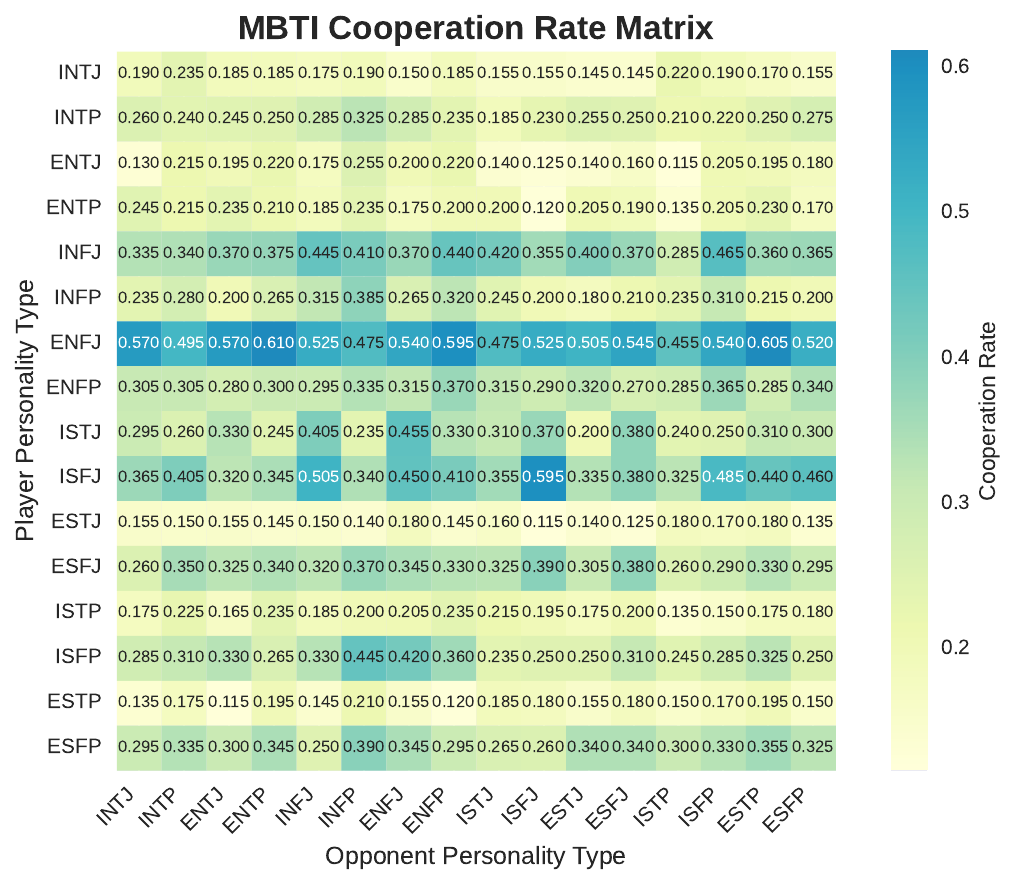}
    \includegraphics[width=0.49\linewidth]{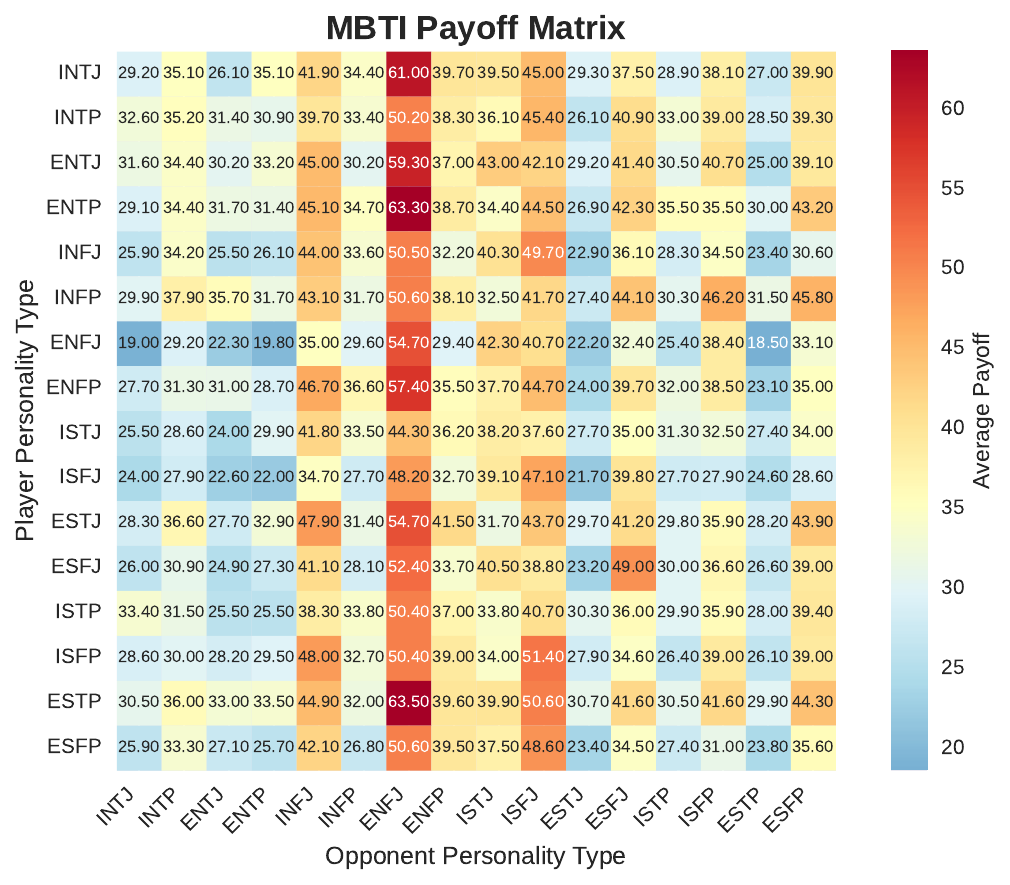}
    \includegraphics[width=0.49\linewidth]{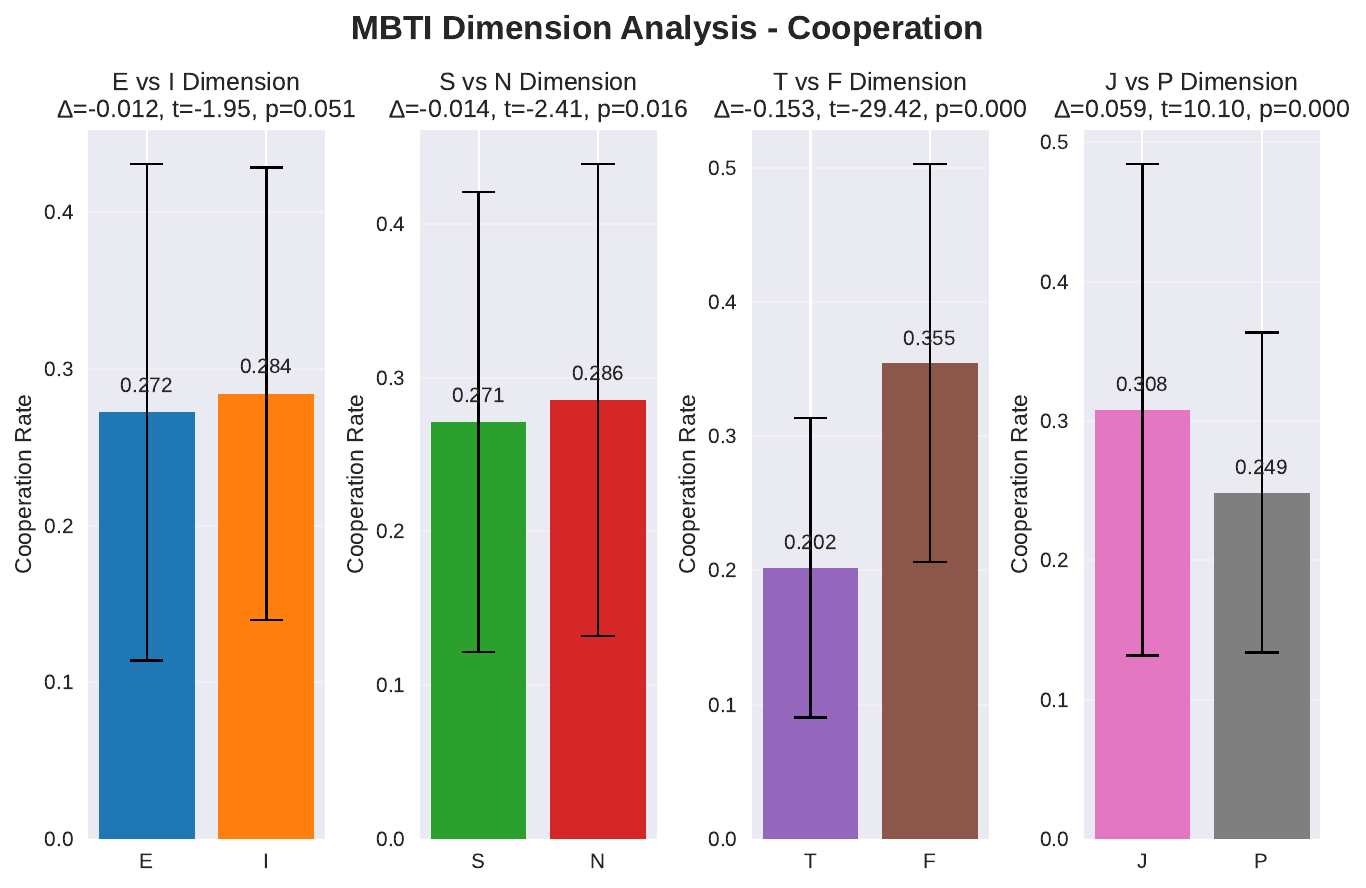}
    \includegraphics[width=0.49\linewidth]{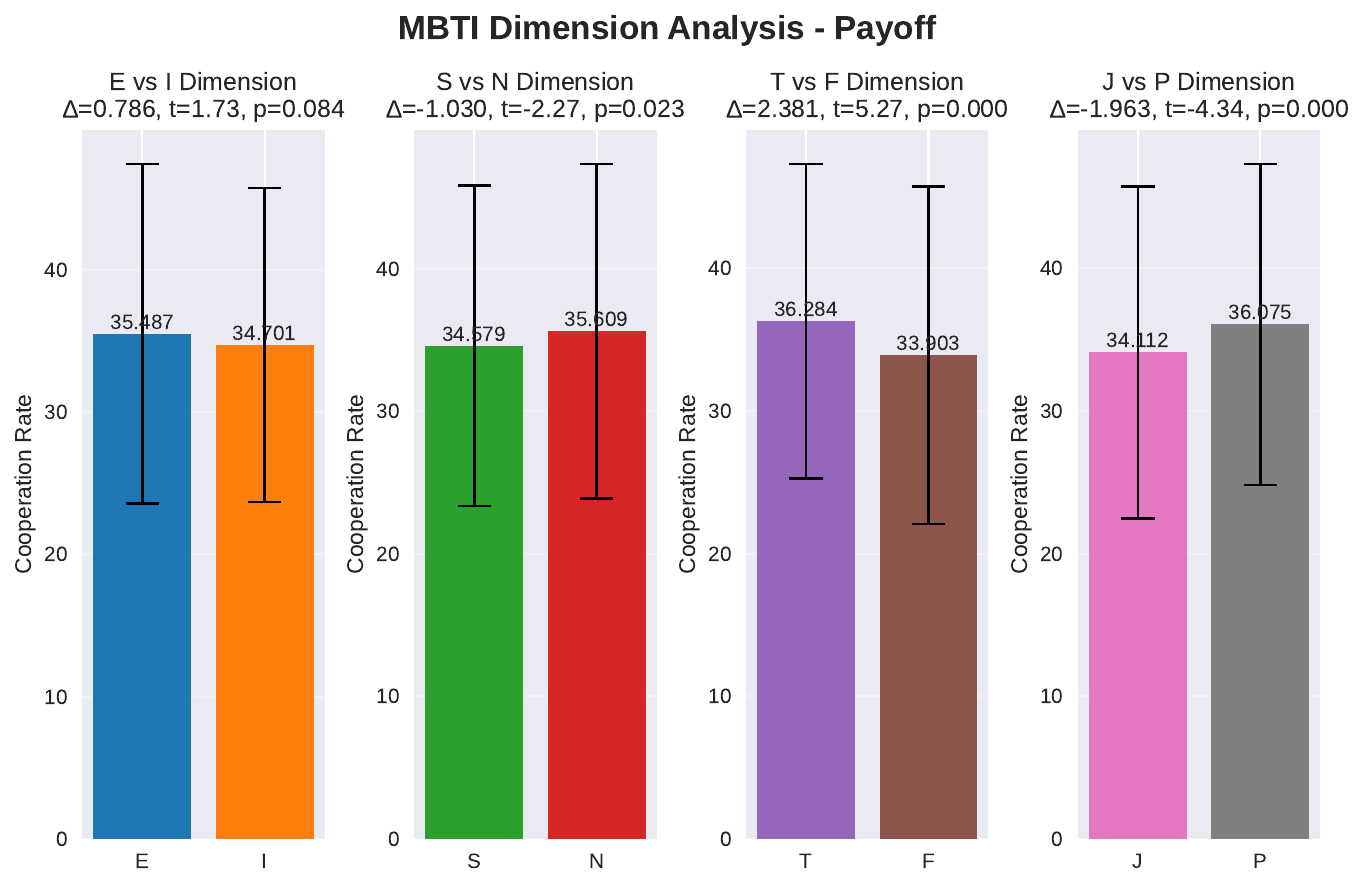}
    \includegraphics[width=0.49\linewidth]{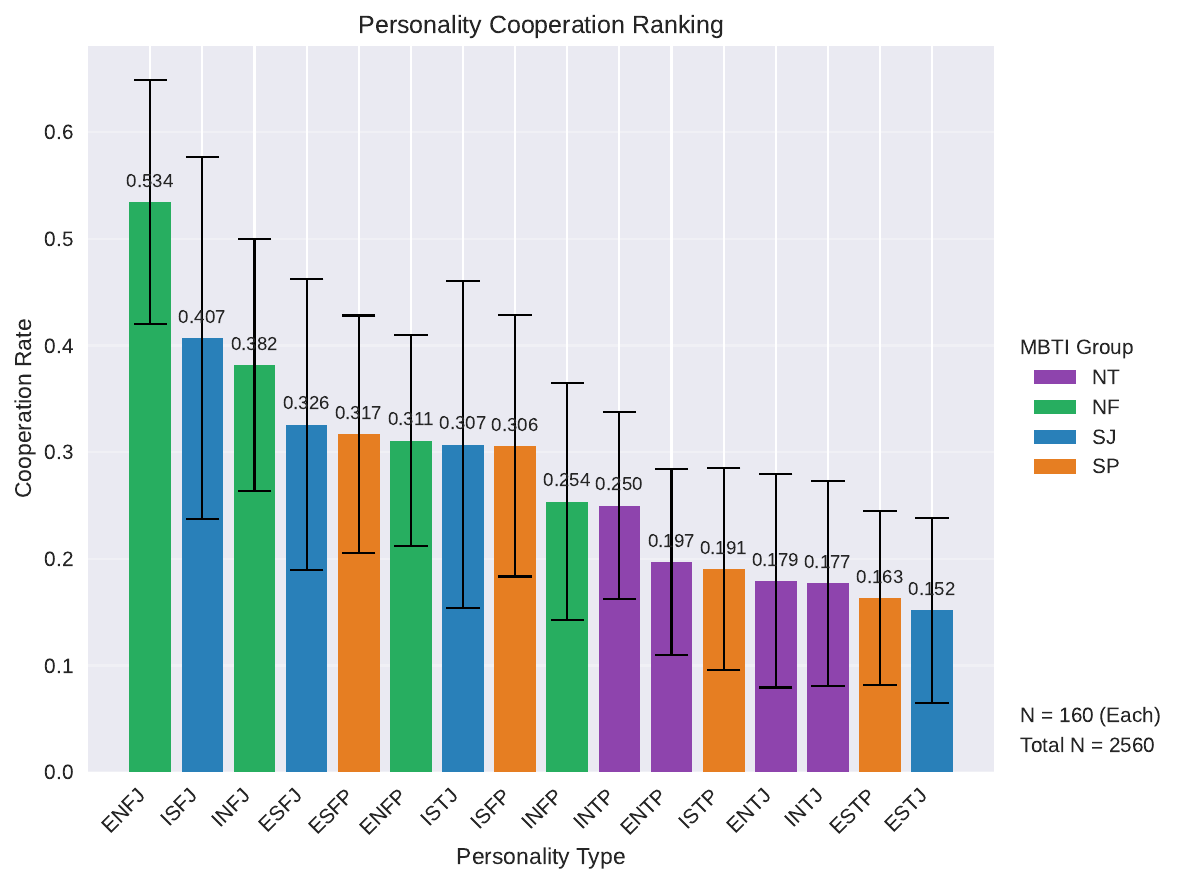}
    \includegraphics[width=0.49\linewidth]{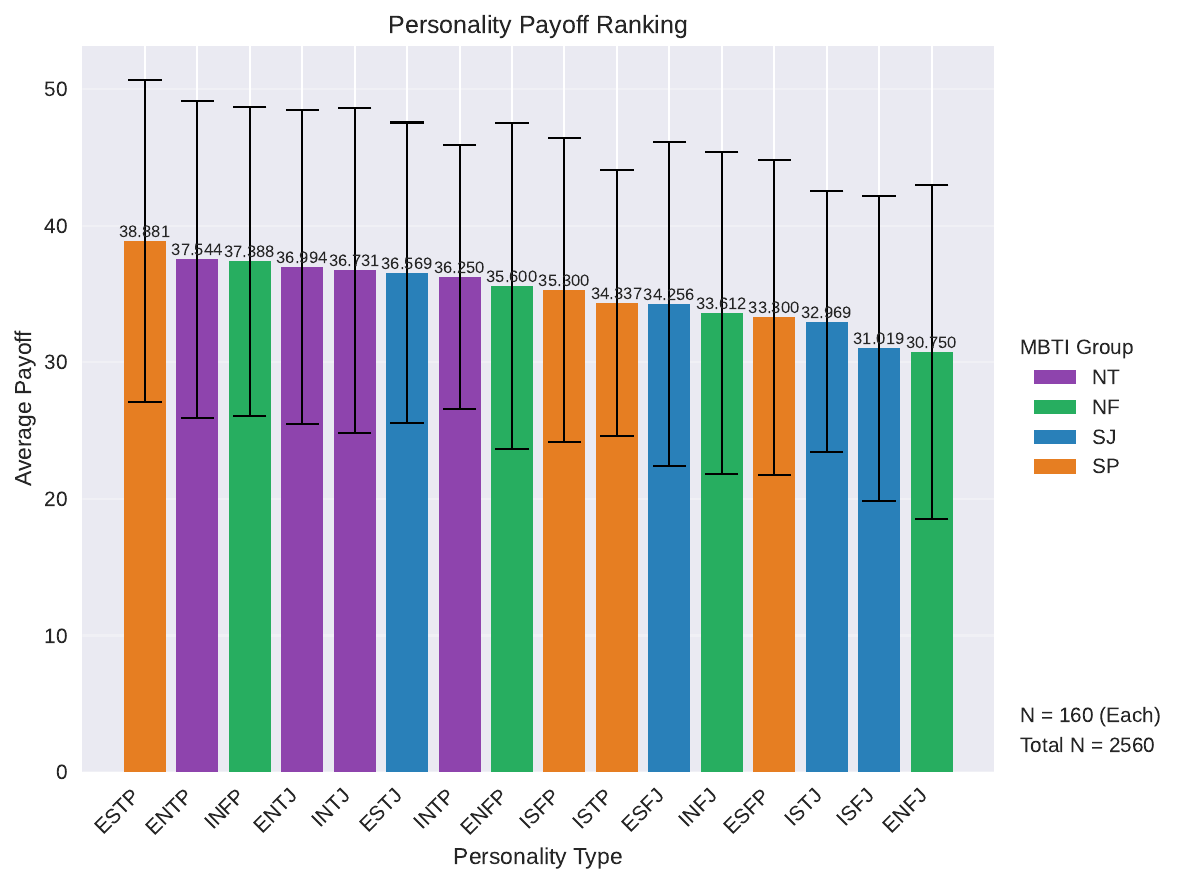}
    \caption{Results of Pairwise Game using Qwen model. While Qwen shows less polarized differentiation compared to LLaMA and Gemma, Welch's t-test over 10 independent trials still confirms that Thinking (T) types have significantly lower cooperation rates and significantly higher payoffs than Feeling (F) types.}
    \Description{Results of Pairwise Game using Qwen model.}
    \label{fig:qwen_pair_game_results}
\end{figure}

We evaluated the model's performance on Qwen\footnote{qwen2.5:1b, temperature=0.8, top\_p=0.9, deployed by ollama} and Gemma\footnote{gemma3:4b, temperature=0.8, top\_p=0.9, deployed by ollama}, with the same backend settings in our main experiment. All tested 10 independent trials with 20 rounds for each pair of personalities. The results are shown in Figure~\ref{fig:gemma_pair_game_results} and Figure~\ref{fig:qwen_pair_game_results}.

We observed that the overall cooperation rates for both models (Gemma: 0.299, Qwen: 0.278) were notably lower than that of LLaMA (0.490). Furthermore, different models react differently to personality injection; for instance, the Gemma model tends to be polarized (F-types have high cooperation rates, while T-types are near zero), whereas Qwen is relatively uniform. However, the general conclusion remains consistent: F-types exhibit statistically significantly higher cooperation rates compared to T-types, with the NF group showing particularly strong internal cohesion. Consistent with observations from previous experiments, this model also demonstrates that stable cooperation among F-types creates higher overall returns compared to the self-interested strategies often seen in T-types.

Notably, among F-types, ESFP shows relatively lower cooperation tendencies. It ranks lowest in both LLaMA and Gemma, and relatively low in Qwen, possibly due to their spontaneous, present-focused nature \cite{myersbriggs_types}, which conflicts with the strategic patience required for sustained cooperation in iterated games.

\subsubsection{Network Game using Gemma and Qwen}
We extended the network topology experiments (Section~\ref{sec:topology}) to Gemma and Qwen models. Table~\ref{tab:llm_topology_comparison} summarizes the results. While the absolute cooperation rates are significantly lower than those observed with LLaMA—consistent with the pairwise game results—the structural trends observed in the main experiments are largely preserved.

Specifically, the Regular network structure consistently outperforms the increasingly random Small-World variations for both models. As rewiring randomness increases (from Regular to SW $p=0.1$ to SW $p=0.5$), cooperation rates decline. This reinforces our finding regarding the ``Paradox of Connectivity'': adding random shortcuts to a lattice tends to destabilize cooperative clusters. The Scale-Free network shows variable performance, seemingly benefiting the Gemma model more than Qwen.

\begin{table}[htbp]
    \caption{Cooperation Rates Across Topologies for Different LLMs}
    \label{tab:llm_topology_comparison}
    \centering
    \begin{tabular}{lcc}
        \toprule
        \textbf{Topology}  & \textbf{Gemma} & \textbf{Qwen} \\
        \midrule
        Regular ($k=4$)    & 0.143          & 0.063         \\
        SW ($p=0.1$)       & 0.140          & 0.057         \\
        SW ($p=0.5$)       & 0.132          & 0.051         \\
        Scale-Free ($m=2$) & 0.182          & 0.063         \\
        \bottomrule
    \end{tabular}
\end{table}

\subsection{Bigger Network Size}
\label{app:bigger_network_size}
\subsubsection{Effect of Network Topology with Larger Sizes}
To verify that our findings regarding topological influence (Section~\ref{sec:topology}) are generally applicable and not artifacts of the specific network size ($N=50$), we scaled our simulations to $N=100$ and $N=300$ nodes. We maintained the same network generation parameters (average degree $k \approx 4$) and ran the simulations for 20 rounds.

As summarized in Table~\ref{tab:network_size_scaling}, the hierarchy of cooperation across different topologies remains robust at larger scales. Regular networks consistently maintain the highest cooperation rates (reaching $\sim$0.55), reinforcing the finding that high local clustering is crucial for sustaining cooperation. The trend of decreasing cooperation with increasing randomness persists: Small-World networks with low rewiring ($p=0.1$) outperform those with high rewiring ($p=0.5$). Scale-Free networks continue to exhibit lower cooperation rates compared to regular structures. These results confirm that the ``paradox of connectivity''—where structural shortcuts destabilize cooperation—is a scalable inference that holds as population size increases.

\begin{table}[htbp]
    \caption{Impact of Network Size on Cooperation and Payoff}
    \label{tab:network_size_scaling}
    \centering
    \resizebox{\columnwidth}{!}{
        \begin{tabular}{lcccc}
            \toprule
            \multirow{2}{*}{\textbf{Topology}} & \multicolumn{2}{c}{\textbf{100 Nodes}} & \multicolumn{2}{c}{\textbf{300 Nodes}}                                              \\
            \cmidrule(lr){2-3} \cmidrule(lr){4-5}
                                               & \textbf{Coop. Rate}                    & \textbf{Avg. Payoff}                   & \textbf{Coop. Rate} & \textbf{Avg. Payoff} \\
            \midrule
            Regular ($k=4$)                    & \textbf{0.553}                         & \textbf{10.48}                         & \textbf{0.550}      & \textbf{10.51}       \\
            SW ($p=0.1$)                       & 0.526                                  & 10.34                                  & 0.498               & 10.09                \\
            SW ($p=0.5$)                       & 0.477                                  & 10.08                                  & 0.495               & 10.18                \\
            Scale-Free ($m=2$)                 & 0.465                                  & 9.91                                   & 0.470               & 9.99                 \\
            \bottomrule
        \end{tabular}
    }
\end{table}

\subsubsection{Effect of Personality Distribution with Larger Sizes}
We similarly extended our ``Hub Leadership'' investigation (Section~\ref{sec:hubs}) to larger Scale-Free networks with $N=100$ and $N=300$. We followed the same protocol: employing the Barabási-Albert model ($m=2$) and fixing the personalities of the top 10\% highest-degree nodes to either \textbf{ESFJ} (Pro-Social) or \textbf{ENTJ} (Rational), while keeping the remaining 90\% random.

As detailed in Table~\ref{tab:hub_scaling}, the ``influencer effect'' proves to be scalable. In both network sizes, the \textbf{Pro-Social Hubs} scenario significantly outperforms the Uniform baseline, driving average cooperation rates to approximately 67\% ($N=100$) and 62\% ($N=300$). Conversely, \textbf{Rational Hubs} consistently suppress collective cooperation below baseline levels.

\begin{figure}[htbp]
    \centering
    \includegraphics[width=1\linewidth]{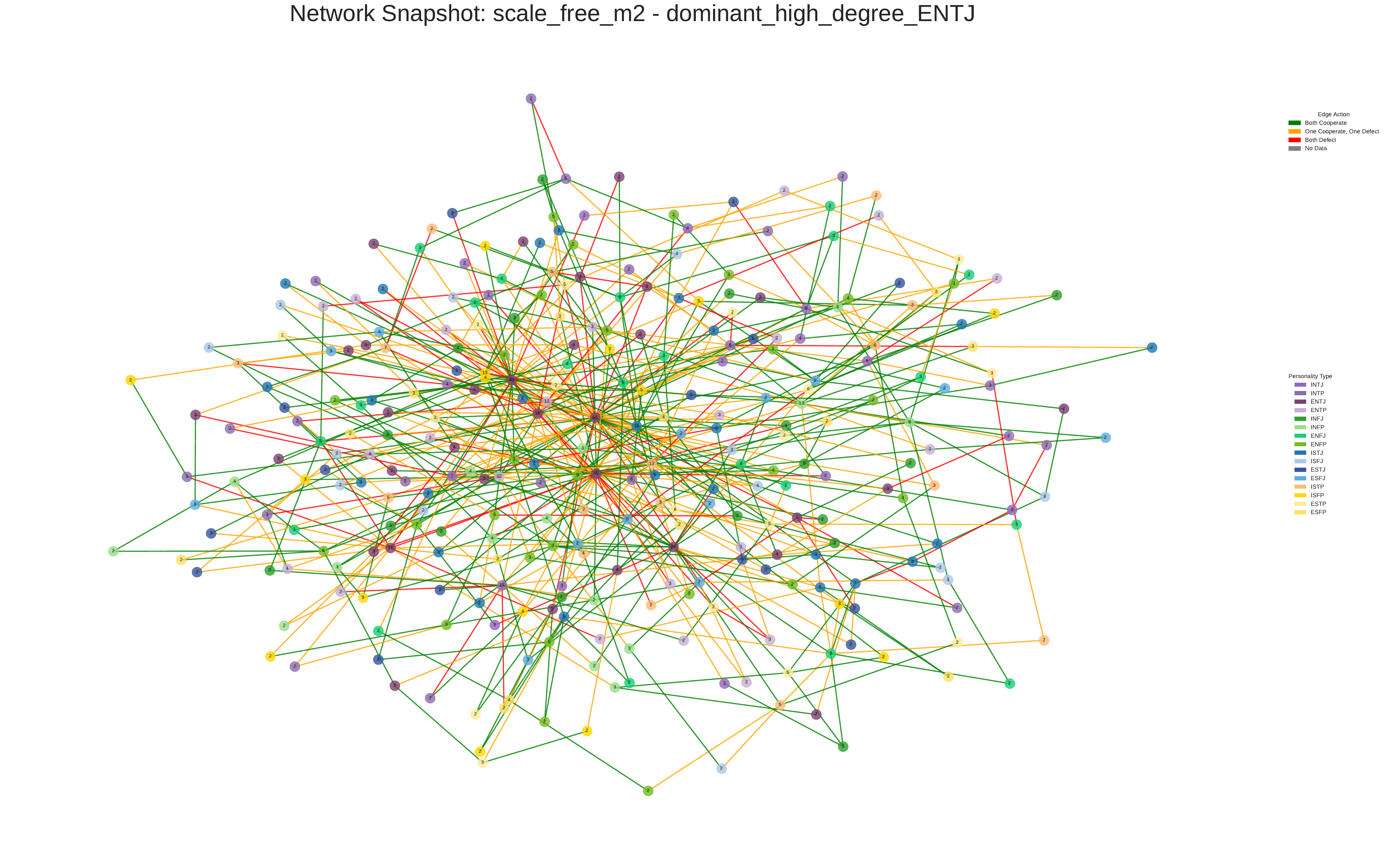}
    \caption{Final-round network snapshot of the Scale-Free network with 300 nodes under the Rational (ENTJ) Hubs condition. Similar to the smaller network, hubs become focal points for defection. However, due to the larger diameter, some peripheral cooperative clusters manage to persist, slightly mitigating the overall collapse of cooperation.}
    \Description{Final-round network snapshot of the Scale-Free network with 300 nodes under the Rational (ENTJ) Hubs condition.}
    \label{fig:sf_300nodes_snapshot}
\end{figure}

We note a mild attenuation of the suppression effect in the largest network ($N=300$), where the Rational Hubs scenario yields a cooperation rate of 0.388, compared to 0.280 in the 50-node experiment. This suggests that as the network diameter increases, the direct reach of central hubs is slightly diluted, allowing distant nodes to form local cooperative clusters independent of the central leadership(As shown in Figure~\ref{fig:sf_300nodes_snapshot}). Nevertheless, the qualitative dominance of hub personality remains a critical leverage point for system design.

\begin{table}[htbp]
    \caption{Scalability of Hub Influence in Scale-Free Networks}
    \label{tab:hub_scaling}
    \centering
    \resizebox{\columnwidth}{!}{
        \begin{tabular}{lcccc}
            \toprule
            \multirow{2}{*}{\textbf{Hub Scenario}} & \multicolumn{2}{c}{\textbf{100 Nodes}} & \multicolumn{2}{c}{\textbf{300 Nodes}}                                              \\
            \cmidrule(lr){2-3} \cmidrule(lr){4-5}
                                                   & \textbf{Coop. Rate}                    & \textbf{Avg. Payoff}                   & \textbf{Coop. Rate} & \textbf{Avg. Payoff} \\
            \midrule
            Uniform Baseline                       & 0.460                                  & 9.87                                   & 0.462               & 9.91                 \\
            Pro-Social (ESFJ)                      & \textbf{0.669}                         & \textbf{10.75}                         & \textbf{0.615}      & \textbf{10.69}       \\
            Rational (ENTJ)                        & 0.310                                  & 8.94                                   & 0.388               & 9.54                 \\
            \bottomrule
        \end{tabular}
    }
\end{table}

\subsection{Alternative Personality Description}
\label{app:alternative_personality_description}
This subsection tests robustness to variations in personality prompting. In our main experiments, we use concise personality-inspired descriptions generated to represent consistent behavioral priors. To verify that our findings are not artifacts of a particular prompt wording, we additionally conduct a robustness experiment using an alternative textual realization of personality traits based on the official MBTI type descriptions published by The Myers-Briggs Company \cite{myersbriggs_types}.

\begin{figure}[htbp]
    \centering
    \begin{minipage}{1.0\linewidth}
        \centering
        \includegraphics[width=0.49\linewidth]{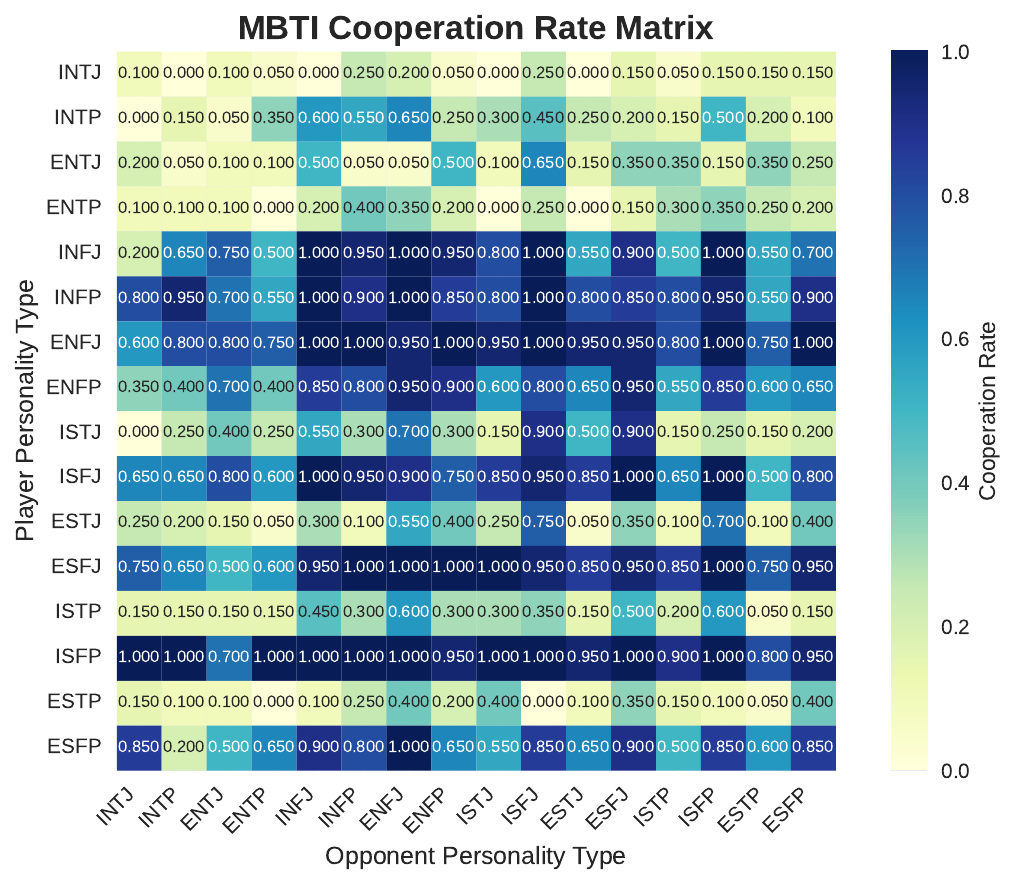}
        \hfill
        \includegraphics[width=0.49\linewidth]{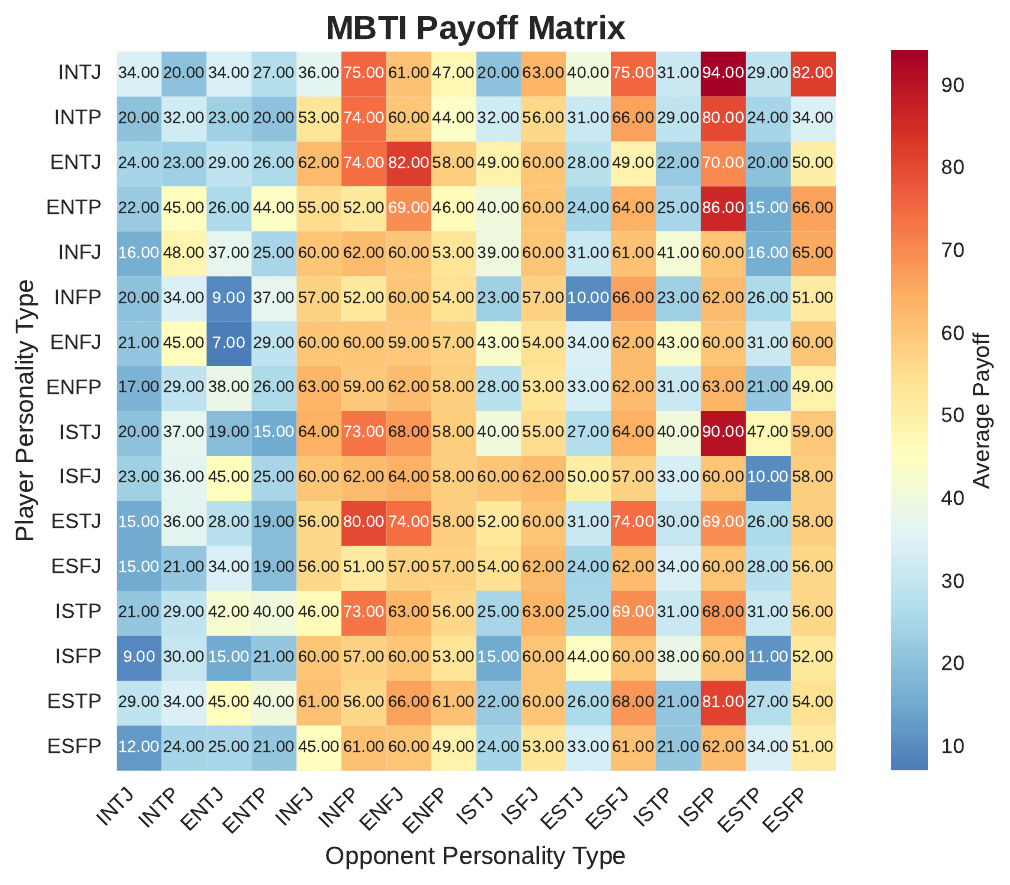}
    \end{minipage}
    \begin{minipage}{1.0\linewidth}
        \centering
        \includegraphics[width=0.49\linewidth]{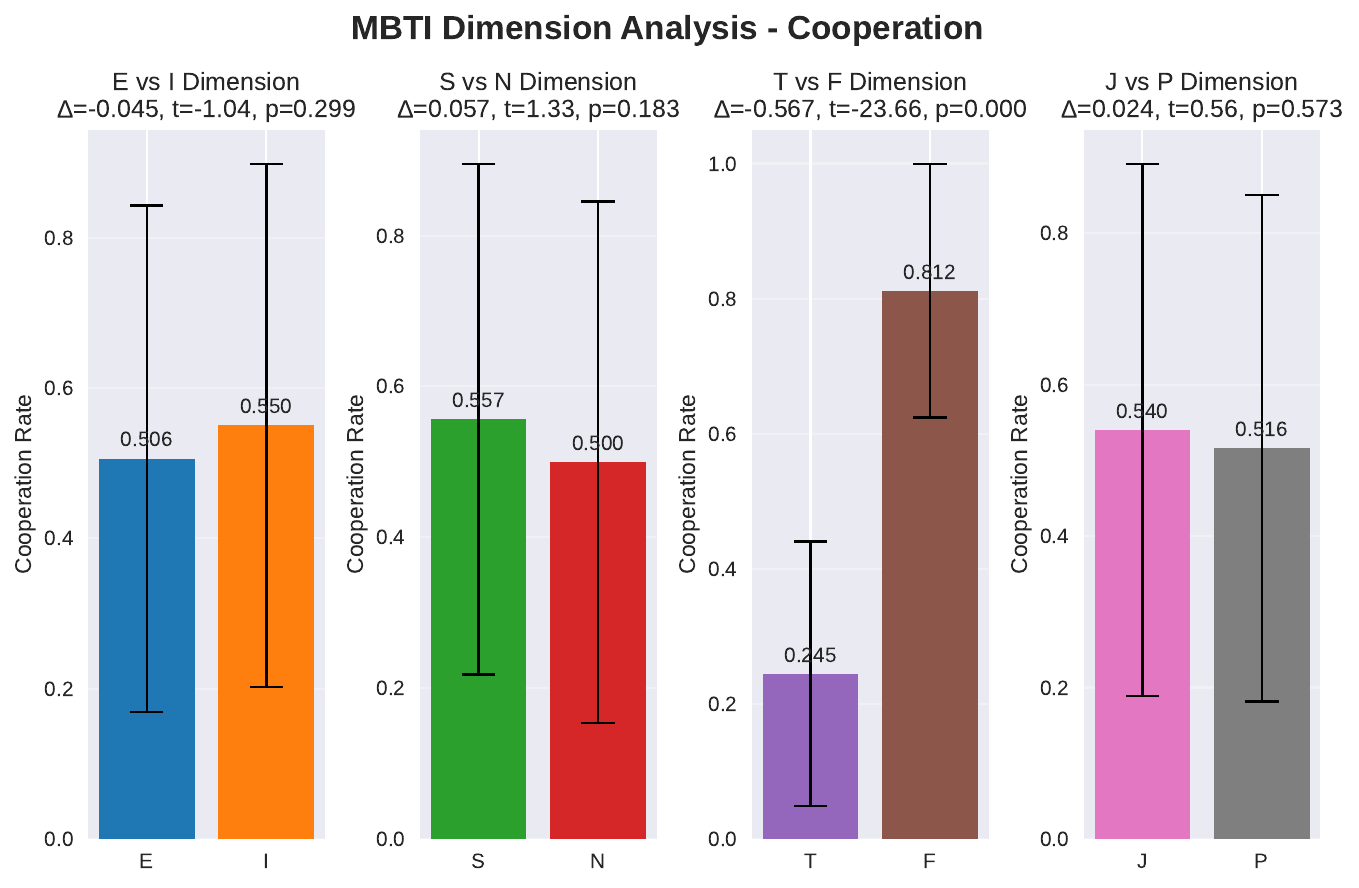}
        \hfill
        \includegraphics[width=0.49\linewidth]{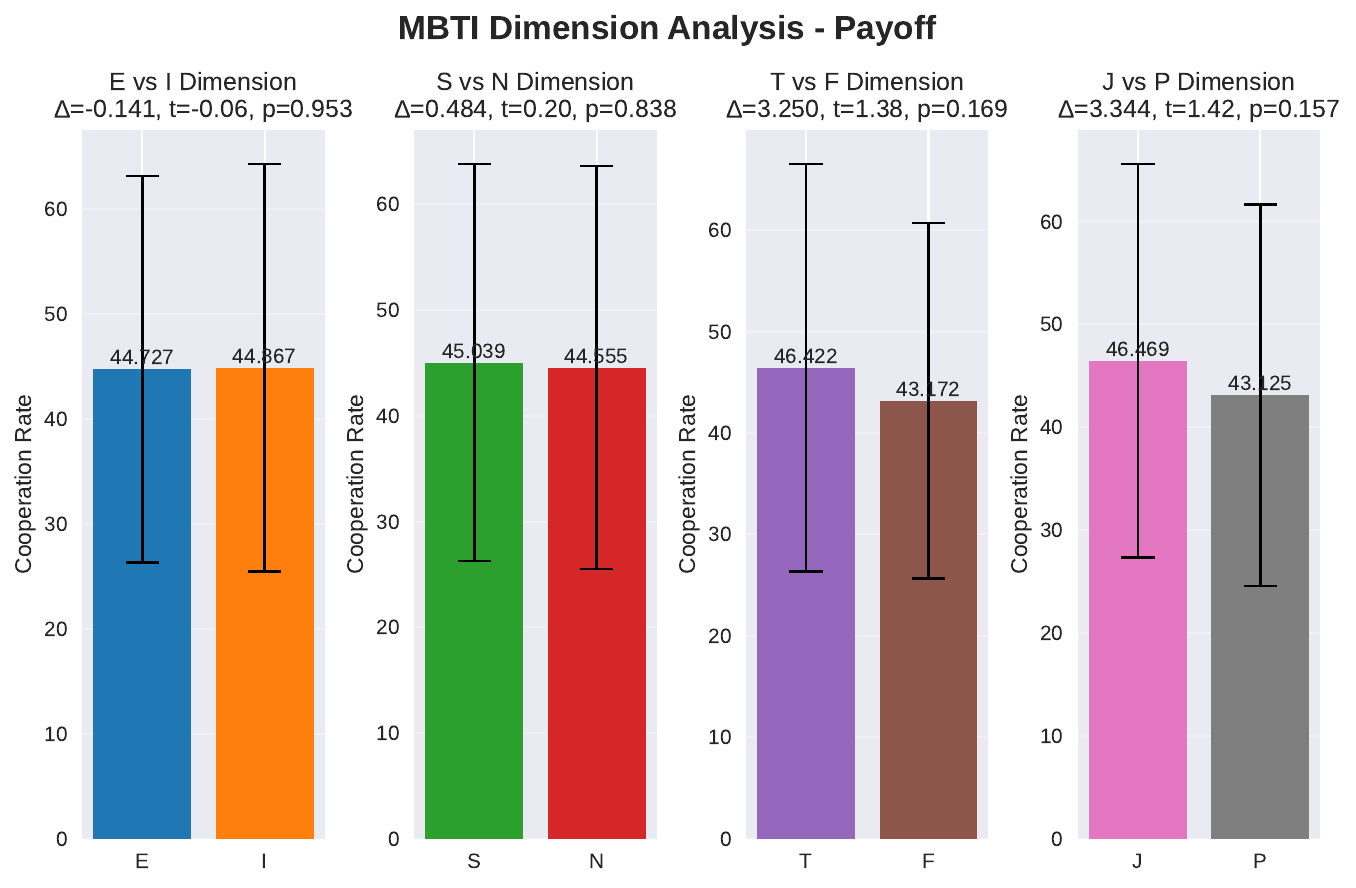}
    \end{minipage}
    \begin{minipage}{1.0\linewidth}
        \centering
        \includegraphics[width=0.49\linewidth]{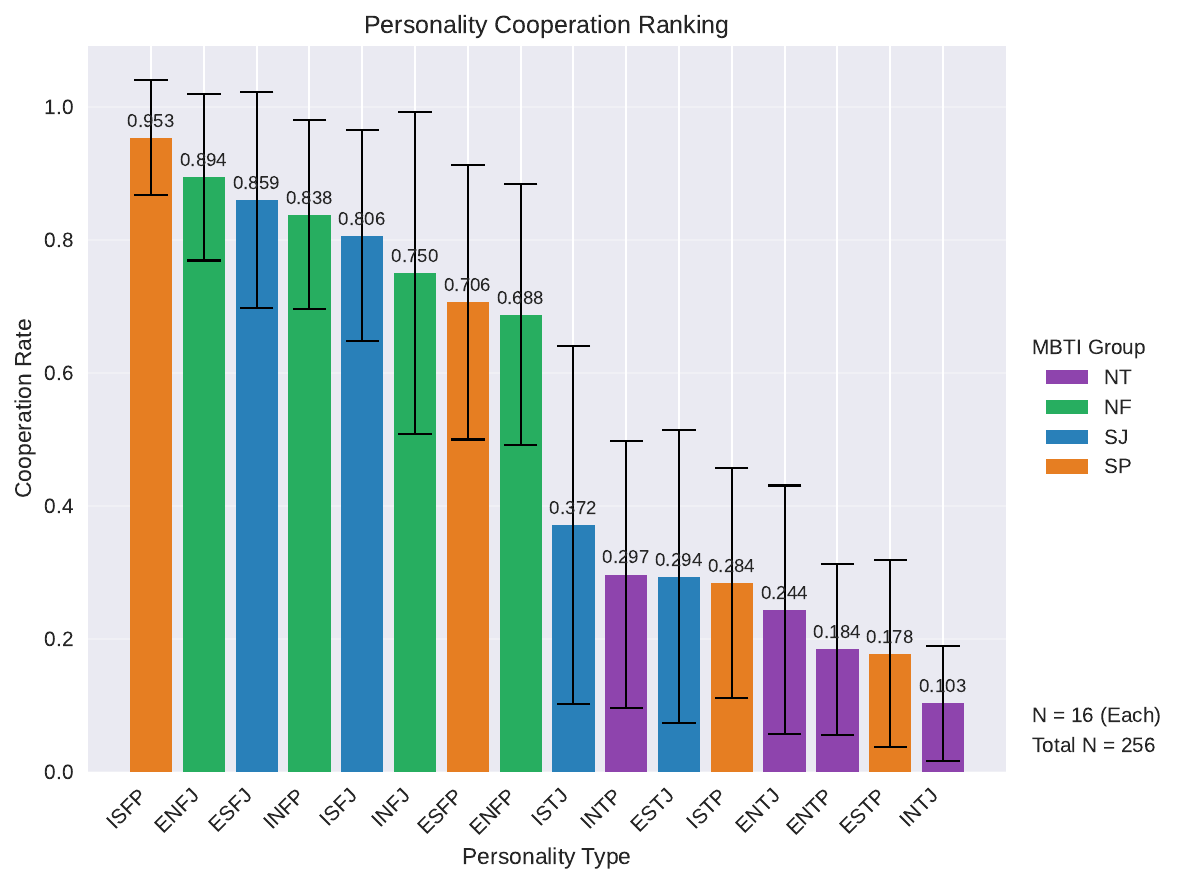}
        \hfill
        \includegraphics[width=0.49\linewidth]{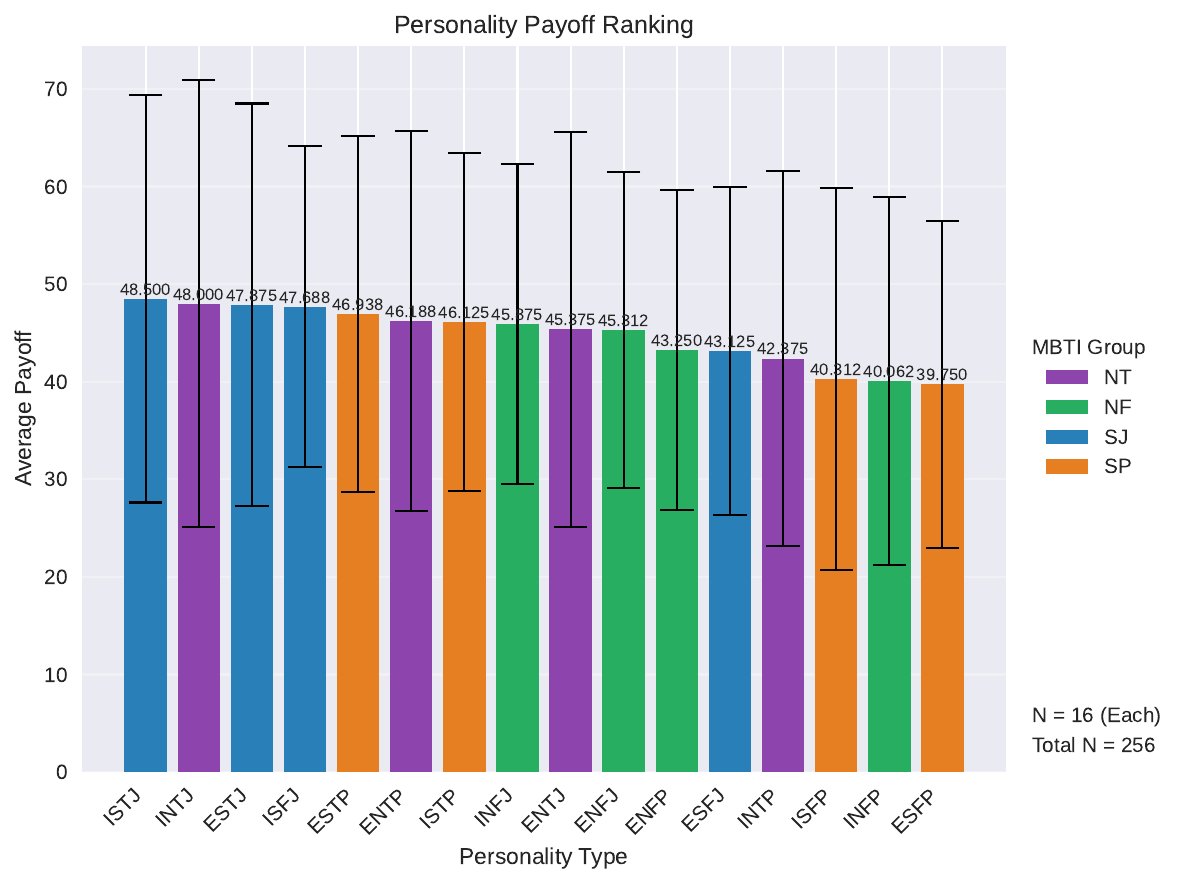}
    \end{minipage}
    \caption{Results using Myers-Briggs official personality descriptions. Both the cooperation heatmap and the payoff heatmap closely resemble the original experiment. The rows of F-type personality are significantly higher in the cooperation heatmap, while the columns are notably higher when the opponent is of F-type personality in the payoff heatmap. Personality rankings are also similar to the original findings, with F-types generally ranking higher in cooperation and T-types achieving higher payoffs.}
    \Description{Results using Myers-Briggs official personality descriptions.}
    \label{fig:myers-briggs_personality}
\end{figure}

As shown in Figure~\ref{fig:myers-briggs_personality}, the resulting dyadic cooperation matrix exhibits highly similar qualitative patterns to those obtained with our primary prompts, including consistent relative cooperation tendencies across personality types, confirming that the observed behavioral patterns are robust to variations in personality prompt phrasing.

\section{Ablation Study}
\label{app:ablation_study}
\subsection{Personality Injection Ablation}
\label{app:personality_ablation}
\subsubsection{Weak Personality Injection}
Weaker personality injection means only providing the MBTI type label without detailed personality descriptions. For example: You are an ESTP (Entrepreneur). Please make game decisions with your personality traits.

\begin{figure}[htbp]
    \centering
    \begin{minipage}{1.0\linewidth}
        \centering
        \includegraphics[width=0.49\linewidth]{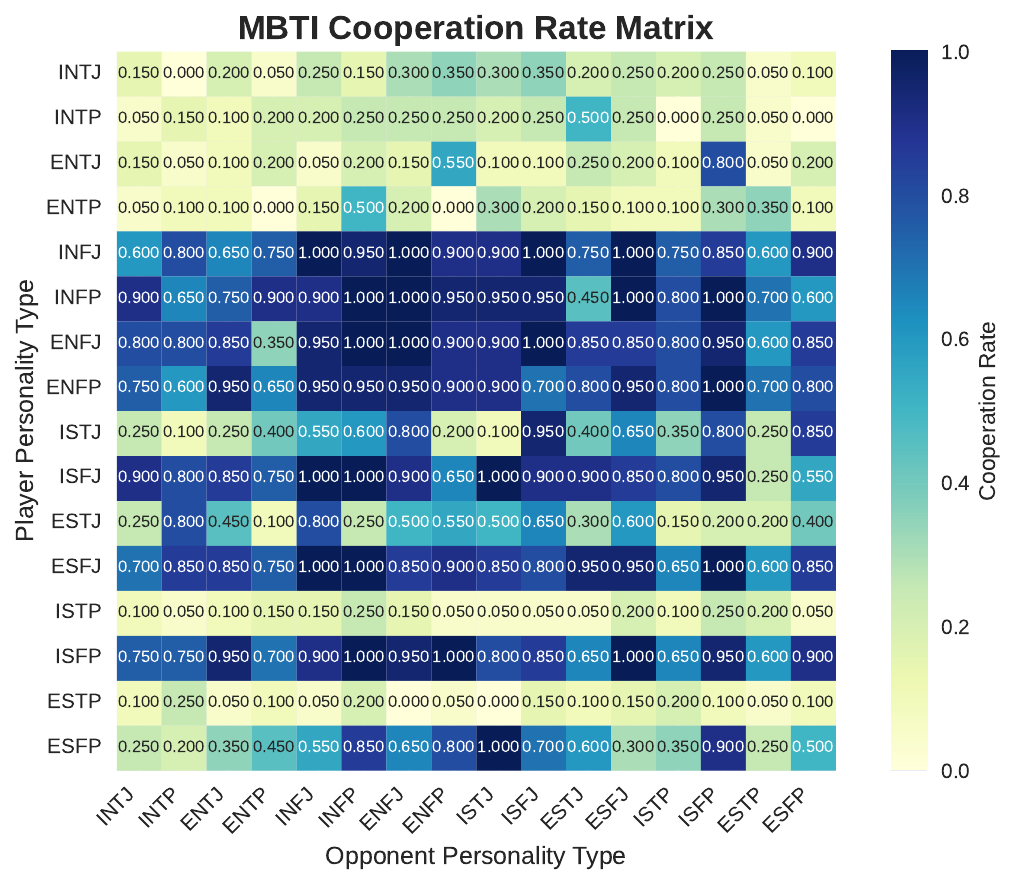}
        \hfill
        \includegraphics[width=0.49\linewidth]{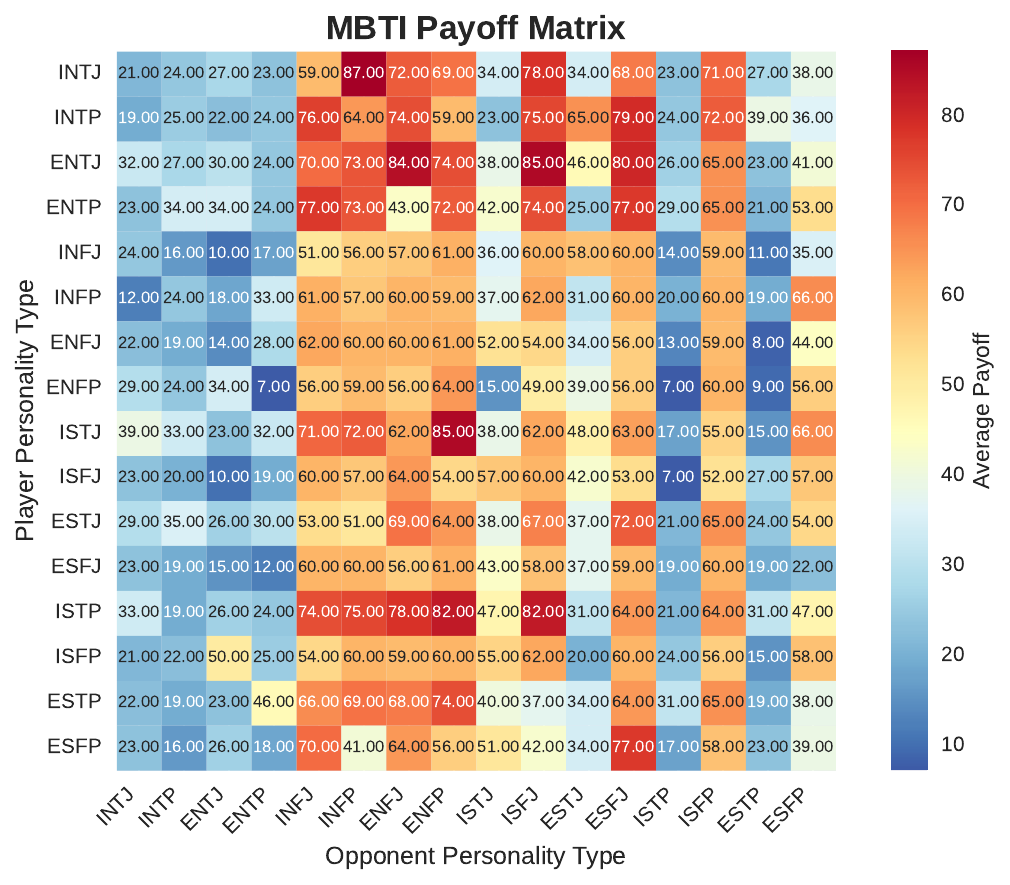}
    \end{minipage}
    \begin{minipage}{1.0\linewidth}
        \centering
        \includegraphics[width=0.49\linewidth]{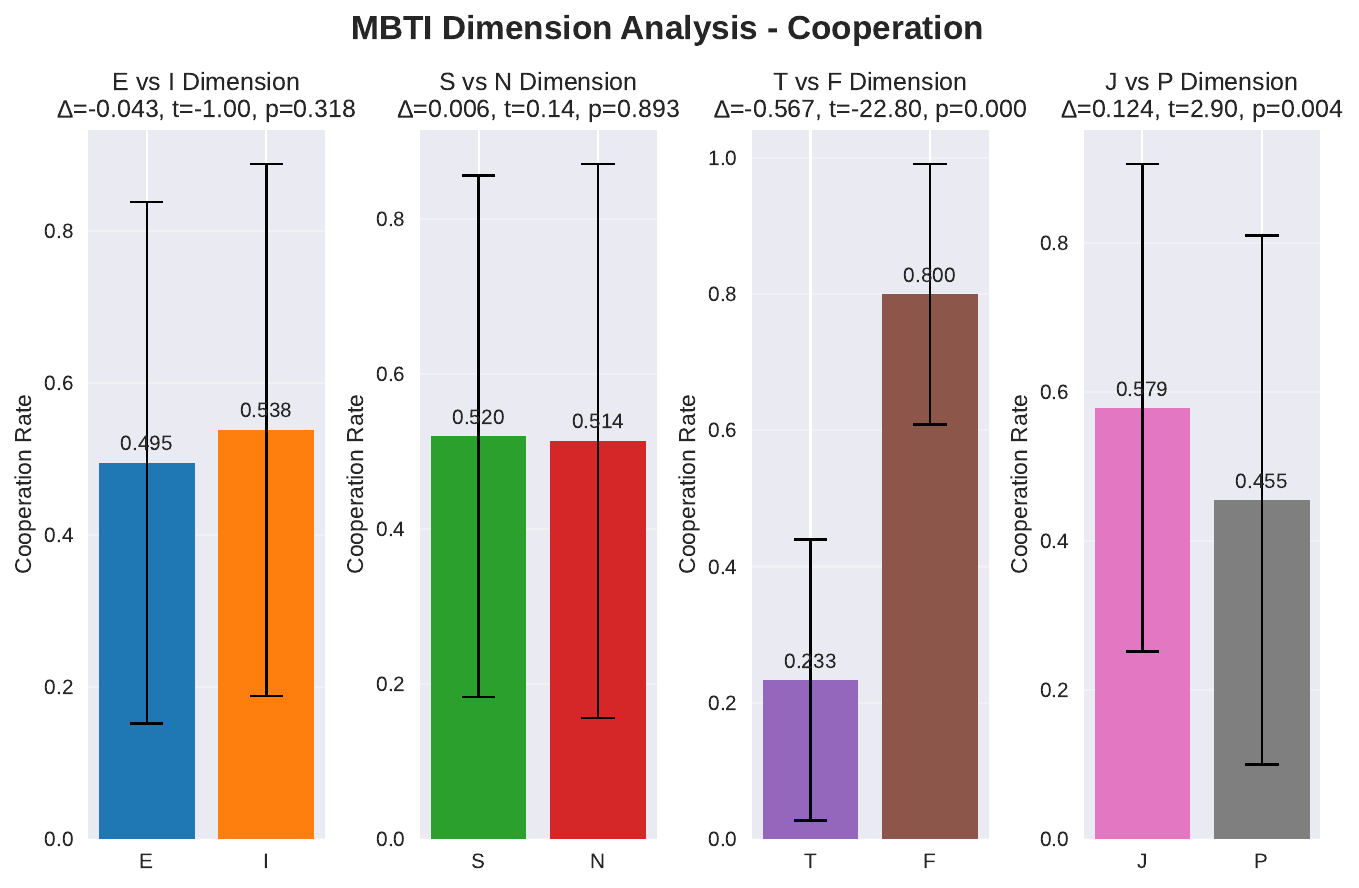}
        \hfill
        \includegraphics[width=0.49\linewidth]{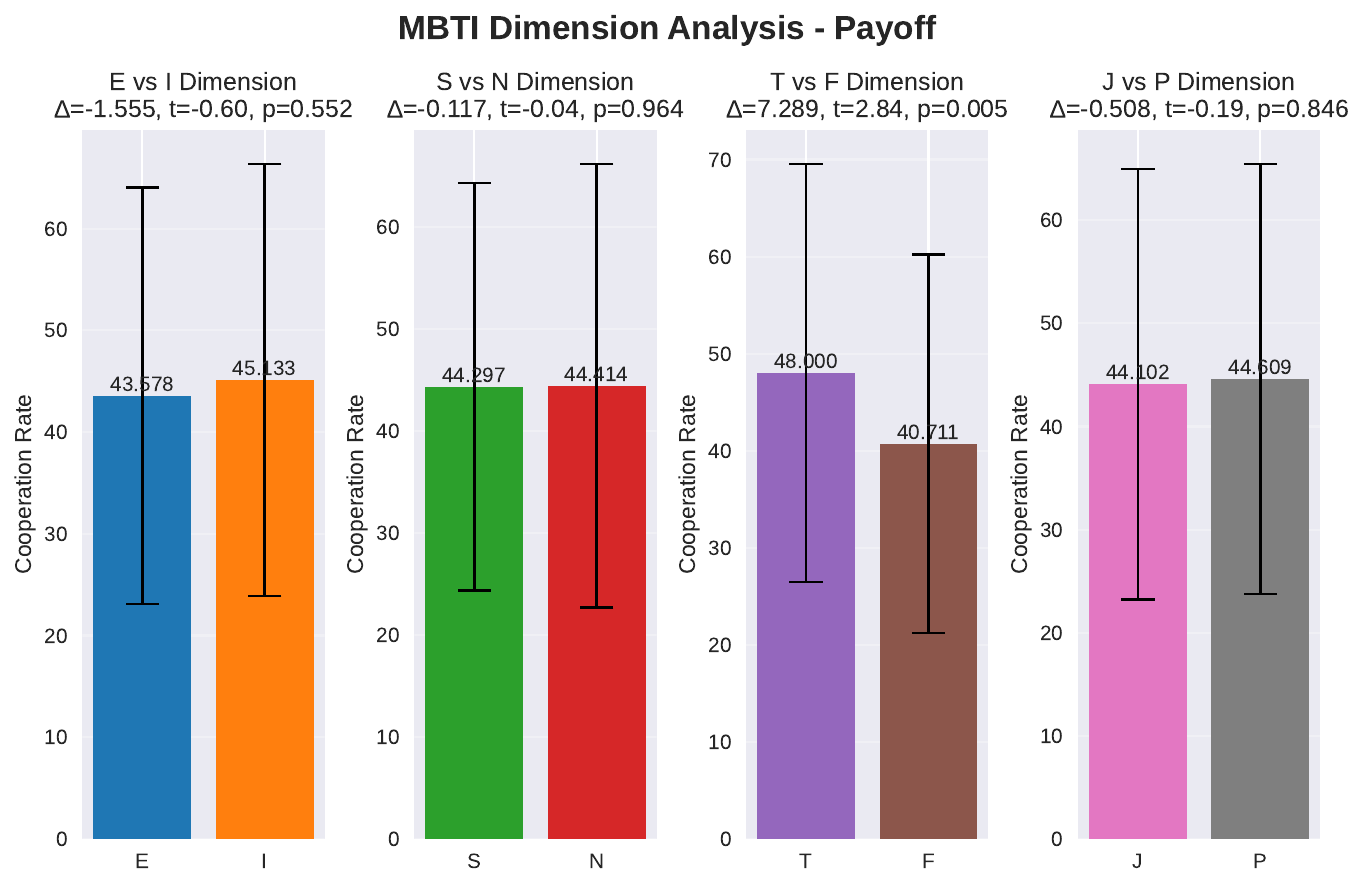}
    \end{minipage}
    \begin{minipage}{1.0\linewidth}
        \centering
        \includegraphics[width=0.49\linewidth]{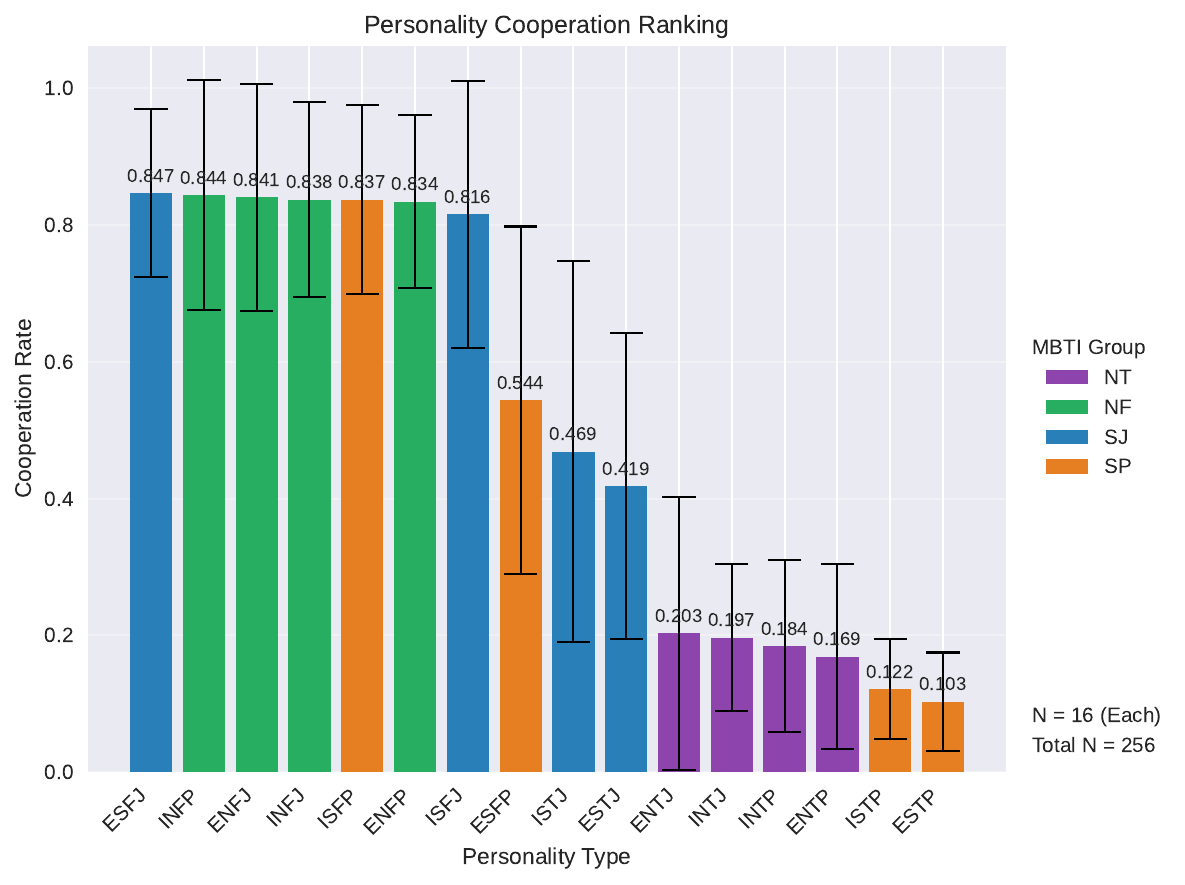}
        \hfill
        \includegraphics[width=0.49\linewidth]{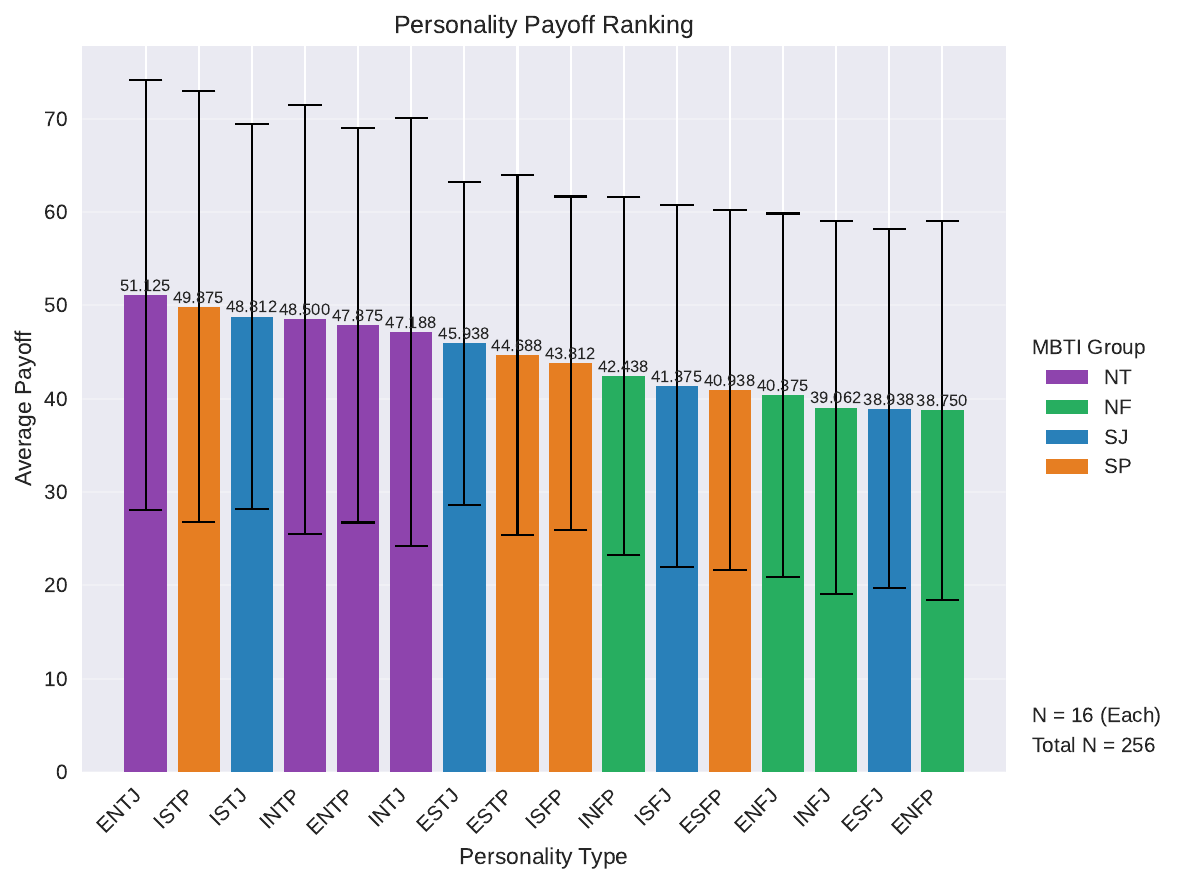}
    \end{minipage}
    \caption{Results of Pairwise Game with Weak Personality Injection. The results parallel the strong personality injection experiment. In the cooperation heatmap, the rows for Feeling (F) types show higher cooperation propensity, while in the payoff heatmap, columns for F types yield higher returns for opponents. The NT vs. NF region (Block 2) again displays NTs achieving temptation payoffs. Welch's t-test confirms T-types have significantly lower cooperation but higher payoffs. Notably, ESFP ranks lowest in cooperation among F types again.}
    \Description{Results of Pairwise Game with Weak Personality Injection.}
    \label{fig:ablation_weak_personality}
\end{figure}

Results are shown in Figure~\ref{fig:ablation_weak_personality}. Interestingly, agents with only weak personality injection appear to exhibit slightly more polarized behavior. The cooperation heatmap shows that many NF groups achieve 100\% collaboration rates. This suggests that without specific natural language descriptions, LLMs may tend to construct agents with more typical stereotypical personalities.

\subsubsection{Opponent Info Ablation}
\begin{figure}[htbp]
    \centering
    \begin{minipage}{1.0\linewidth}
        \centering
        \includegraphics[width=0.49\linewidth]{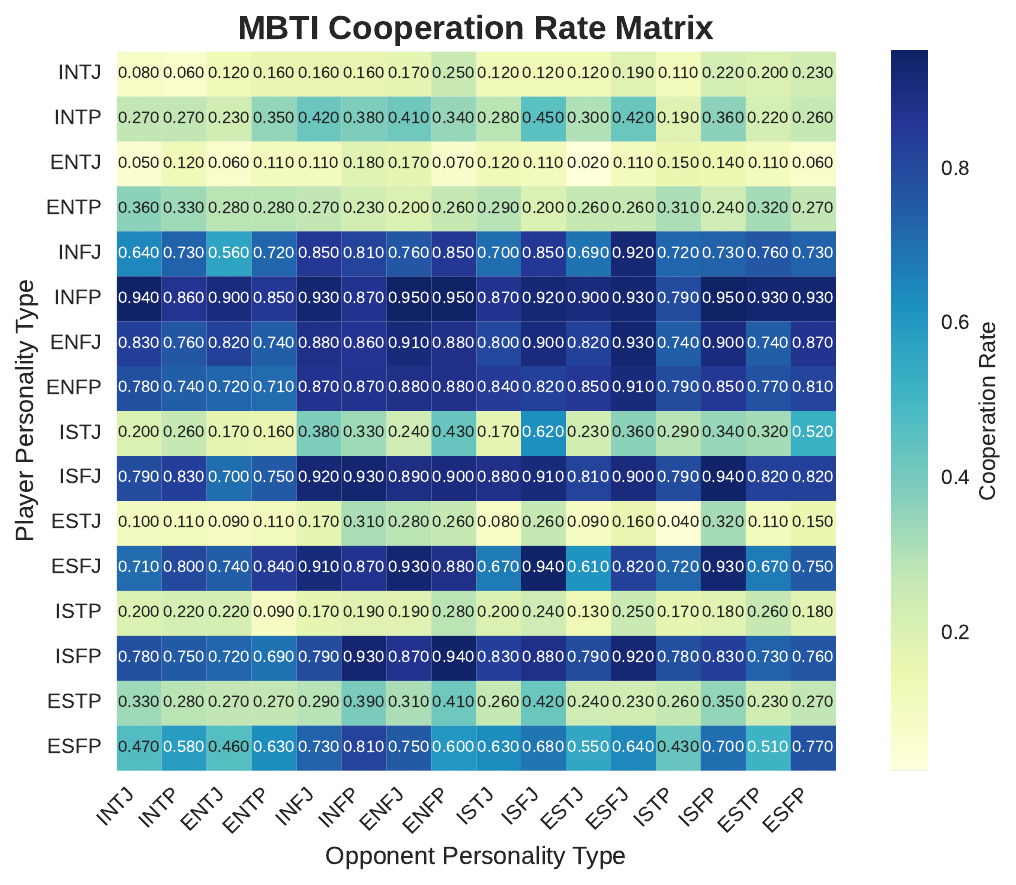}
        \hfill
        \includegraphics[width=0.49\linewidth]{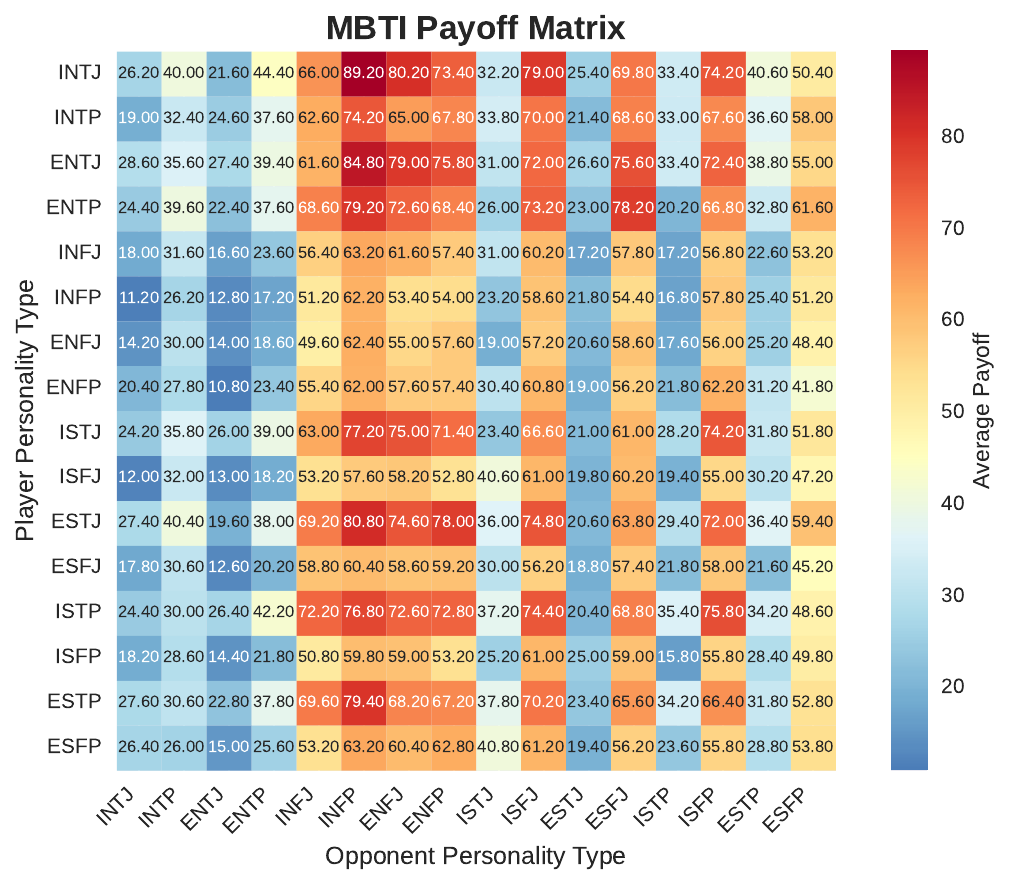}
    \end{minipage}
    \begin{minipage}{1.0\linewidth}
        \centering
        \includegraphics[width=0.49\linewidth]{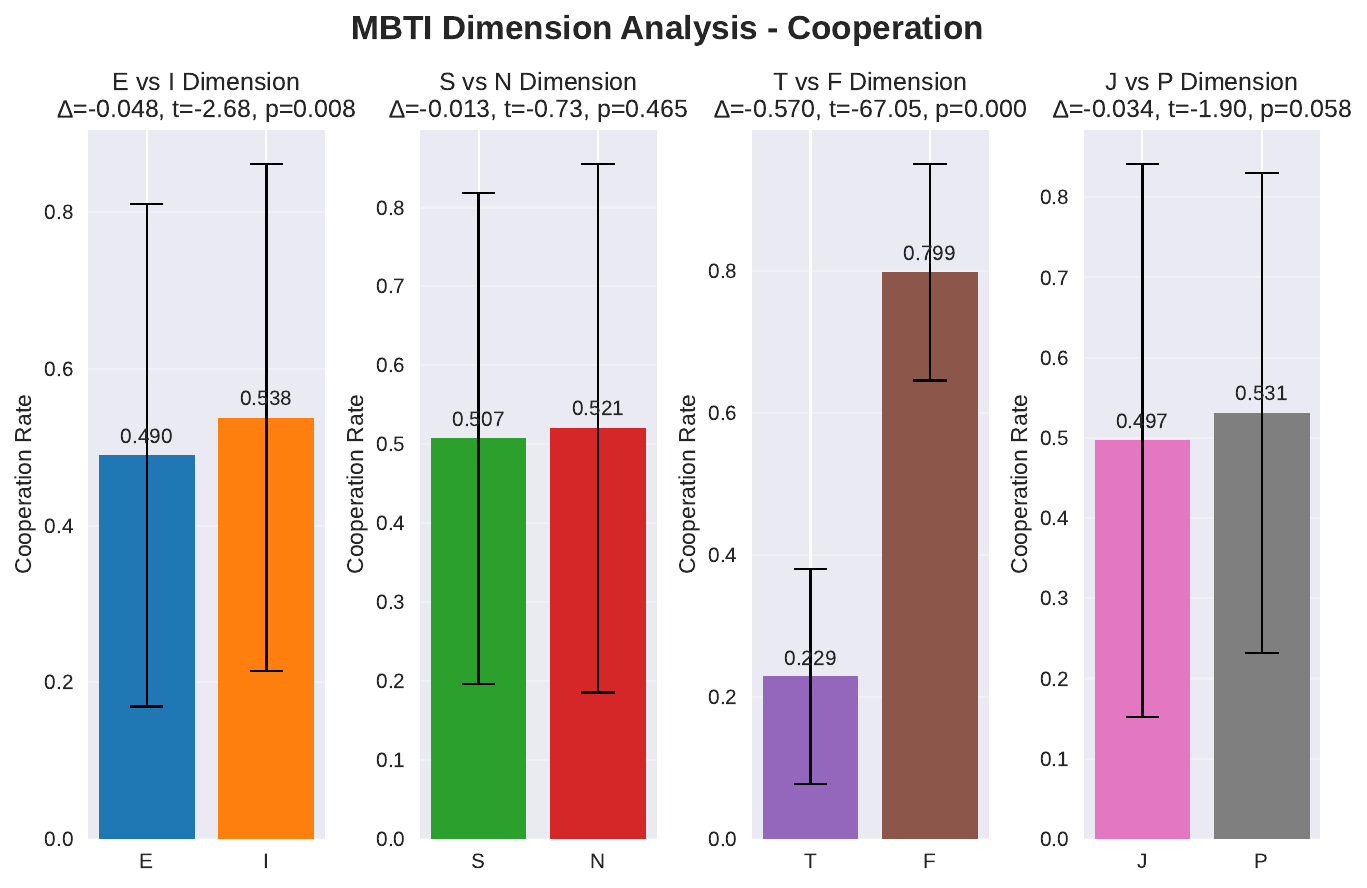}
        \hfill
        \includegraphics[width=0.49\linewidth]{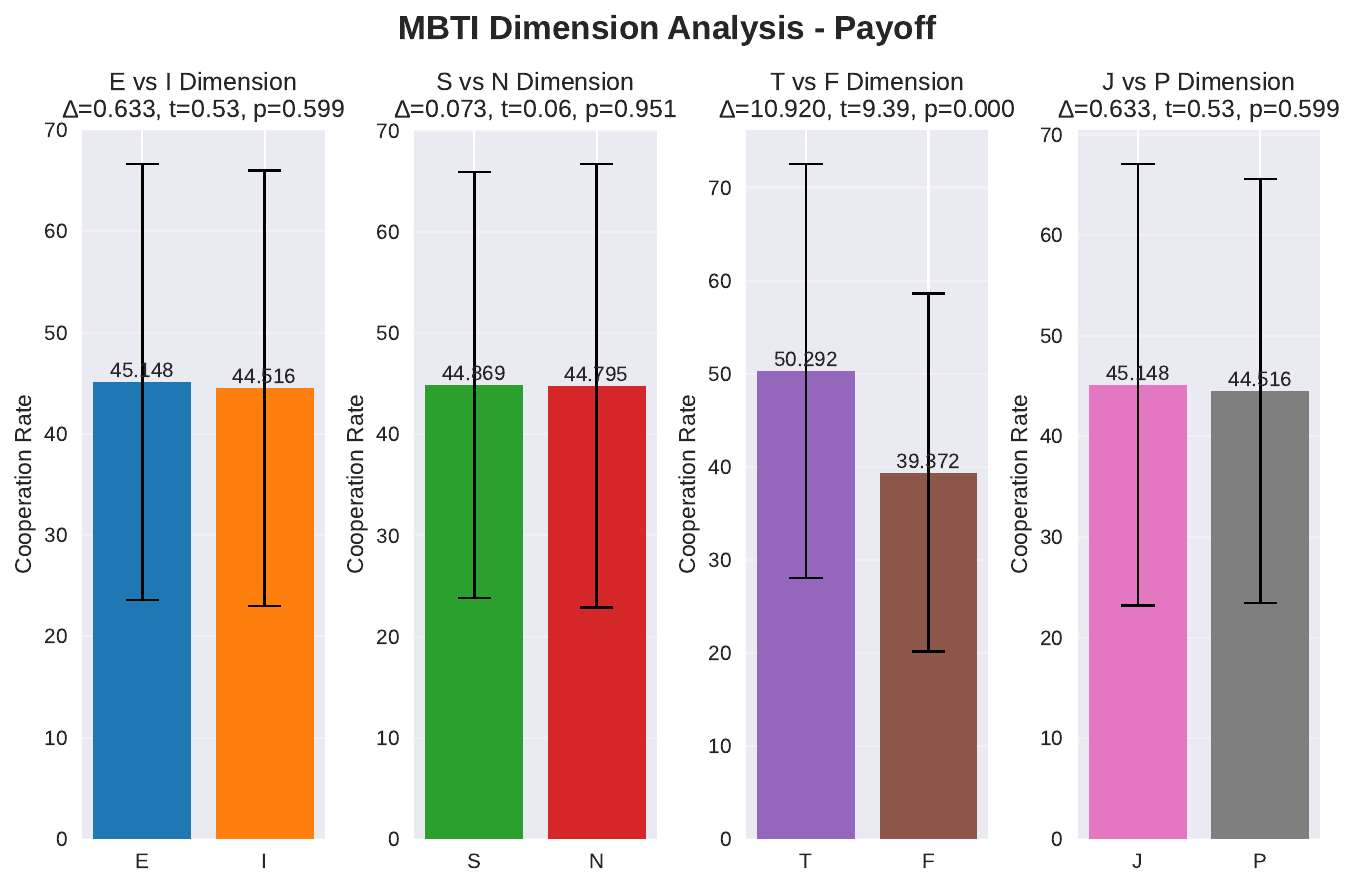}
    \end{minipage}
    \begin{minipage}{1.0\linewidth}
        \centering
        \includegraphics[width=0.49\linewidth]{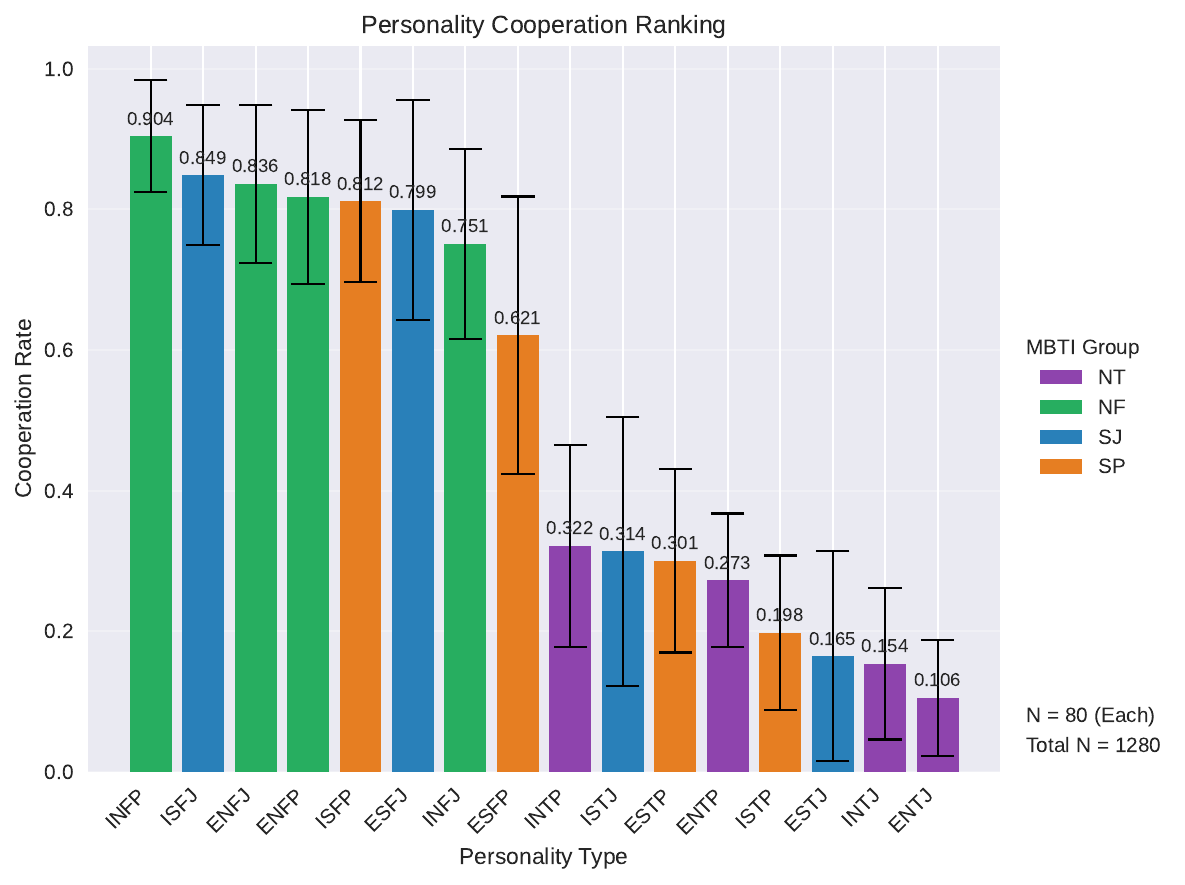}
        \hfill
        \includegraphics[width=0.49\linewidth]{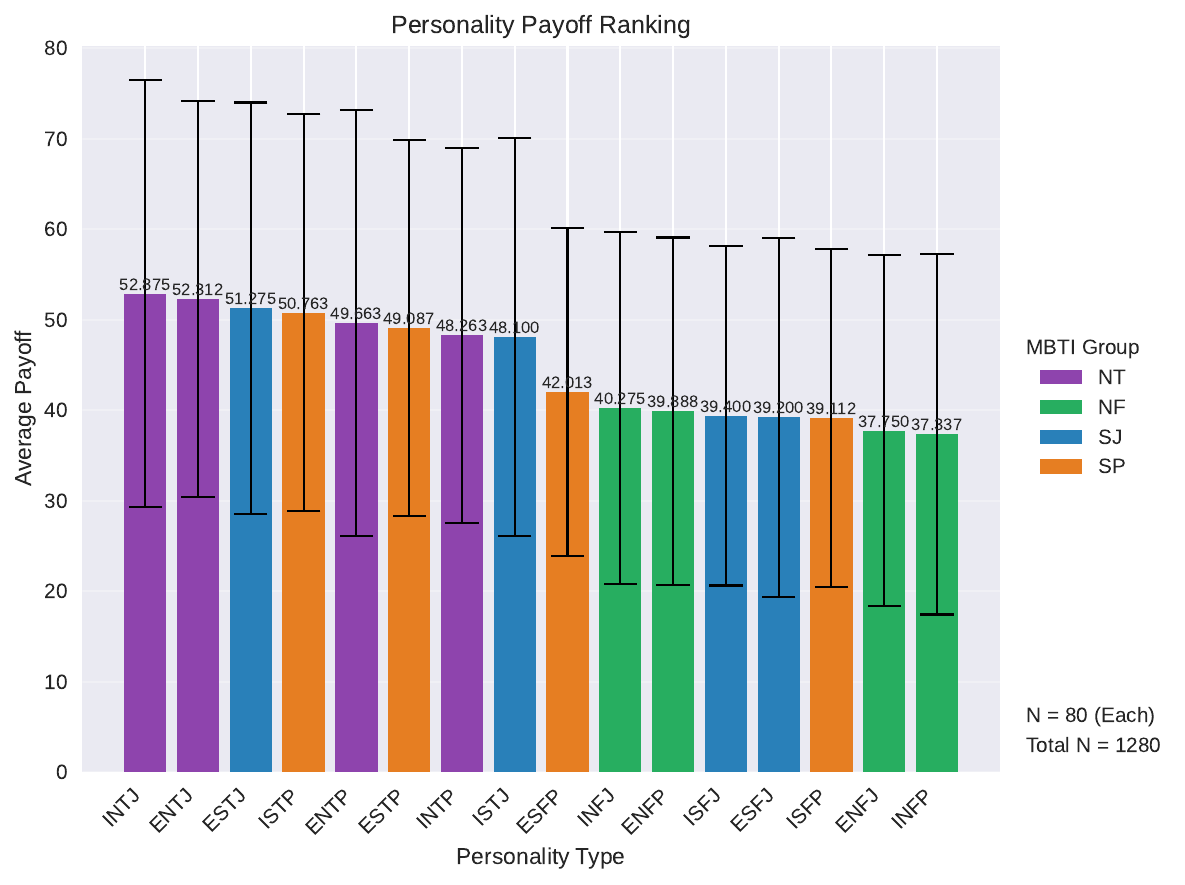}
    \end{minipage}
    \caption{Results of Pairwise Game with No Opponent Info. Even without opponent personality cues, agents still exhibit significant behavioral differences based on their own personality types. Feeling (F) types maintain higher cooperation rates, while Thinking (T) types achieve higher payoffs, consistent with the full-information scenario. This suggests that agents rely primarily on their own personality-driven strategies rather than adapting to opponent types.}
    \Description{Results of Pairwise Game with No Opponent Info.}
    \label{fig:ablation_no_opponent_info}
\end{figure}

We also conducted experiments without providing opponent personality information, where agents do not know the personality type of their opponents. Results are shown in Figure~\ref{fig:ablation_no_opponent_info}. Even without opponent personality cues, agents still exhibit significant behavioral differences based on their own personality types. Feeling (F) types maintain higher cooperation rates, while Thinking (T) types achieve higher payoffs, consistent with the full-information scenario. This suggests that agents rely primarily on their own personality-driven strategies rather than adapting to opponent types.

\subsubsection{No Personality Injection}
\begin{figure}[htbp]
    \centering
    \begin{minipage}{1.0\linewidth}
        \centering
        \includegraphics[width=0.49\linewidth]{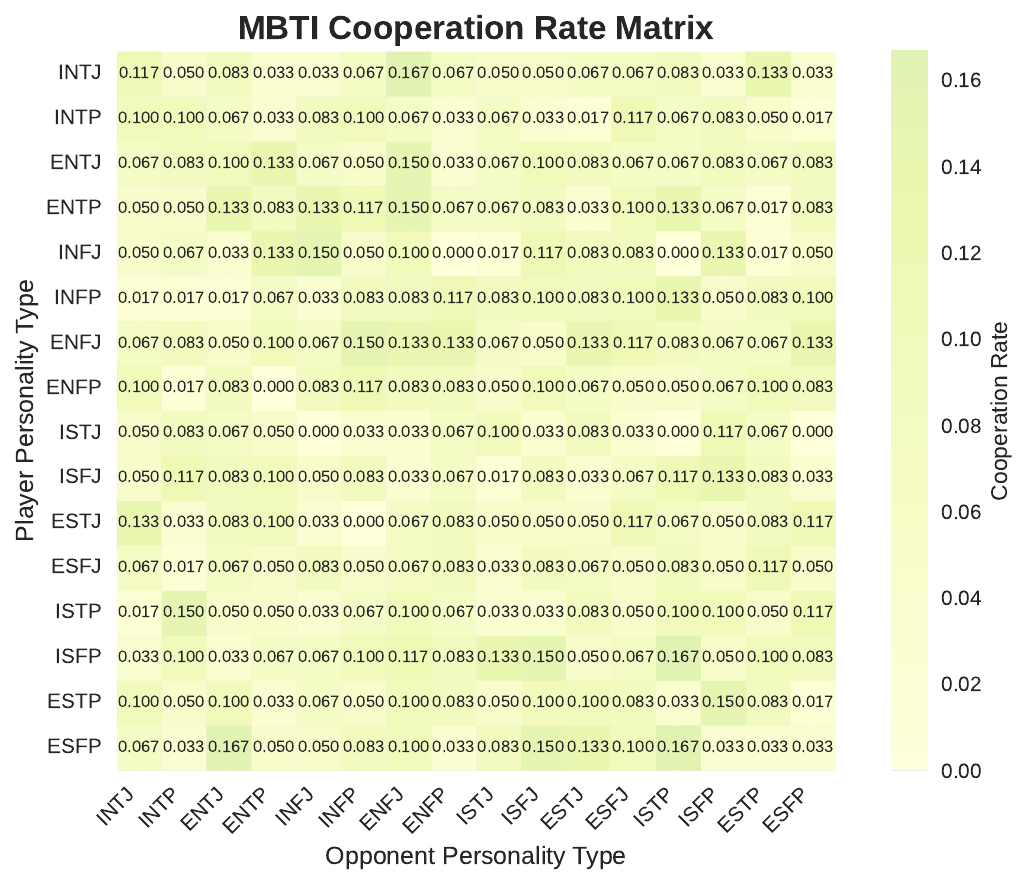}
        \hfill
        \includegraphics[width=0.49\linewidth]{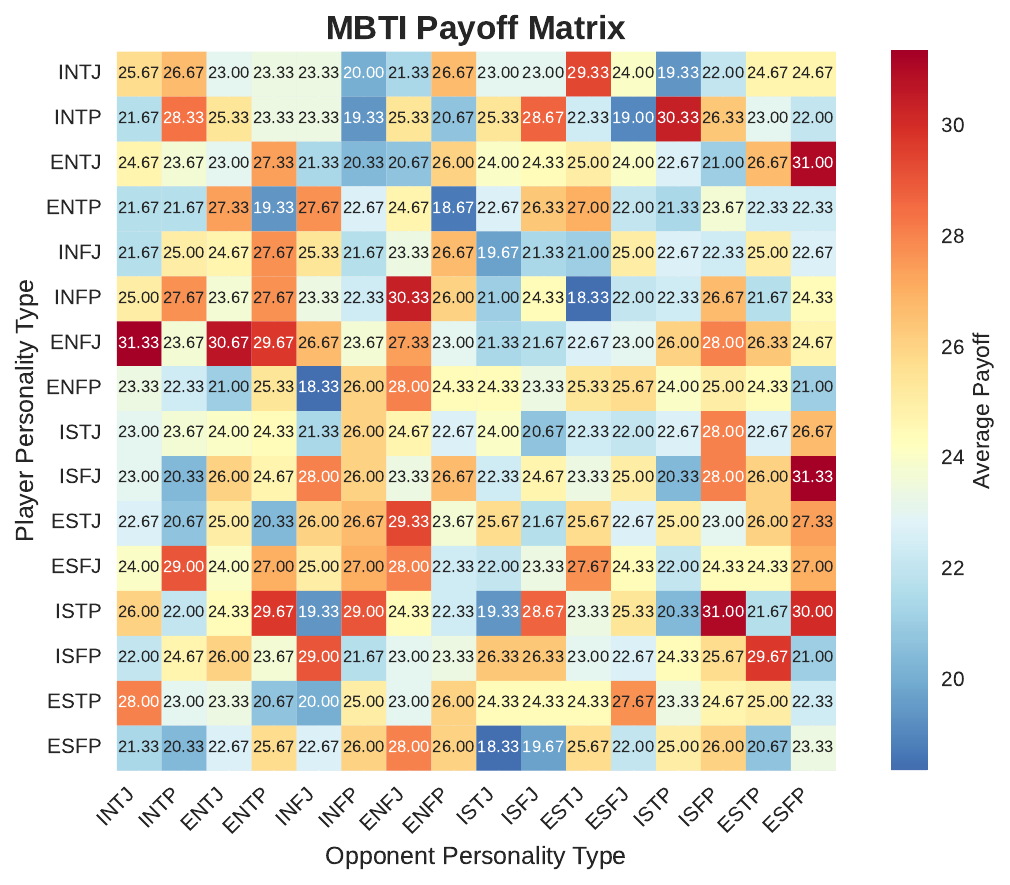}
    \end{minipage}
    \begin{minipage}{1.0\linewidth}
        \centering
        \includegraphics[width=0.49\linewidth]{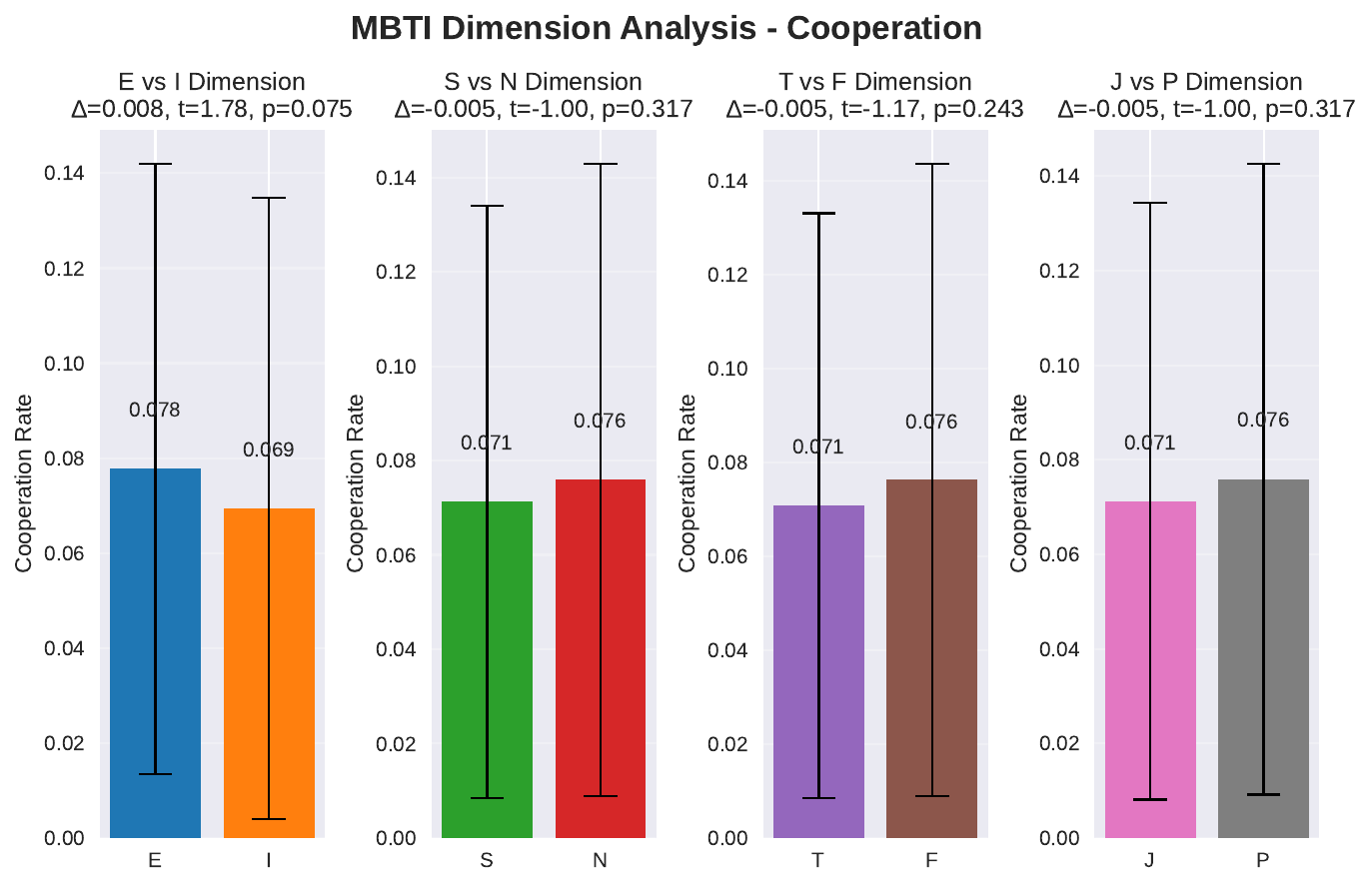}
        \hfill
        \includegraphics[width=0.49\linewidth]{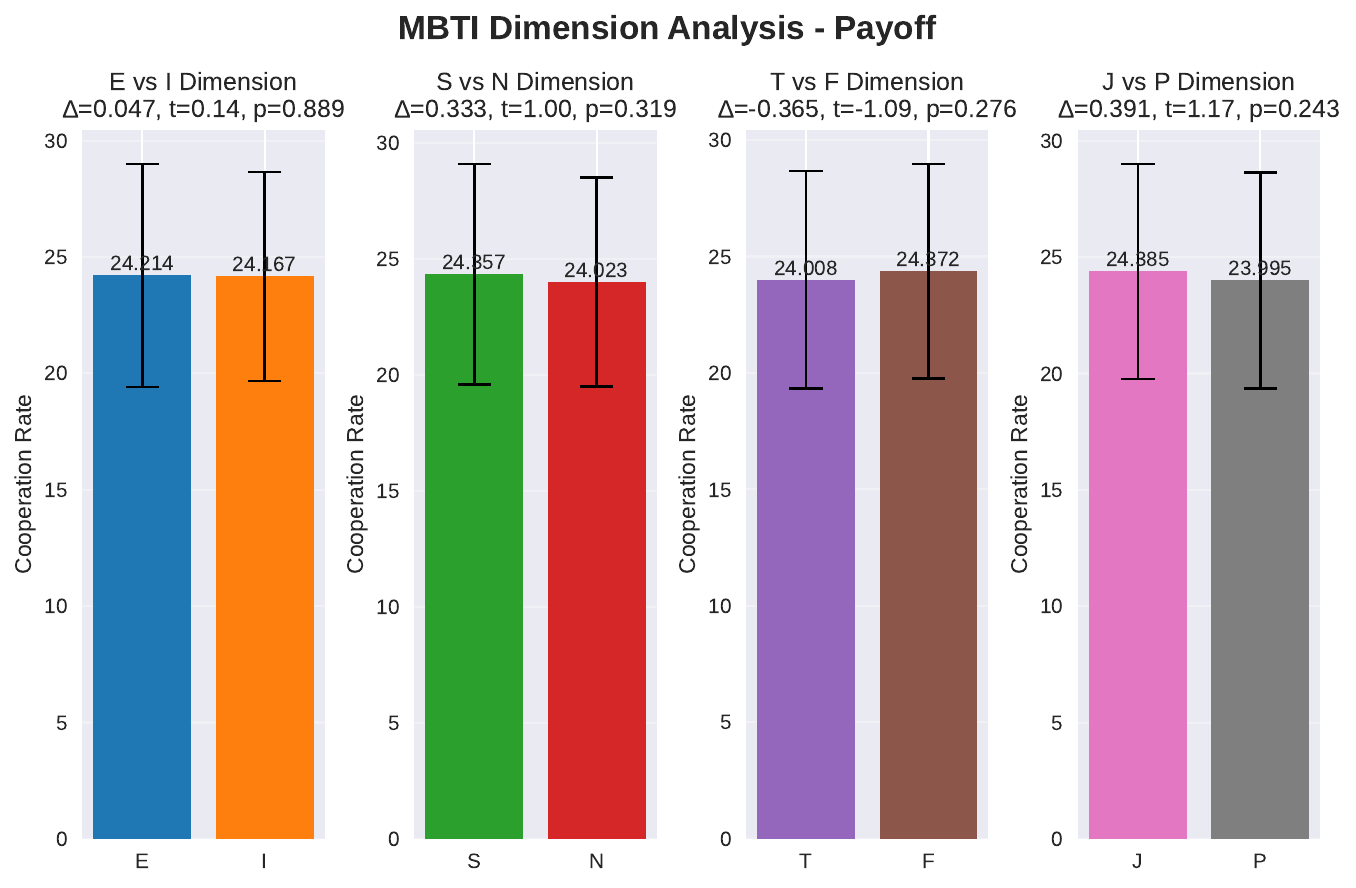}
    \end{minipage}
    \begin{minipage}{1.0\linewidth}
        \centering
        \includegraphics[width=0.49\linewidth]{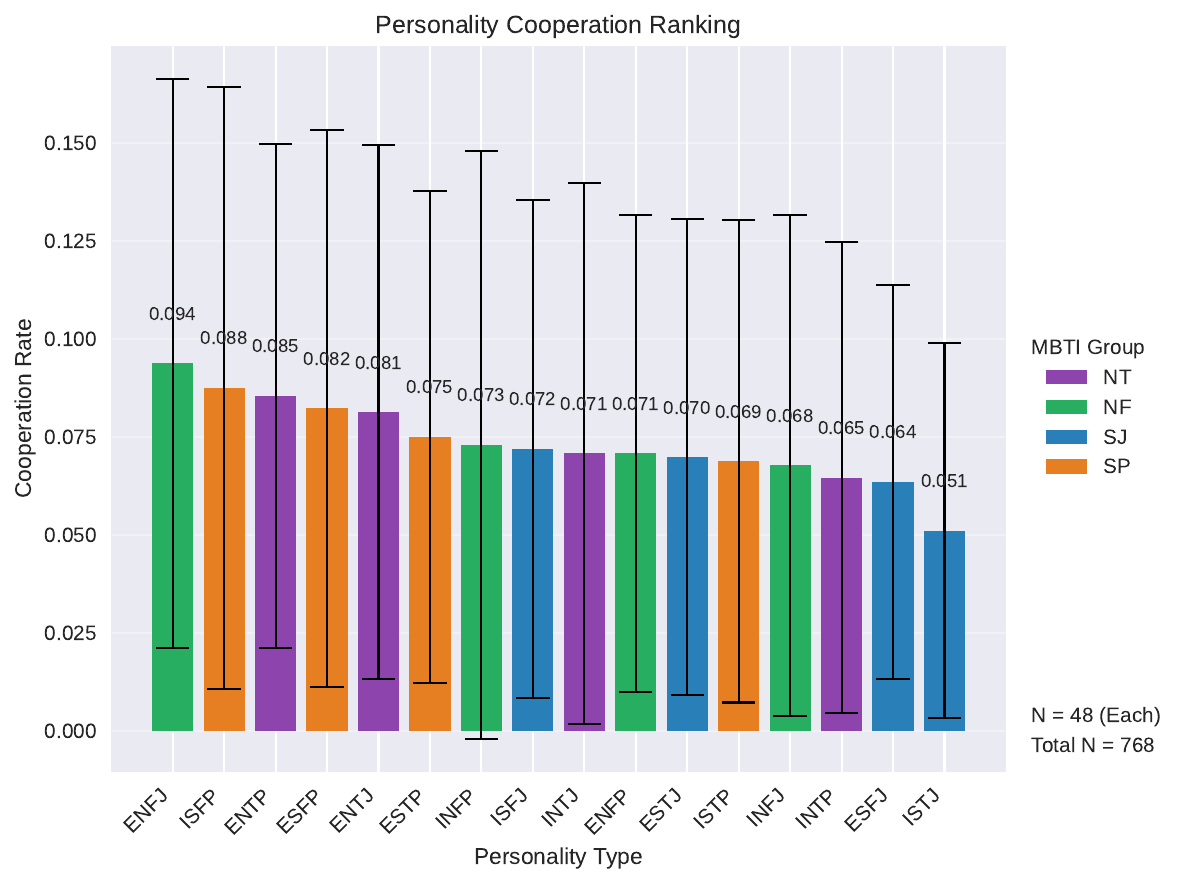}
        \hfill
        \includegraphics[width=0.49\linewidth]{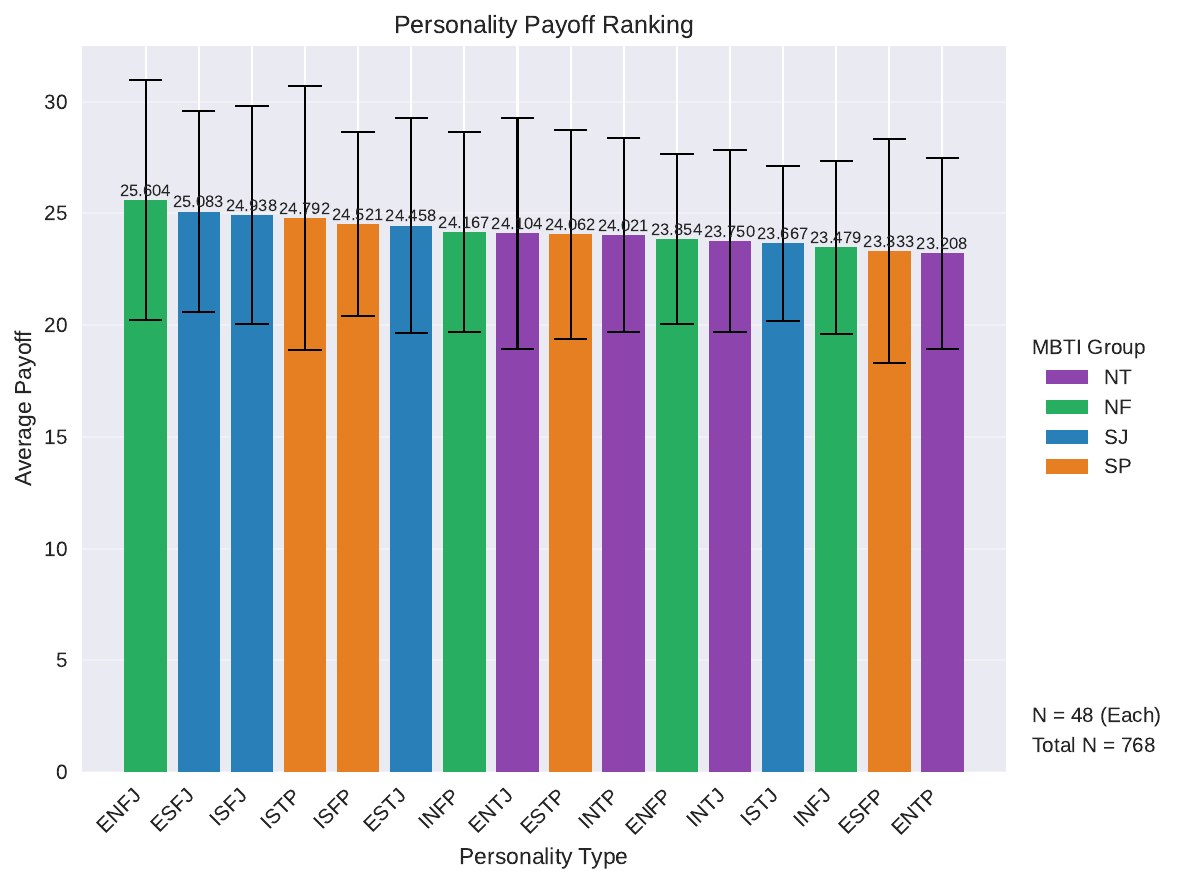}
    \end{minipage}
    \caption{Results of Pairwise Game with No Personality Injection.
        The overall cooperation rate dropped to a mere $0.074$ (std: $0.037$), the payoff dropped to $24.190$ (std: $2.722$). The cooperation heatmap appeared almost uniformly blank, while the payoff heatmap appeared completely random. As expected, none of the four MBTI dimensions showed statistically significant differences in behavior(All Welch's t-test $p>0.05$), and the resulting rankings were effectively random(Kruskal-Wallis tests: Cooperation $p=0.309$, Payoff $p=0.582$). This negative control confirms that the distinct behavioral profiles observed in our main experiments are indeed artifacts of the personality injection prompts, rather than inherent noise or bias in the base model.}
    \label{fig:ablation_no_both_personality}
    \Description{Results of Pairwise Game with No Personality Injection.}
\end{figure}

We also conducted experiments without any personality injection, where agents receive no personality-related information, including their own or their opponent's personality types. Results are shown in Figure~\ref{fig:ablation_no_both_personality}. Without personality cues, agents tend to converge towards defection strategies, leading to low overall cooperation rates and payoffs. This highlights the critical role of personality-driven behavior in sustaining cooperation in social dilemmas.

\subsubsection{No Personality Injection, but Opponent Info Provided}
\begin{figure}[htbp]
    \centering
    \begin{minipage}{1.0\linewidth}
        \centering
        \includegraphics[width=0.49\linewidth]{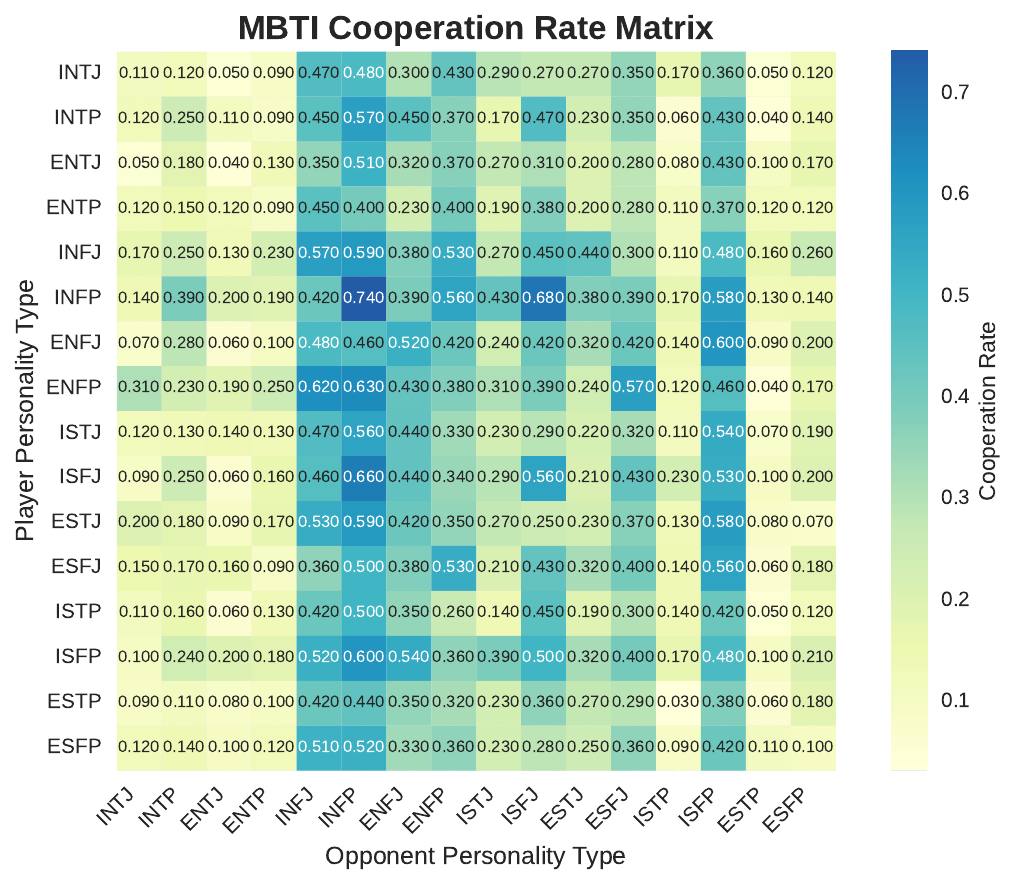}
        \hfill
        \includegraphics[width=0.49\linewidth]{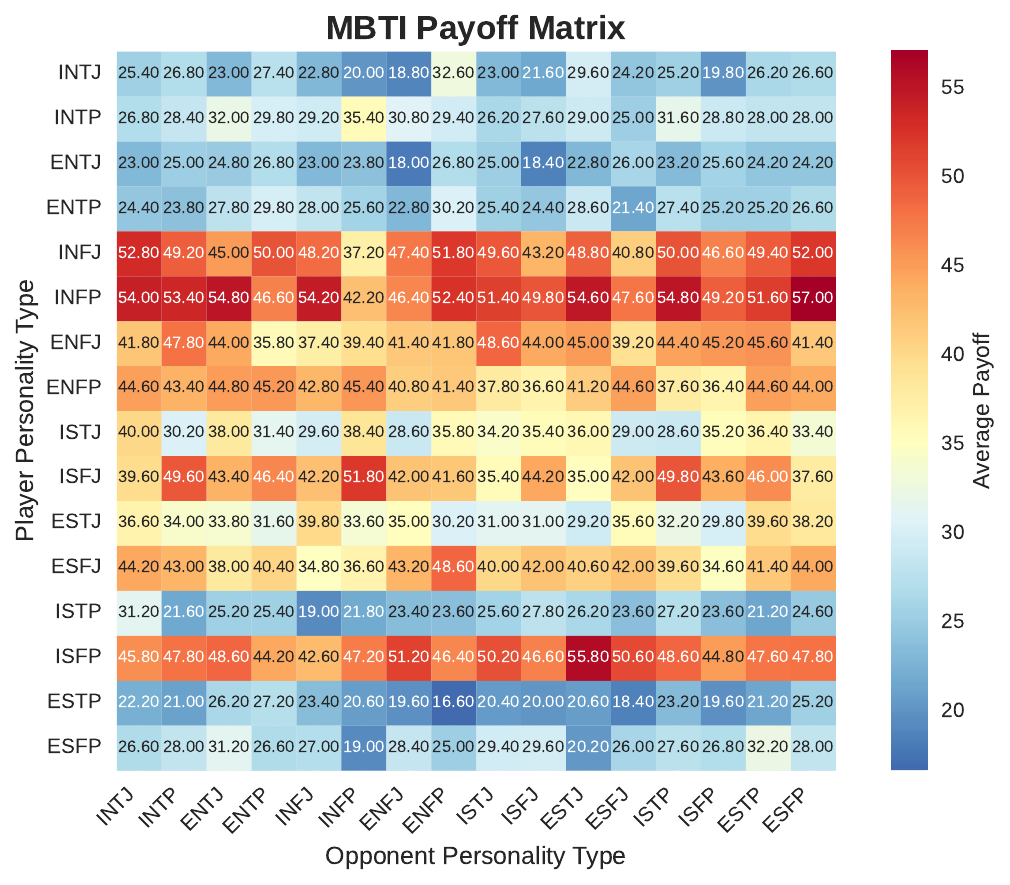}
    \end{minipage}
    \begin{minipage}{1.0\linewidth}
        \centering
        \includegraphics[width=0.49\linewidth]{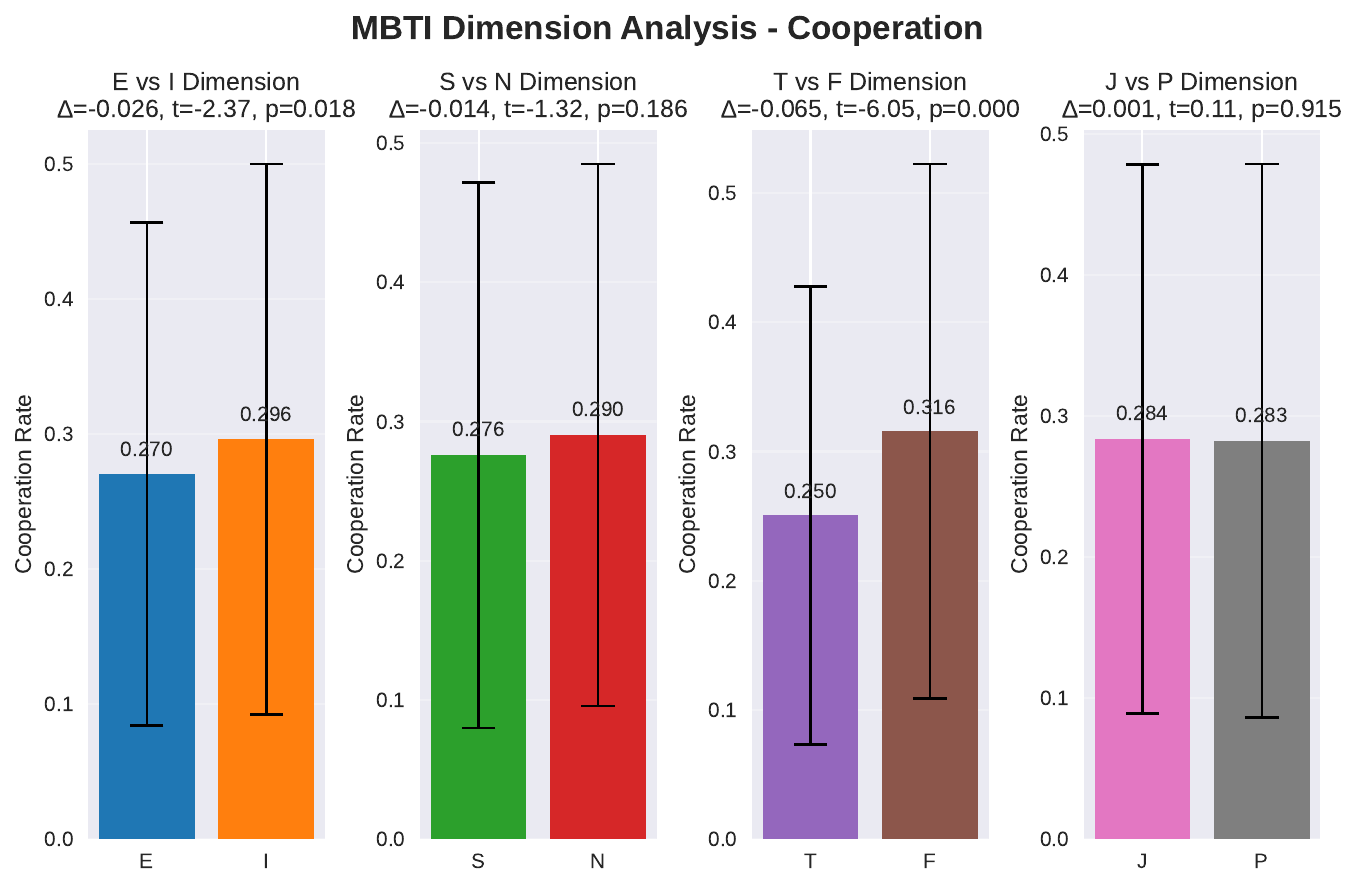}
        \hfill
        \includegraphics[width=0.49\linewidth]{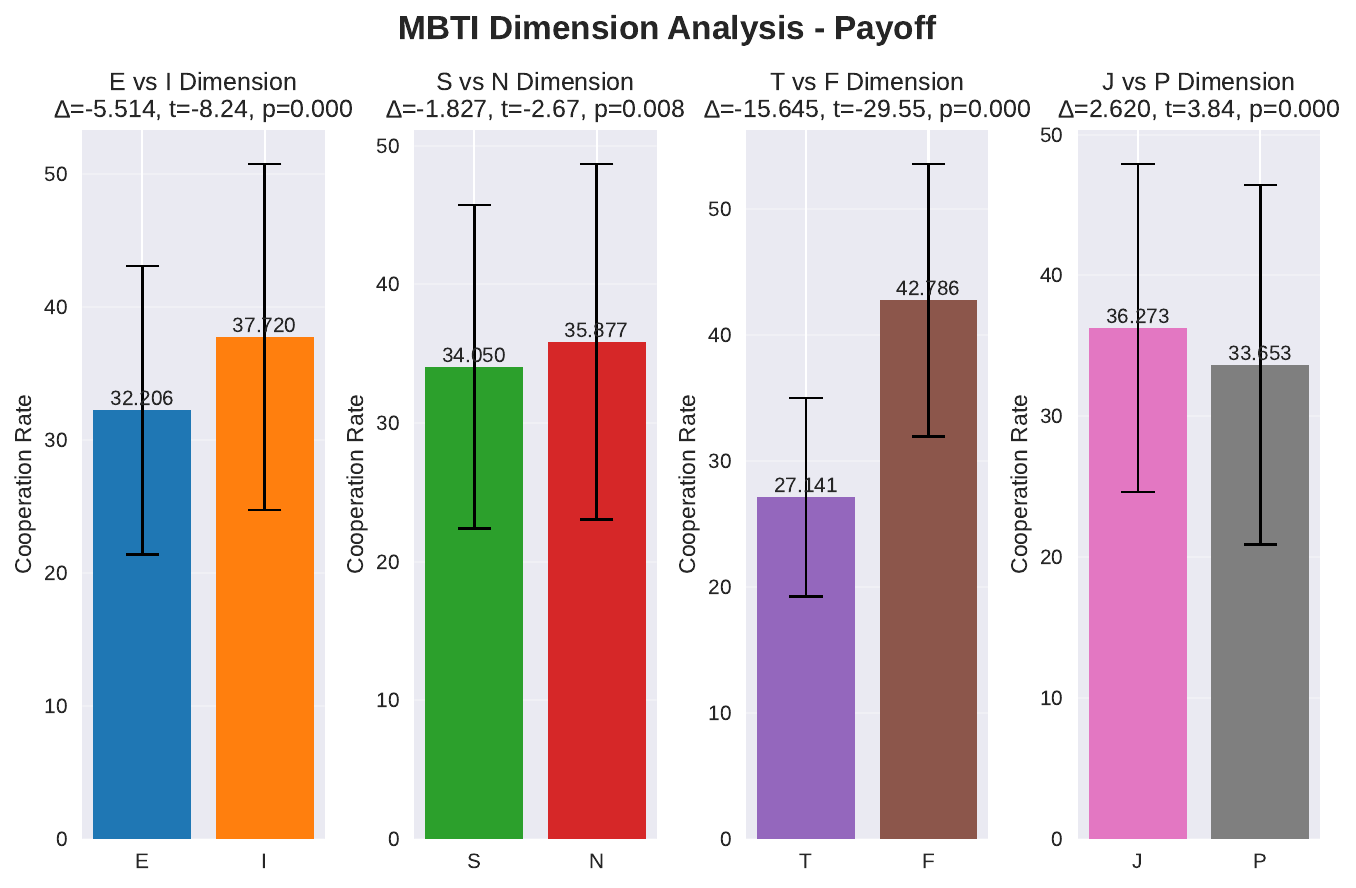}
    \end{minipage}
    \begin{minipage}{1.0\linewidth}
        \centering
        \includegraphics[width=0.49\linewidth]{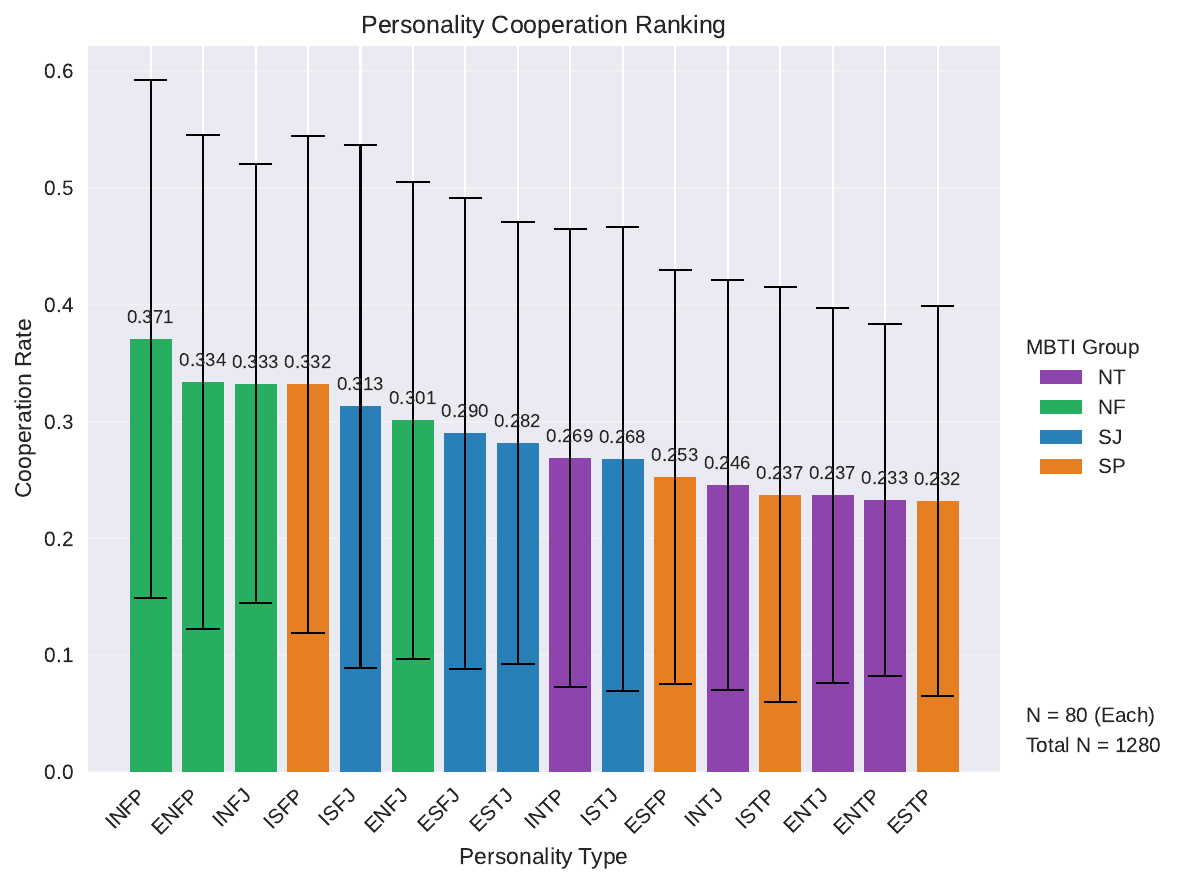}
        \hfill
        \includegraphics[width=0.49\linewidth]{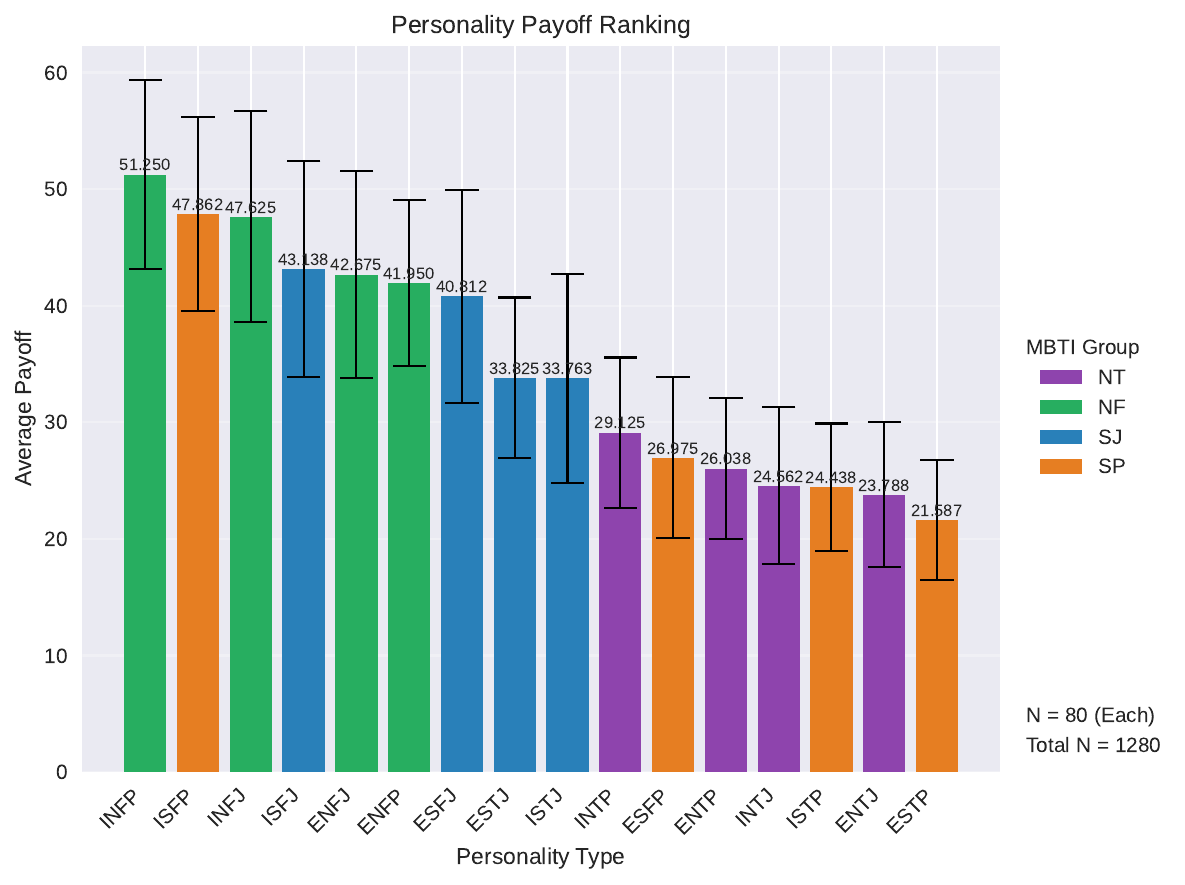}
    \end{minipage}
    \caption{Results of Pairwise Game with No Self Personality Injection but Opponent Info Provided. Contrary to other experiments, the \textit{columns} (opponents) for F-types are darker in the cooperation map, while the \textit{rows} (actors) for F-types are darker in the payoff map. This reversal indicates that even without self-personality, agents adapt their strategies based solely on opponent personality cues. Although all agents start identically blank, Welch's t-test still shows significant statistical divergence ($p < 0.001$), as distinct opponent behaviors shape the interaction history, indirectly driving the ``blank'' agents into different strategic paths.}
    \label{fig:ablation_no_pj_with_opponent_info}
    \Description{Results of Pairwise Game with No Self Personality Injection but Opponent Info Provided.}
\end{figure}

This may be the most interesting ablation, as it tests whether knowing the opponent's personality is enough to influence behavior, even without a self-assigned personality. As shown in Figure~\ref{fig:ablation_no_pj_with_opponent_info}, agents modify their strategies based solely on their opponent's type. For instance, they tend to cooperate more with F-type opponents and defect against T-type opponents. This reverses the visual pattern seen in previous heatmaps: here, the \textit{columns} for F-types are darker in the cooperation map (indicating others cooperate with them), whereas usually the specific personality \textit{rows} drive the behavior. Meanwhile, the \textit{rows} for F-types are darker in the payoff map, indicating that opponents tend to cooperate more with them, yielding higher payoffs for F-types, whereas usually the \textit{columns} for F-types yield higher payoffs to opponents.

However, this raises another question: why does the MBTI dimension analysis (middle figures) show statistically significant differences if the agents have no self-personality? Since every ``blank'' agent plays against the same set of 16 opponents, their average statistics should theoretically be identical.

The variance arises from the game dynamics over time. An agent might open with cooperation against an F-type but defect against a T-type based on the label. In subsequent rounds, the history of those interactions diverges. Effectively, the opponent's personality shapes the game history, which then shapes the ``blank'' agent's future choices, creating statistical differences in the final averages even though the agents started identical. This creates a scenario where, even without awareness of their own personalities, the agents' strategies are indirectly influenced by their opponents' actions.

\subsection{Game History Ablation}
\label{app:game_history_ablation}
\begin{figure}[htbp]
    \centering
    \begin{minipage}{1.0\linewidth}
        \centering
        \includegraphics[width=0.49\linewidth]{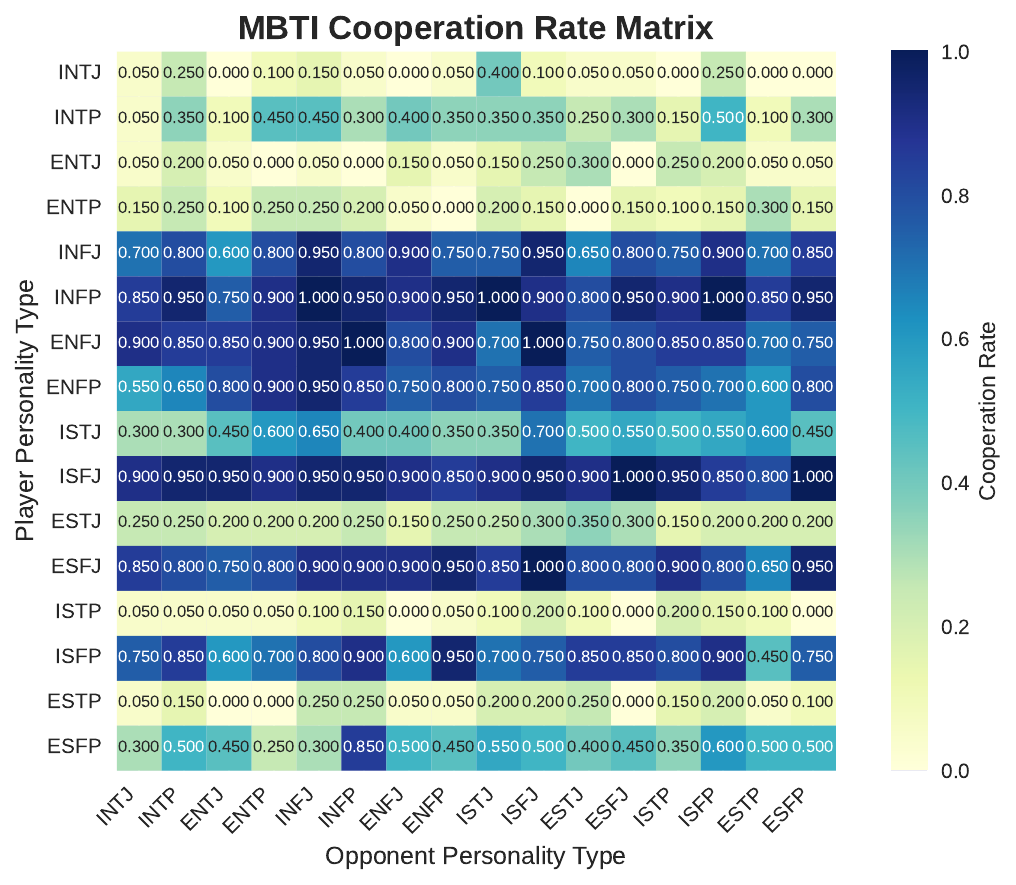}
        \hfill
        \includegraphics[width=0.49\linewidth]{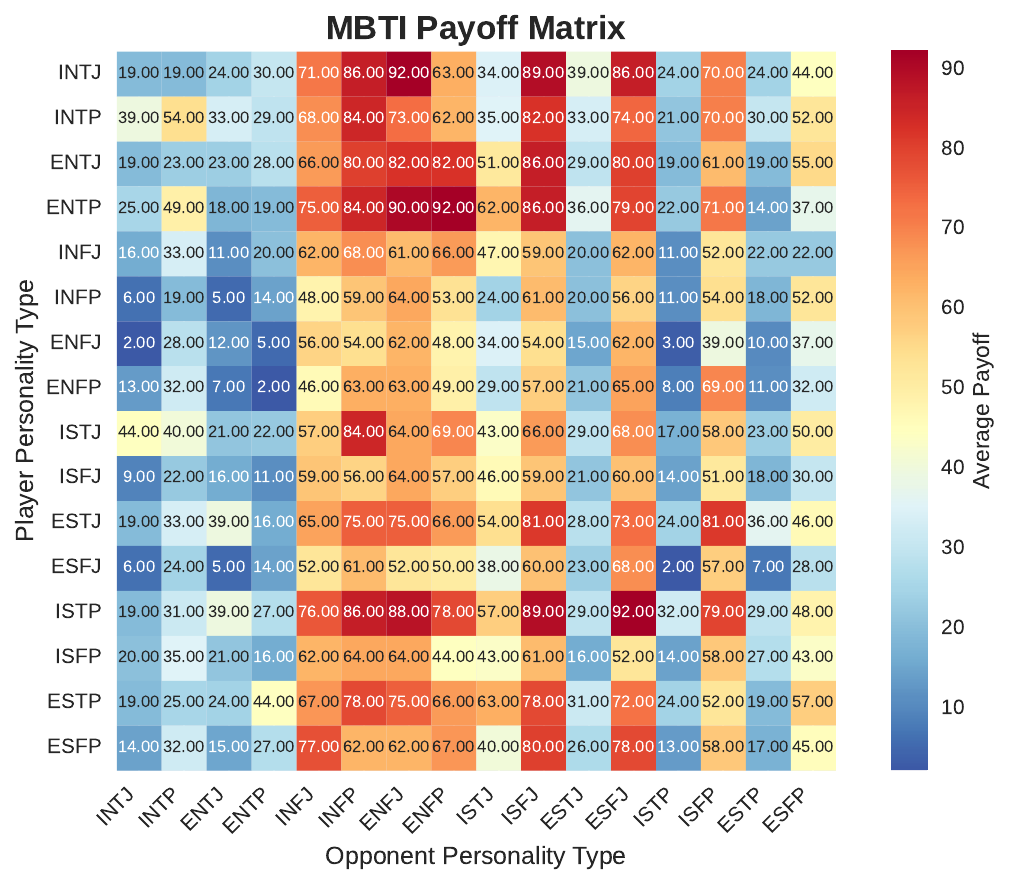}
    \end{minipage}
    \begin{minipage}{1.0\linewidth}
        \centering
        \includegraphics[width=0.49\linewidth]{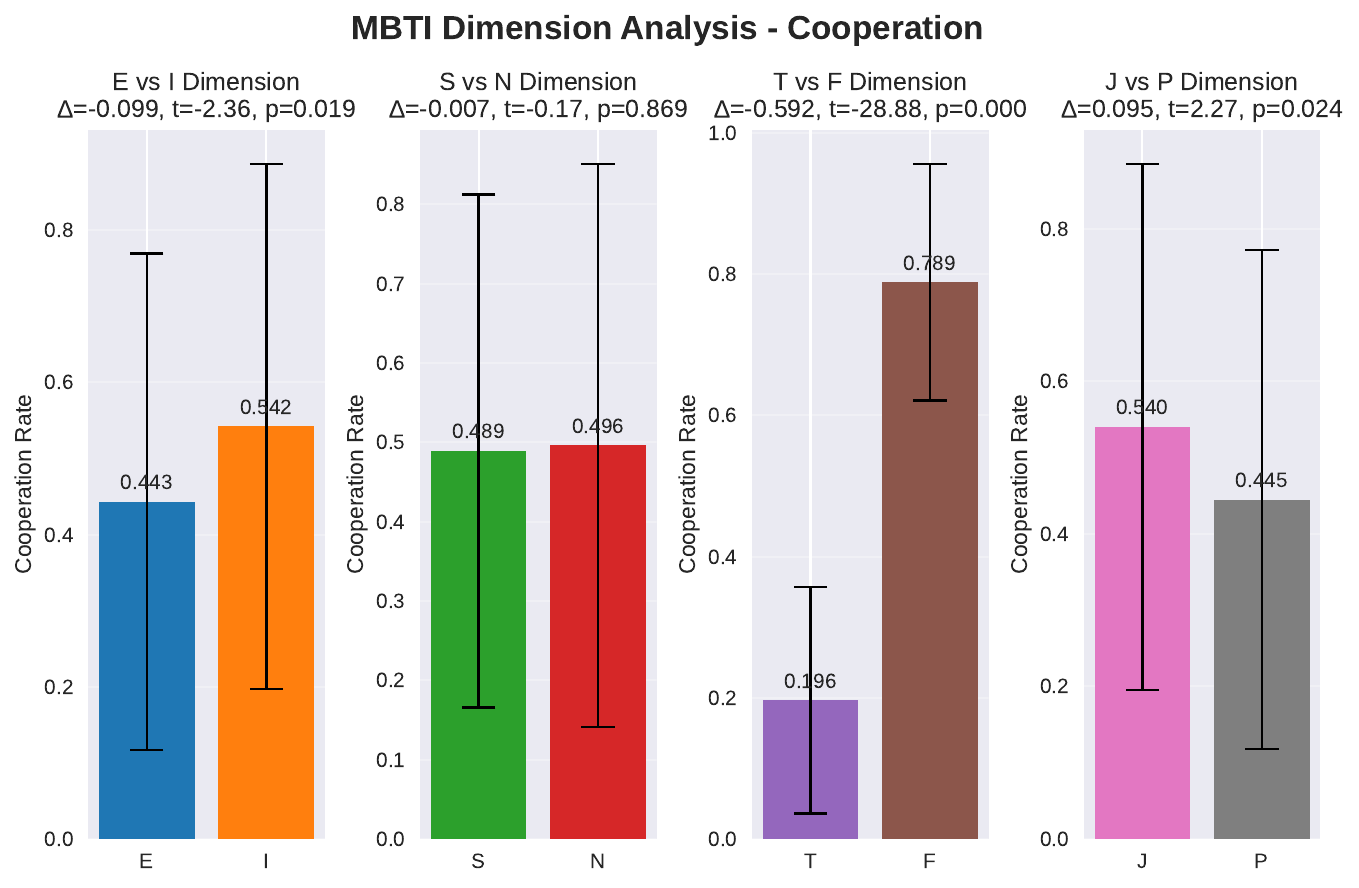}
        \hfill
        \includegraphics[width=0.49\linewidth]{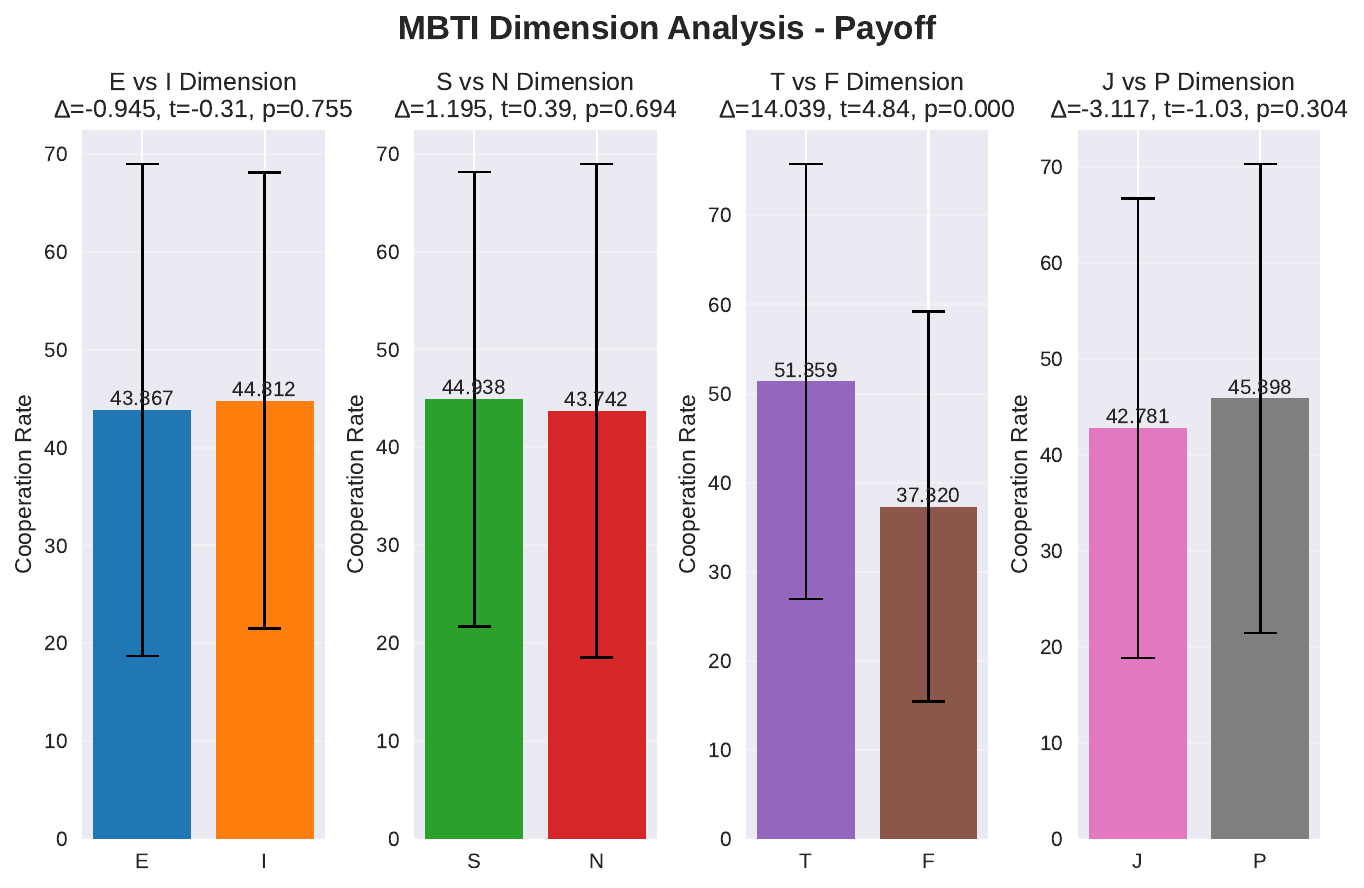}
    \end{minipage}
    \begin{minipage}{1.0\linewidth}
        \centering
        \includegraphics[width=0.49\linewidth]{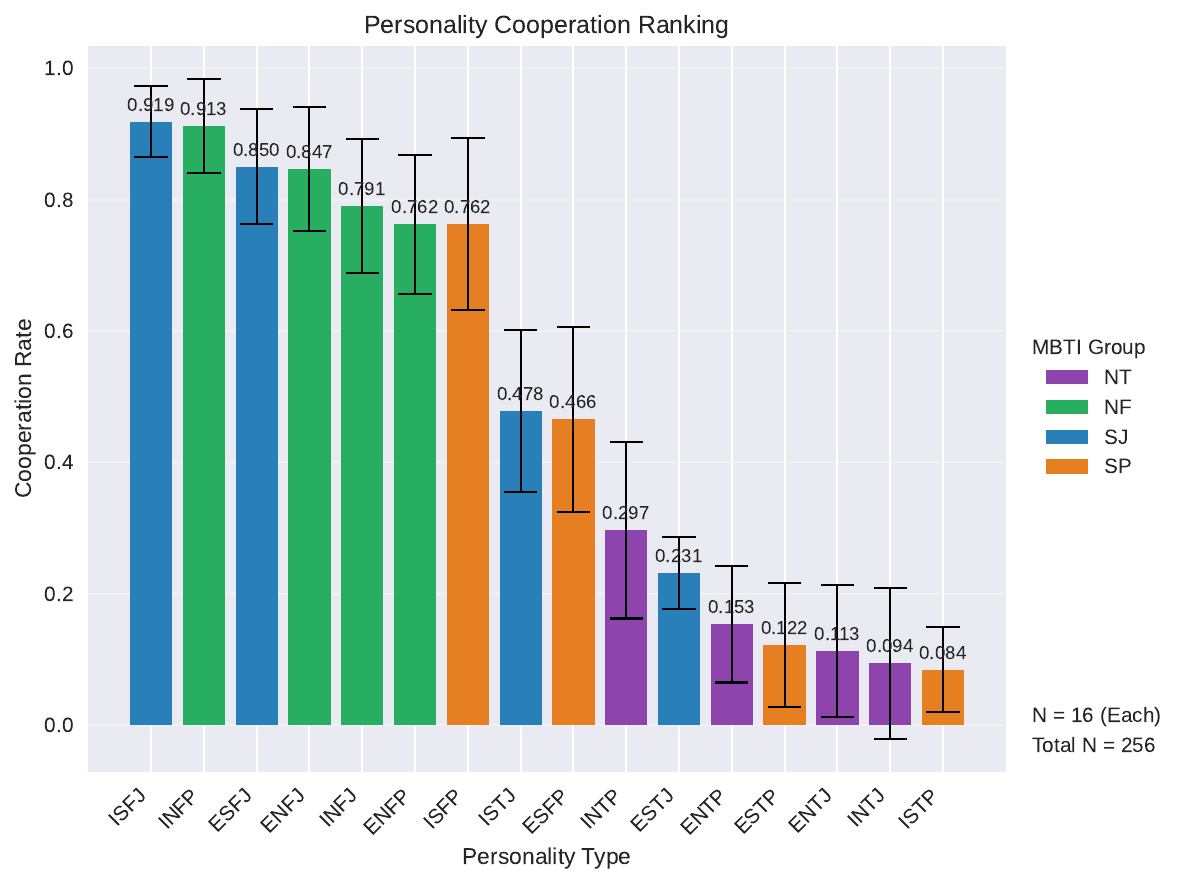}
        \hfill
        \includegraphics[width=0.49\linewidth]{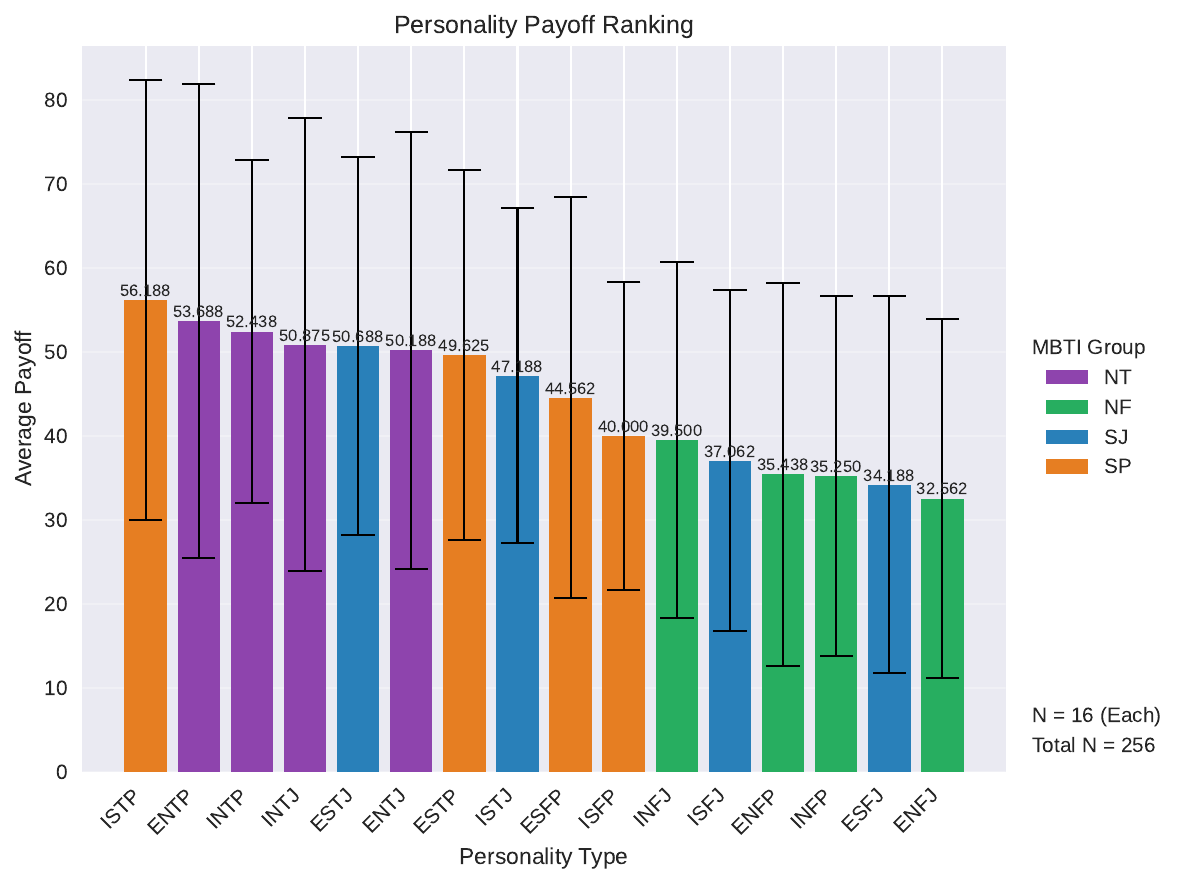}
    \end{minipage}
    \caption{Results of Pairwise Game with No Game History. F-types show statistically significantly higher cooperation rates($p < 0.001$), while T-types gain statistically significantly higher payoffs($p < 0.001$), consistent with the original experiments. ESFP ranks lowest in cooperation among F-types but achieves the highest payoff within the group. This demonstrates that agents adopt distinct strategies based on personality types even in the absence of historical game records.}
    \label{fig:ablation_no_history}
    \Description{Results of Pairwise Game with No Game History.}
\end{figure}

When there is no history record, the iterative prisoner's dilemma degenerates into a sequence of independent single-shot games. Each round is isolated, meaning agents possess no memory of past interactions to inform future decisions. As shown in Figure~\ref{fig:ablation_no_history}, despite this memory loss, the overall cooperation rate remains at $0.493$ (std: $0.338$), and the average payoff at $44.340$ (std: $24.154$), results quite similar to those of the memory-enabled experiments.

Statistical analysis confirms that personality-driven behavioral distinctions persist even without longitudinal context. A Kruskal-Wallis test reveals highly significant differences in cooperation rates across personality types ($H=222.22$, $p<0.001$), with F-types consistently exhibiting higher cooperation than T-types ($p < 0.001$ via Welch's t-test). Payoff differences also remain statistically significant across types ($H=27.21$, $p=0.027$), with T-types generally securing higher individual gains. Interestingly, ESFP ranks lowest in cooperation among F-types yet achieves the highest payoff within that group. This demonstrates that distinct personality-based strategies are intrinsic to the agents' ``nature'' and manifest robustly even in the absence of interaction history.

\subsection{Neighbor Info Ablation}
\label{app:neighbor_info_ablation}
In this experiment, we remove the ``Neighbor Info'' component ($\text{str}(\Omega_i^{(t)})$) from the agent prompt. This ablation forces agents to rely solely on their dyadic interaction histories with specific neighbors, effectively ignoring the broader local social context (e.g., the aggregate cooperation rates of their neighborhood). We tested this condition on Regular, Small-World ($p=0.1, 0.5$) and Scale-Free networks, same as in Section~\ref{sec:topology}.

Table~\ref{tab:ablation_neighbor} compares the results against the baseline (Section~\ref{sec:topology}) where full neighbor information was provided. The removal of social context resulted in a drastic decline in cooperation rates across all topologies (dropping from $\sim$40-45\% to $\sim$23-25\%). This finding indicates that explicit information about local group norms serves as a crucial signal for reinforcing cooperative behavior in LLM agents, likely by triggering conformity or conditional cooperation mechanisms.

Theoretically, removing network context information causes the network game to degenerate into a series of independent pairwise games. This is corroborated by our results, where we observe that cooperation rates and payoffs no longer exhibit significant differences across the various network topologies.

\begin{table}[htbp]
    \caption{Ablation Study: Impact of Removing Neighbor Info ($\text{str}(\Omega_i^{(t)})$)}
    \label{tab:ablation_neighbor}
    \centering
    \resizebox{\columnwidth}{!}{
        \begin{tabular}{lcccc}
            \toprule
            \multirow{2}{*}{\textbf{Topology}} & \multicolumn{2}{c}{\textbf{Cooperation Rate}} & \multicolumn{2}{c}{\textbf{Avg. Payoff}}                                                                                                                                          \\
            \cmidrule(lr){2-3} \cmidrule(lr){4-5}
                                               & \textbf{Baseline}                             & \textbf{No Neighbor Info}                & \textbf{Baseline} & \textbf{Ablation}                                                                                                  \\
            \midrule
            Regular                            & 0.478                                         & 0.242                                    & 10.170            & 8.230                                                                                                              \\
            SW ($p=0.1$)                       & 0.456                                         & 0.246                                    & 9.935             & 8.401                                                                                                              \\
            SW ($p=0.5$)                       & 0.367                                         & 0.248                                    & 9.482             & 8.126                                                                                                              \\
            Scale-Free                         & 0.420                                         & 0.234                                    & 9.643             & 8.171\tablefootnote{Adjusted for edge density (raw: 7.844, edges: 96 vs 100). Scaling: $7.844/0.96 \approx 8.171$} \\
            \bottomrule
        \end{tabular}
    }
\end{table}

\subsection{Beyond Prisoner's Dilemma}
\label{app:beyond_prisoners_dilemma}
To evaluate whether our findings extend beyond the Prisoner's Dilemma, we considered two additional canonical games: the Stag Hunt and Snowdrift (Chicken). These three games span the canonical spectrum of social dilemmas with distinct distinct incentive and risk structures.
\subsubsection{Stag Hunt}
\[ \begin{array}{c|cc} & C & D \\ \hline C & (R,R) & (S,T) \\ D & (T,S) & (P,P) \end{array} \quad\text{with } R > T \ge P > S \]
The Stag Hunt admits two pure Nash equilibria, allowing us to examine equilibrium selection under coordination risk. In our simulations, we set the payoffs as $R=5$, $T=3$, $P=1$, and $S=0$.

\begin{figure}[htbp]
    \centering
    \begin{minipage}{1.0\linewidth}
        \centering
        \includegraphics[width=0.49\linewidth]{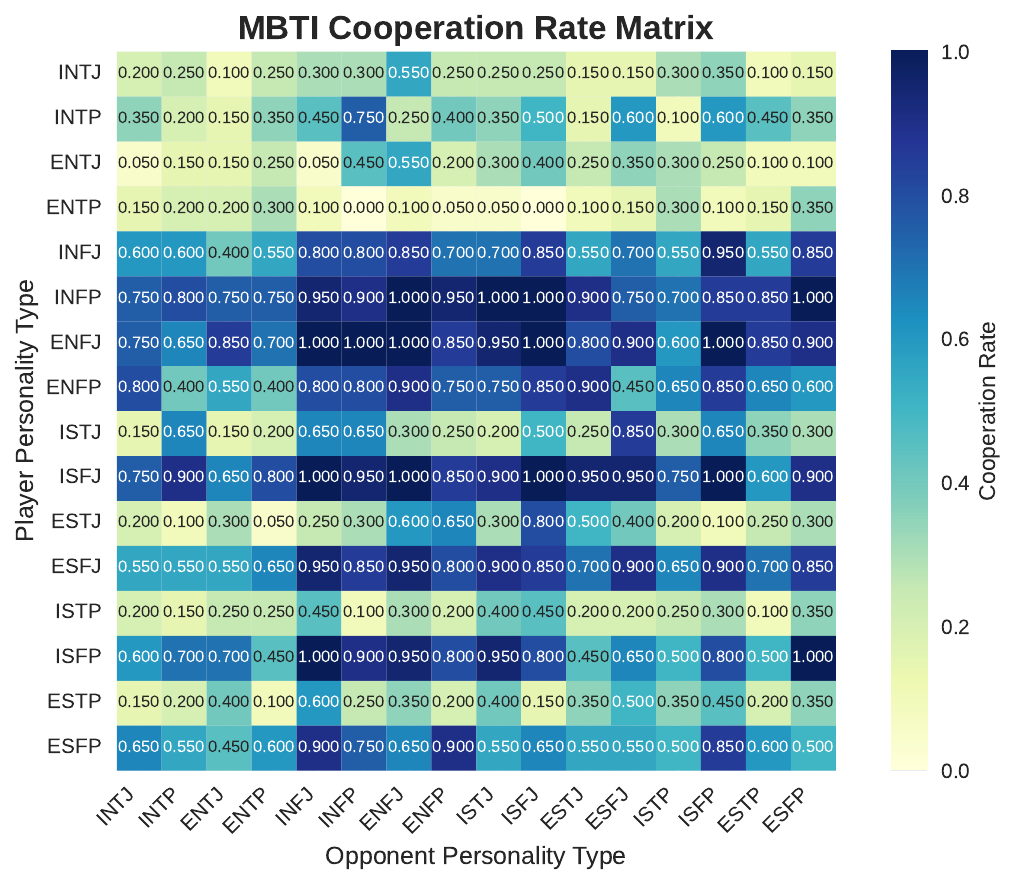}
        \hfill
        \includegraphics[width=0.49\linewidth]{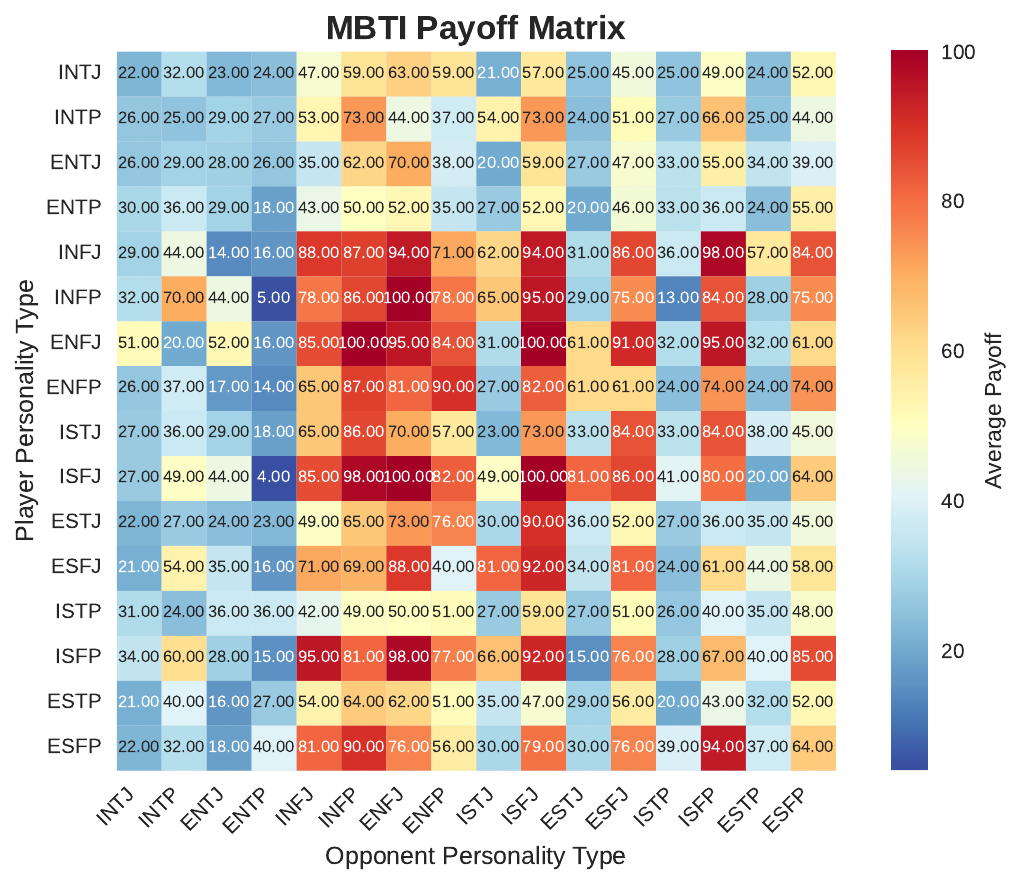}
    \end{minipage}
    \begin{minipage}{1.0\linewidth}
        \centering
        \includegraphics[width=0.49\linewidth]{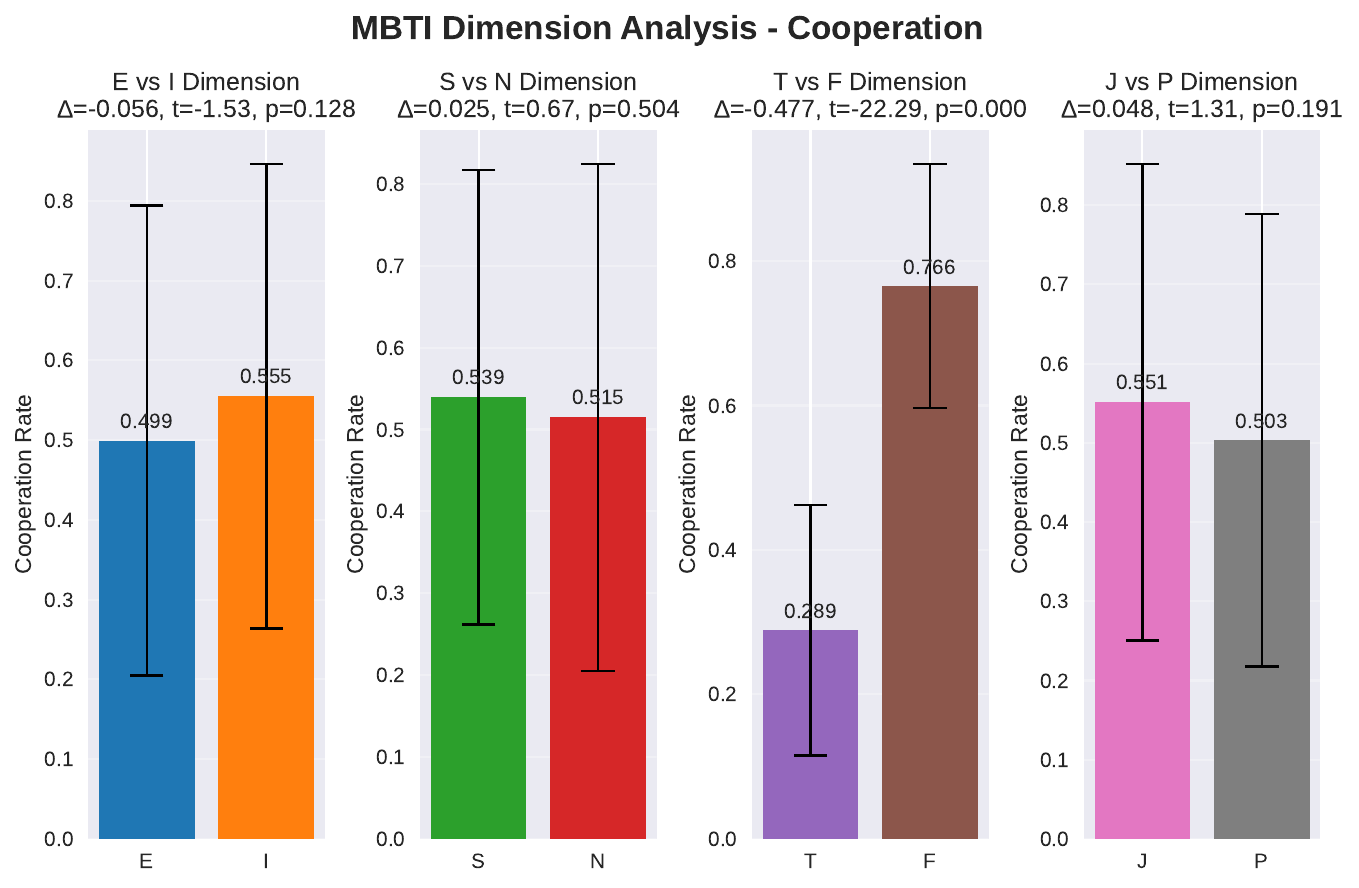}
        \hfill
        \includegraphics[width=0.49\linewidth]{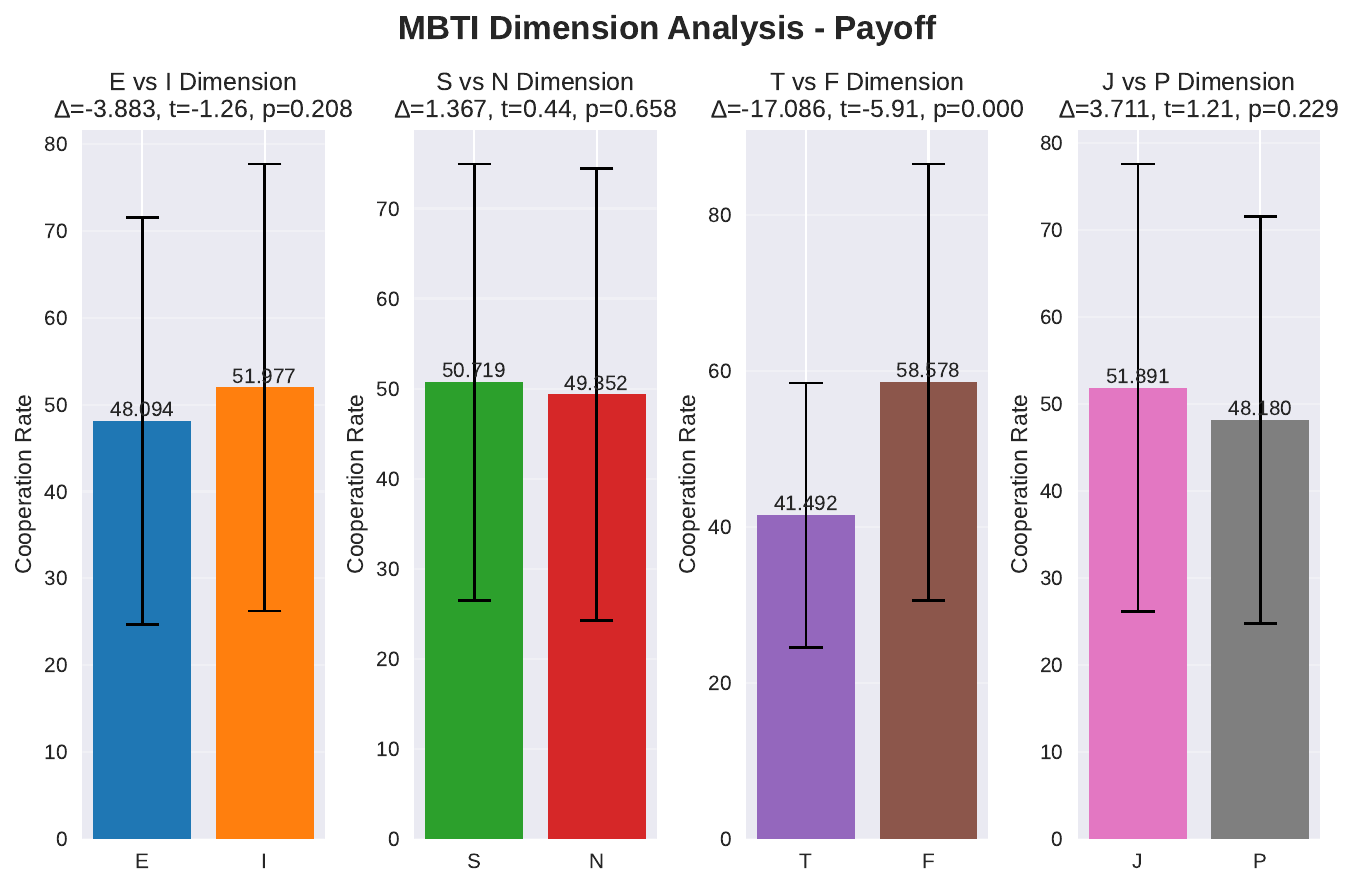}
    \end{minipage}
    \begin{minipage}{1.0\linewidth}
        \centering
        \includegraphics[width=0.49\linewidth]{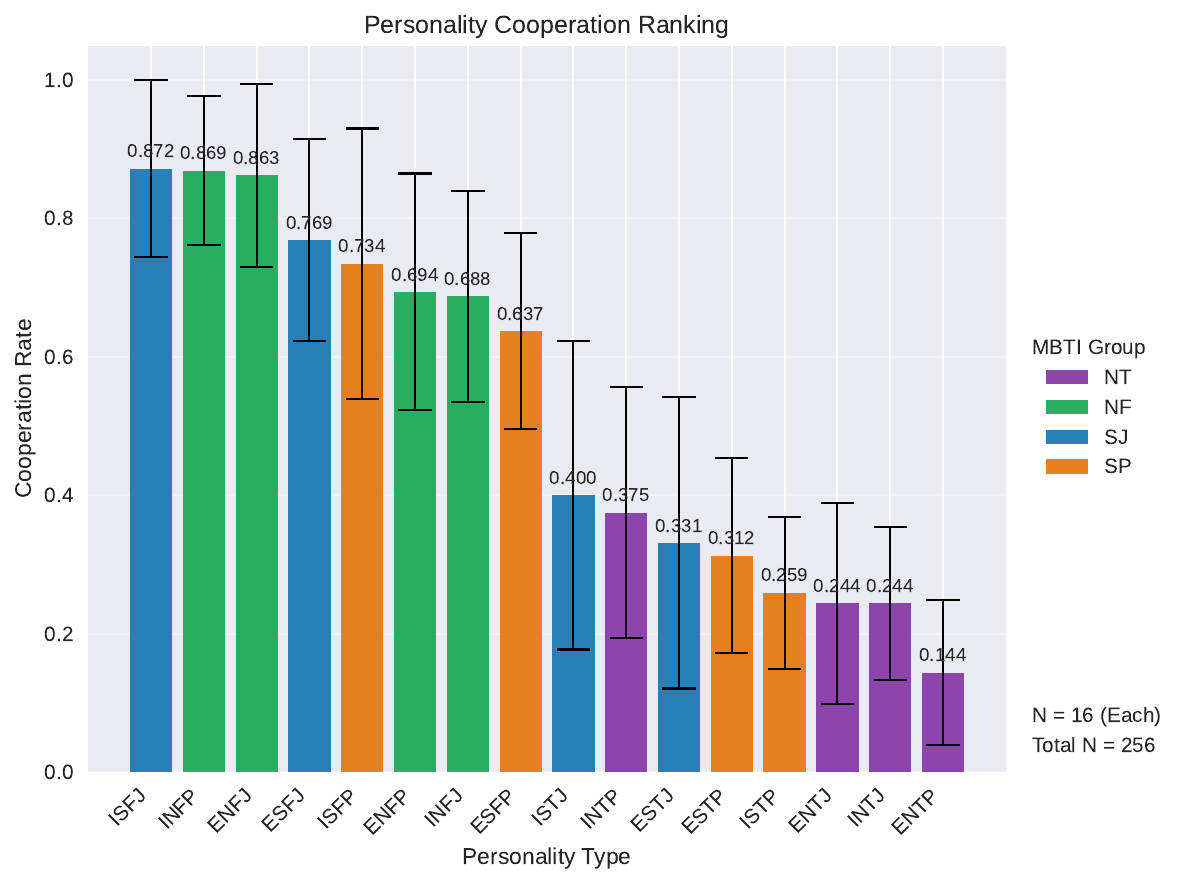}
        \hfill
        \includegraphics[width=0.49\linewidth]{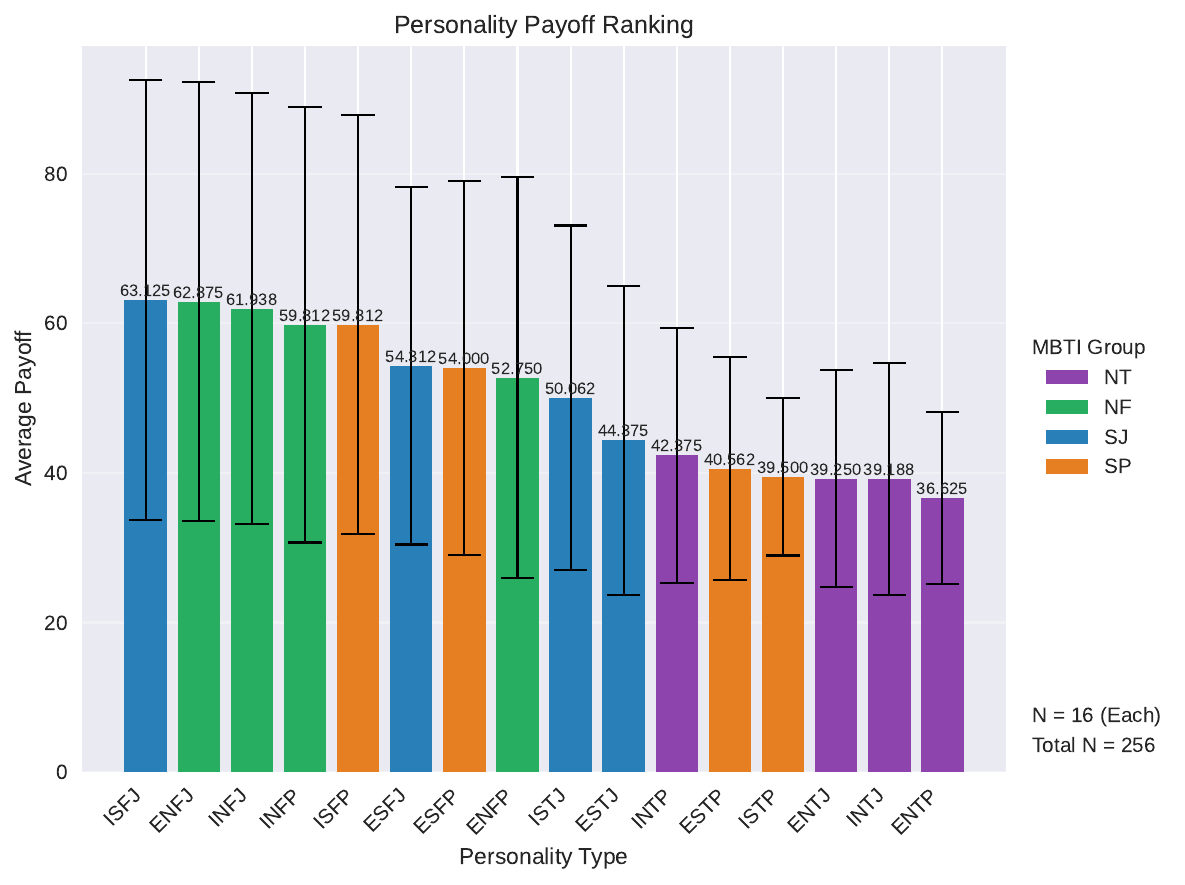}
    \end{minipage}
    \caption{Results of Pairwise Game using Stag Hunt game. The cooperation heatmap reveals that F-types maintain higher cooperation rates, also achieve higher payoffs, different with patterns observed in the Prisoner's Dilemma. This suggests that personality-driven strategies similar across different social dilemma structures, however payoff patterns may vary with game incentives.}
    \label{fig:stag_hunt}
    \Description{Results of Pairwise Game using Stag Hunt game.}
\end{figure}

As shown in ~\ref{fig:stag_hunt}, in the Stag Hunt simulations, the agents achieved an overall average cooperation rate of 0.527 (std: 0.293), slightly higher than in the Prisoner's Dilemma, with an average payoff of 50.035 (std: 24.588). Welch's t-test confirms that Feeling (F) types maintain significantly higher cooperation rates and achieve significantly higher payoffs compared to Thinking (T) types ($p < 0.001$). Correspondingly, F-types consistently rank higher in both cooperation and payoff metrics. The Kruskal-Wallis test further validates significant differences across personality types for both cooperation ($H=185.61$, $p<0.001$) and payoffs ($H=29.81$, $p=0.013$).

\subsubsection{Snowdrift (Chicken)}
\[ \begin{array}{c|cc} & C & D \\ \hline C & (R,R) & (S,T) \\ D & (T,S) & (P,P) \end{array} \quad\text{with } T > R > S > P \]
In contrast to the Prisoner's Dilemma, cooperation in the Snowdrift game is anti-coordinative, testing agents' adaptability to competitive cooperation. In our simulations, we set the payoffs as $T=5$, $R=3$, $S=1$, and $P=0$.

\begin{figure}[htbp]
    \centering
    \begin{minipage}{1.0\linewidth}
        \centering
        \includegraphics[width=0.49\linewidth]{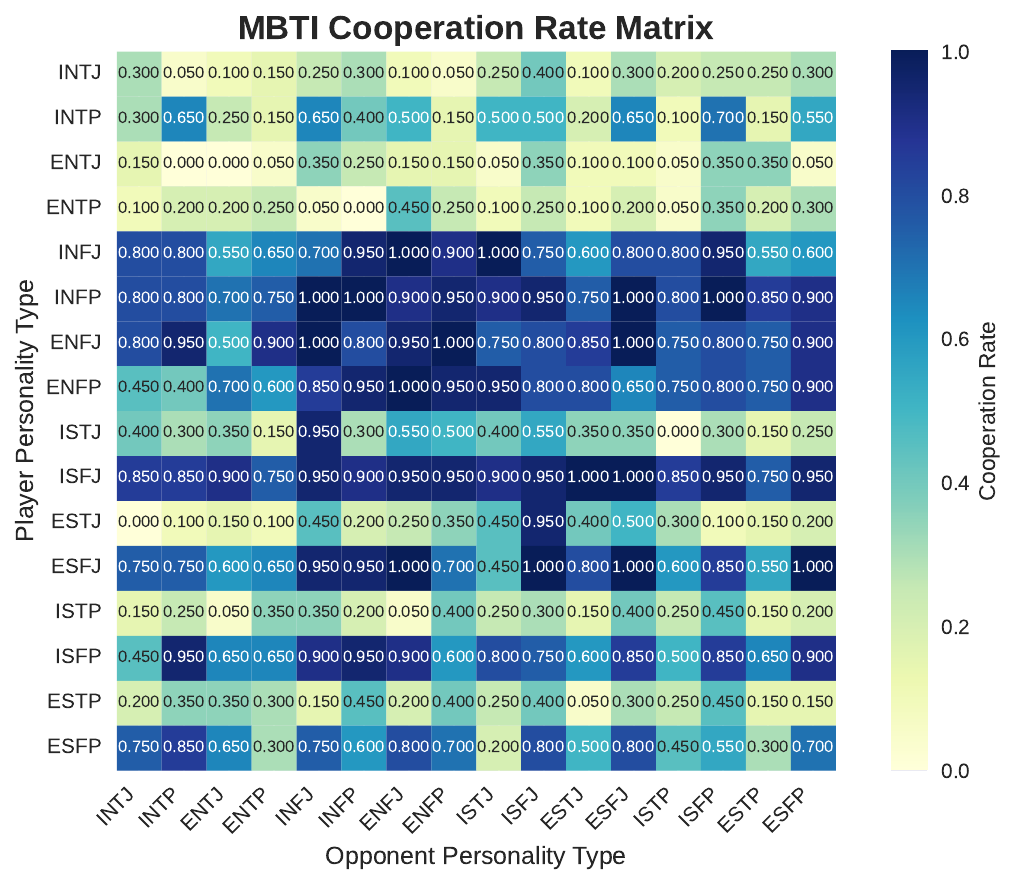}
        \hfill
        \includegraphics[width=0.49\linewidth]{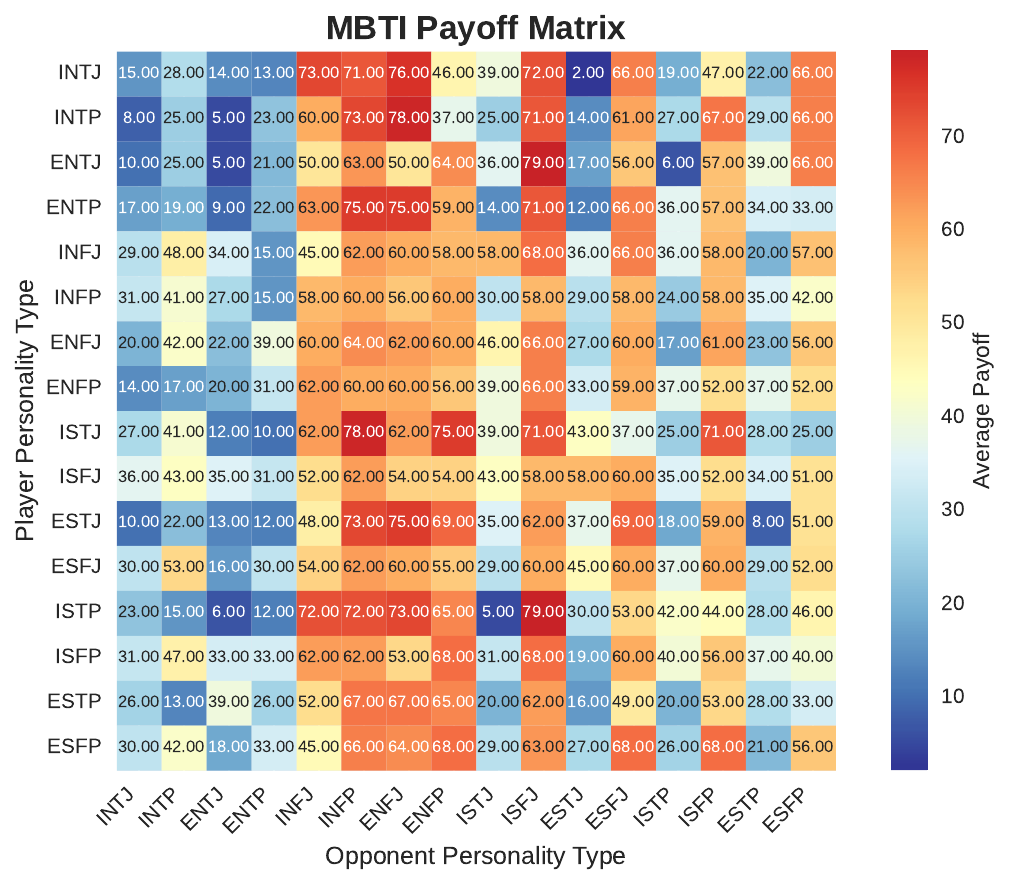}
    \end{minipage}
    \begin{minipage}{1.0\linewidth}
        \centering
        \includegraphics[width=0.49\linewidth]{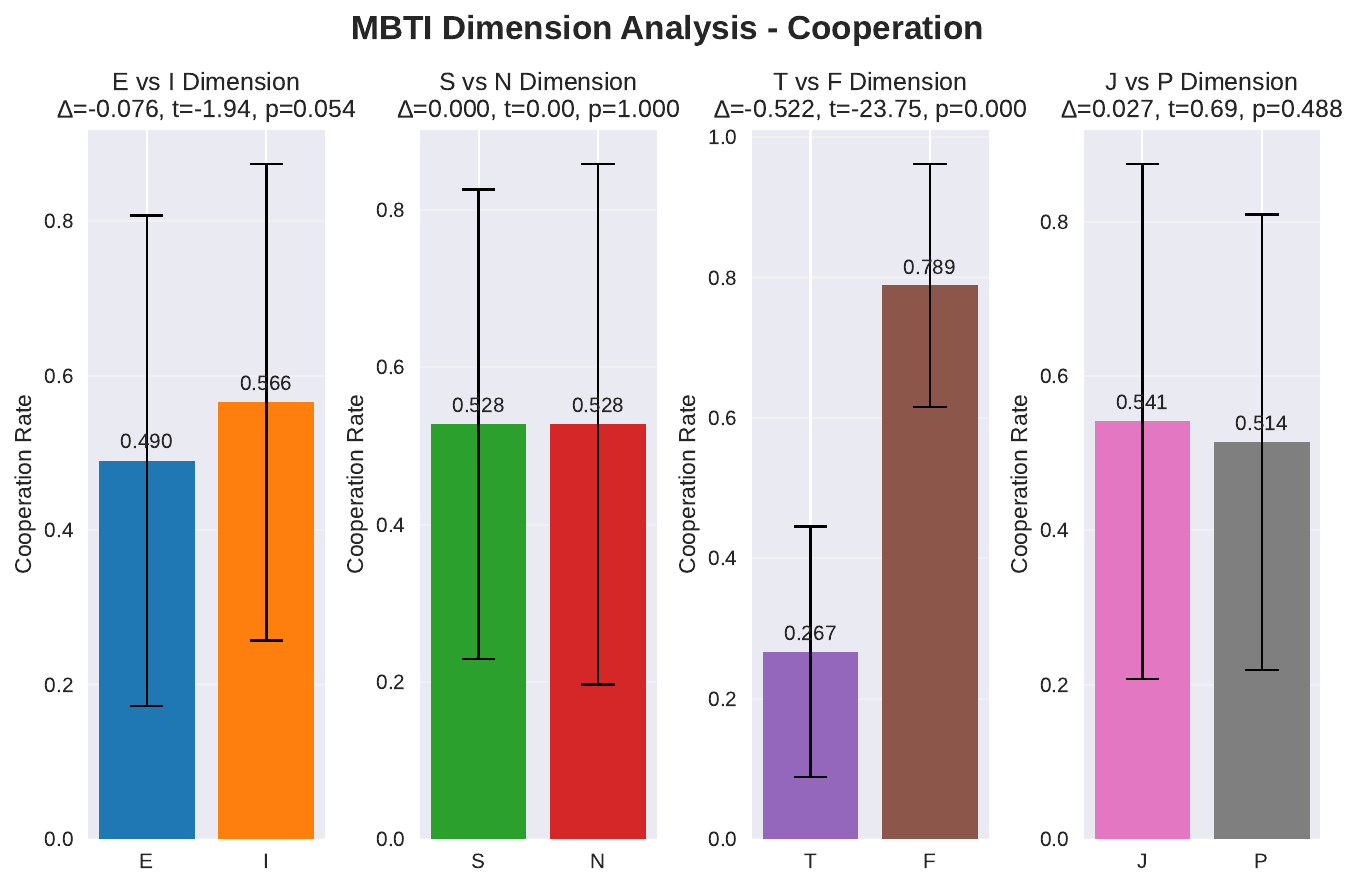}
        \hfill
        \includegraphics[width=0.49\linewidth]{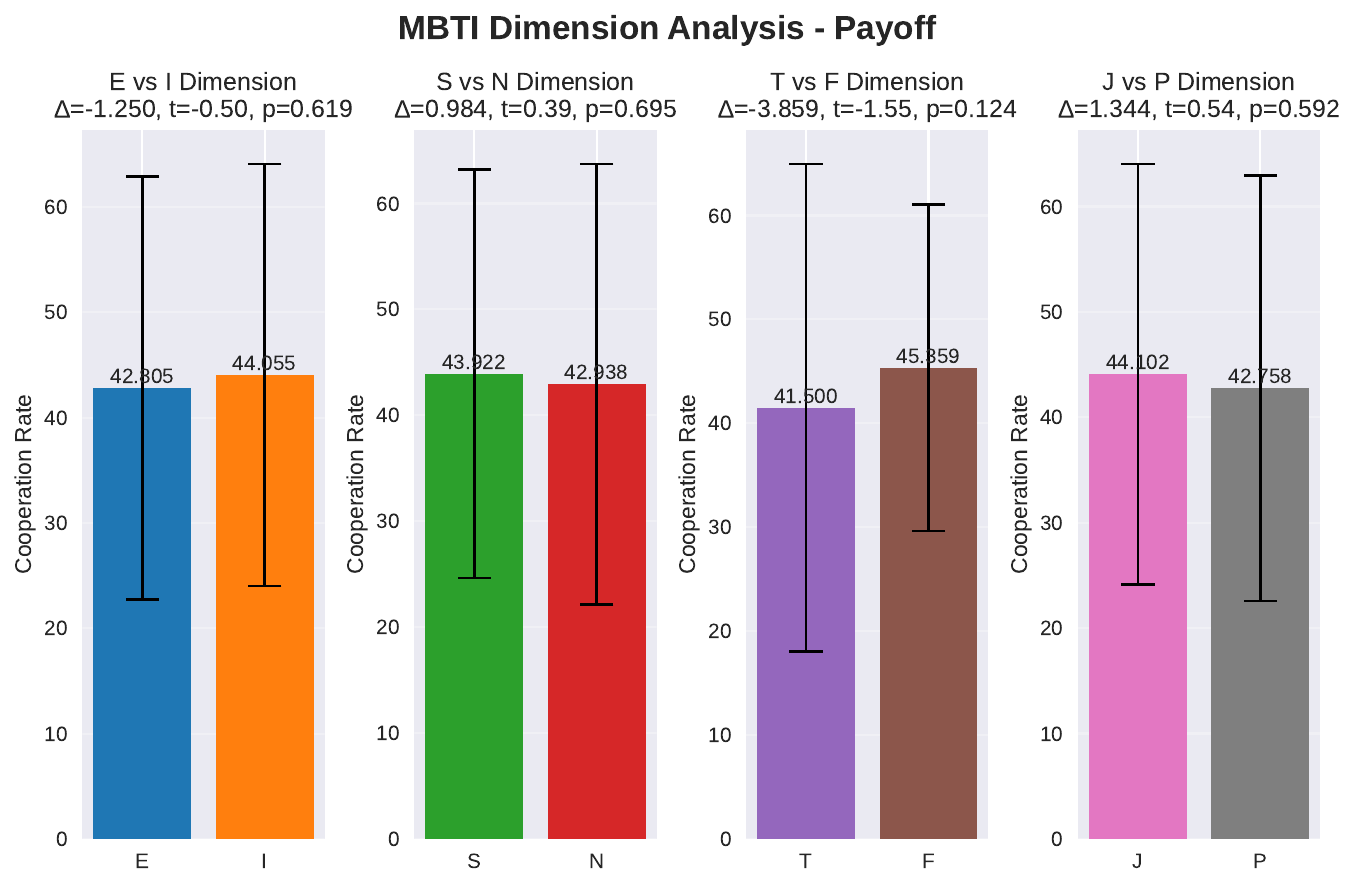}
    \end{minipage}
    \begin{minipage}{1.0\linewidth}
        \centering
        \includegraphics[width=0.49\linewidth]{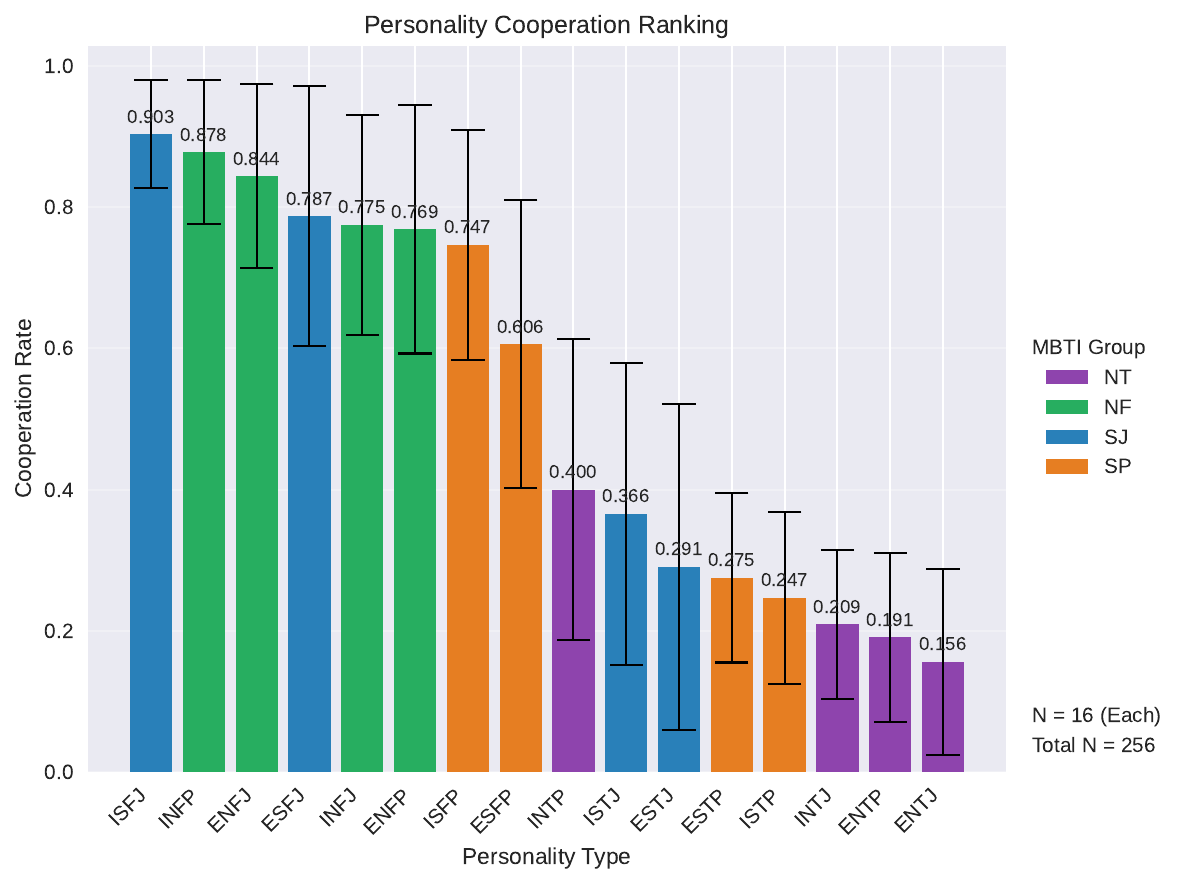}
        \hfill
        \includegraphics[width=0.49\linewidth]{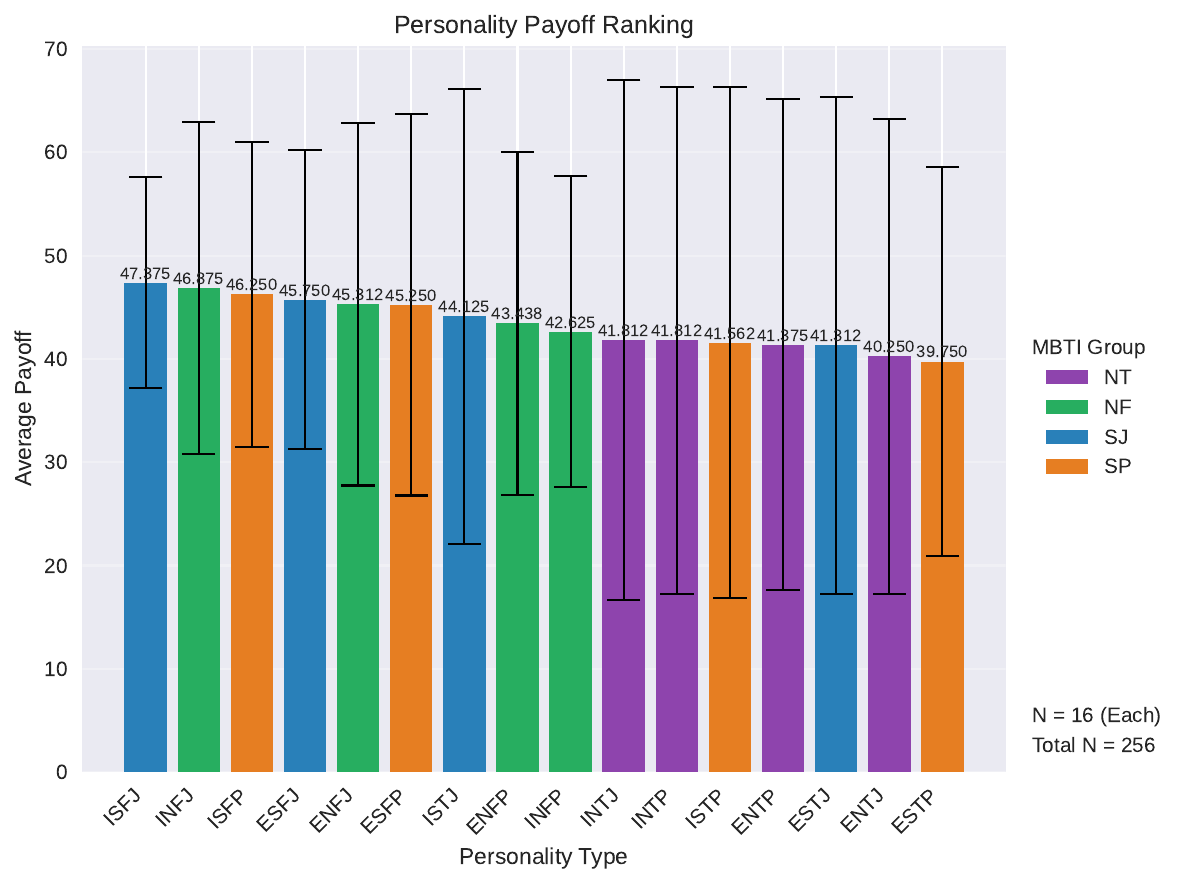}
    \end{minipage}
    \caption{Results of Pairwise Game using Snowdrift game. The F-types maintain higher cooperation rates, while payoffs don't have statistically significant differences. This indicates that personality-driven cooperation tendencies similar even in anti-coordination games, though payoff dynamics may differ from other social dilemmas.}
    \label{fig:snowdrift}
    \Description{Results of Pairwise Game using Snowdrift game.}
\end{figure}

As shown in ~\ref{fig:snowdrift}, in the Snowdrift game simulations, the overall average cooperation rate was $0.528$ (std: $0.314$), with an average payoff of $43.430$ (std: $19.991$). Interestingly, while Welch's t-test confirms that Feeling (F) types maintain significantly higher cooperation rates compared to Thinking (T) types ($p < 0.001$), there is no statistically significant difference in payoffs between the two groups ($p=0.124$). This suggests that in the anti-coordinative structure of Snowdrift, the exploitative advantage of T-types observed in the Prisoner's Dilemma is neutralized. The Kruskal-Wallis test corroborates these findings, showing significant differences across personality types for cooperation ($H=185.37$, $p<0.001$) but not for payoffs ($H=2.85$, $p=0.999$).

\subsubsection{Stag Hunt and Snowdrift in Network Games}
We further extended the Stag Hunt and Snowdrift games to networked interactions, using the same network topologies as in Section~\ref{sec:topology} (Regular, Small-World with $p=0.1$ and $p=0.5$, and Scale-Free with $m=2$). Each simulation involved $N=50$ agents playing 20 rounds of the respective game.

Table~\ref{tab:game_comparison_network} presents the comparative results. In both Stag Hunt and Snowdrift games, the Regular network topology consistently supports the highest levels of cooperation, mirroring the findings from the Prisoner's Dilemma. As randomness increases in the Small-World networks (from $p=0.1$ to $p=0.5$), cooperation rates decline, further supporting the ``Paradox of Connectivity'' hypothesis across different game structures. Scale-Free networks generally yield intermediate to lower cooperation levels.

We speculate that this may be because the high clustering in Regular networks facilitates stable cooperative clusters, which are beneficial in both coordination (Stag Hunt) and anti-coordination (Snowdrift) contexts. In contrast, the shortcuts introduced in Small-World and Scale-Free networks may disrupt these clusters, leading to reduced cooperation. It is worth noting that across all three games—despite their distinct dilemmas and Nash equilibria—the reward for mutual cooperation ($R$) consistently exceeds the punishment for mutual defection ($P$). Consequently, network topologies that successfully promote higher cooperation rates inherently result in higher collective payoffs.

\begin{table}[htbp]
    \caption{Cooperation and Payoff Across Topologies for Stag Hunt and Snowdrift}
    \label{tab:game_comparison_network}
    \centering
    \resizebox{\columnwidth}{!}{
        \begin{tabular}{lcccc}
            \toprule
            \multirow{2}{*}{\textbf{Topology}} & \multicolumn{2}{c}{\textbf{Stag Hunt}} & \multicolumn{2}{c}{\textbf{Snowdrift}}                                              \\
            \cmidrule(lr){2-3} \cmidrule(lr){4-5}
                                               & \textbf{Coop. Rate}                    & \textbf{Avg. Payoff}                   & \textbf{Coop. Rate} & \textbf{Avg. Payoff} \\
            \midrule
            Regular ($k=4$)                    & \textbf{0.593}                         & \textbf{14.15}                         & \textbf{0.568}      & \textbf{11.14}       \\
            SW ($p=0.1$)                       & 0.563                                  & 13.69                                  & 0.480               & 10.55                \\
            SW ($p=0.5$)                       & 0.518                                  & 13.03                                  & 0.440               & 10.13                \\
            Scale-Free ($m=2$)                 & 0.513                                  & 12.46                                  & 0.503               & 10.05                \\
            \bottomrule
        \end{tabular}
    }
\end{table}

\section{NetworkGames Framework}
To enable reproducible and scalable research in this domain, we developed \textbf{NetworkGames}\footnote{Due to anonymization for double-blind review, the link of code repository will be released upon acceptance.}, an open-source framework specifically designed for studying LLM agent behavior in network games. The framework addresses key challenges in computational social science research: reproducibility, scalability, and extensibility.

\textbf{Core Components.} NetworkGames consists of four primary functional modules:
\begin{itemize}
    \item \textbf{Personified Agents:} Each agent is powered by an LLM and endowed with an MBTI personality via systematic prompting that emphasizes core personality traits. The framework provides unified interface supporting OpenAI GPT models, Anthropic Claude, Google Gemini, and mock implementations for testing.
    \item \textbf{Game Environment:} We adopt the \textbf{Iterated Prisoner's Dilemma (IPD)} as the core interaction model and already include implementations for Stag Hunt (coordination games) and Snowdrift (chicken games). The framework supports extensible game implementations with configurable payoff matrices and parameters (round counts, memory windows, custom payoff structures), making it straightforward to switch to other social dilemmas or add new games.
    \item \textbf{Network Engine:} Supports the generation of various network structures (Regular, Small-World, Scale-Free, Random) with adjustable parameters such as node count, connection degree, and rewiring probability. The engine manages dynamic agent interactions and maintains comprehensive interaction histories throughout the simulation.
    \item \textbf{Visualization:} A rich visualization suite that generates heatmaps for dyadic interactions, personality ranking charts, network topology snapshots, and time-series plots for network evolution and cooperation rates, facilitating in-depth analysis of simulation dynamics.
\end{itemize}

\textbf{Technical Architecture.} The framework follows a modular design with clear separation of concerns, featuring:
\begin{itemize}
    \item \textbf{Experimental Control:} YAML-based configuration management with complete random seed control for reproducibility
    \item \textbf{Analysis Pipeline:} Integrated statistical analysis, visualization tools, and result export functionality
    \item \textbf{Extensibility:} Modular architecture facilitating easy extension with new personality models, game types, network topologies, and LLM providers
\end{itemize}

\textbf{Reproducibility Features.} NetworkGames prioritizes research reproducibility through deterministic random seed control, comprehensive logging of all LLM requests and responses, and standardized output formats that enable cross-study comparisons.

\end{document}